\documentclass[useAMS, usenatbib, fleqn]{mn2e}

\usepackage{times}
\usepackage{url}
\usepackage{amsmath}
\usepackage{graphicx}
\usepackage{subfig}
\usepackage{float}
\usepackage{caption}
\usepackage{amsfonts}
\usepackage{amssymb}
\usepackage{multirow}
\usepackage{color}
\usepackage[breaklinks, colorlinks, citecolor=blue, linkcolor=black, urlcolor=black]{hyperref}

\newcommand{\ang}{{\bmath{n}}}

\newcommand{\ellmax}{{\ell_{\rm max}}}
\newcommand{\mCl}{{\mathcal{C}_\ell}}
\newcommand{\Pl}{{\mat{P}^\ell}}

\newcommand{\npix}{{\rm N_{pix}}}
\newcommand{\nside}{{\rm N_{side}}}
\newcommand{\tr}{{\textrm{Tr}\ }}

\newcommand{\equ}[1]{\begin{equation}#1\end{equation}}

\newcommand{\mat}[1]{\mathbfss{#1}}
\newcommand{\est}[1]{\tilde{#1}}
\newcommand{\bra}{\langle}
\newcommand{\degree}{\ensuremath{^\circ}}
\newcommand{\ket}{\rangle}
\newcommand{\fsky}{f_{\rm sky}}
\renewcommand{\vec}[1]{\bmath{#1}}
\newcommand{\mpt}{\hspace*{-2pt}}

\usepackage{color}
\newcommand{\bl}[1]{{#1}}

\pagerange{\pageref{firstpage}--\pageref{lastpage}} \pubyear{2013}
\def\LaTeX{L\kern-.36em\raise.3ex\hbox{a}\kern-.15em
    T\kern-.1667em\lower.7ex\hbox{E}\kern-.125emX}

\title[The large-scale angular power spectrum in the presence of systematics:  a case study of SDSS quasars]
  {Estimating the large-scale angular power spectrum in the presence of systematics: a case study of Sloan Digital Sky Survey quasars}
\author[Leistedt et al]
  {Boris~Leistedt$^1$, Hiranya~V.~Peiris$^1$, Daniel~J.~Mortlock$^{2,3}$,  \newauthor Aur\'elien~Benoit-L\'evy$^1$ and Andrew~Pontzen$^{1,4,5}$ \\
  $^1$Department of Physics and Astronomy, University College London, London WC1E 6BT, U.K \\ 
  $^2$Astrophysics Group, Imperial College London, Blackett Laboratory, Prince Consort Road, London, SW7 2AZ, U.K\\
  $^3$Department of Mathematics, Imperial College London, London, SW7 2AZ, U.K.\\
  $^4$Oxford Astrophysics, Denys Wilkinson Building, Keble Road, Oxford, OX1 3RH, U.K.\\
  $^5$Balliol College, Broad Street, Oxford, OX1 3BJ, U.K.\\
  \hspace*{-2mm} Email: boris.leistedt.11@ucl.ac.uk}

\begin{document}

\maketitle 

\begin{abstract}
The angular power spectrum is a powerful statistic for analysing cosmological signals imprinted in the clustering of matter. However, current galaxy and quasar surveys cover limited portions of the sky, and are contaminated by systematics that can mimic cosmological signatures and jeopardise the interpretation of the measured power spectra. We provide a framework for obtaining unbiased estimates of the angular power spectra of large-scale structure surveys at the largest scales using quadratic estimators. The method is tested by analysing the 600 CMASS mock catalogues constructed by \cite{manera2012} for the Baryon Oscillation Spectroscopic Survey (BOSS). We then consider the \cite{Richards2008rqcat} catalogue of photometric quasars from the Sixth Data Release (DR6) of the Sloan Digital Sky Survey (SDSS), which is known to include significant stellar contamination and systematic uncertainties. Focusing on the sample of ultraviolet-excess (UVX) sources, we show that the excess clustering power present on the largest-scales can be largely mitigated by making use of improved sky masks and projecting out the modes corresponding to the principal systematics. In particular, we find that the sample of objects with photometric redshift $1.3 < \tilde{z}_p < 2.2$ exhibits no evidence of contamination when using our most conservative mask and mode projection. This indicates that any residual systematics are well within the statistical uncertainties. We conclude that, using our approach, this sample can be used for cosmological studies.
\end{abstract}

\section{Introduction}

The cosmic microwave background (CMB) and the large-scale structure (LSS) of galaxies contain a wealth of physical information that can be used to test models of the origin and evolution of the Universe. Both are well-described by correlated Gaussian random fields, and therefore can be characterised by two-point statistics (see, e.g., \mbox{\citealt{BJK98b, BJK98, Tegmark2002earlysdss}}). In particular, the angular power spectrum is a natural tool for  CMB data analysis, and has also proved useful for the study of the clustering properties of galaxy surveys in redshift bins. Such tomographic approaches will be essential for exploiting next generation surveys such as the Dark Energy Survey\footnote{\url{www.darkenergysurvey.org}} (DES), which will provide large photometric catalogues where the uncertainties on the redshift estimates complicate a full three-dimensional analysis. 

However, data unavoidably contain non-cosmological contributions, for example due to instrumental errors and systematic uncertainties. These contaminants result in additional correlations in the measured power spectra, and can compromise our interpretation of the observables if not correctly treated. In the context of galaxy surveys, observational systematics and calibration errors can result in extra clustering power over a wide range of scales (see e.g., \citealt{Huterer2012calibrationerrors, ross2011weights, ross2012systematics, Thomas2010lrgs1,Thomas2010lrgs2}). This proves especially problematic at the largest scales, since the corresponding modes need to be constrained from partial sky data. These modes are nonetheless crucial for testing early universe theories such as cosmological inflation (e.g., \citealt{Guth1981inflation, Linde1982inflation, AlbrechtSteinhardt1982inflation}), the standard paradigm for describing the origin of structure in the universe. Future galaxy surveys will be able to test this paradigm very precisely, particularly through the search for signatures of primordial non-Gaussianity (PNG). PNG creates a scale-dependent galaxy bias affecting the 2-point clustering properties of LSS tracers at the largest scales \citep{Dalal2008png, matarrese2008}. Hence, these scales, which can be strongly affected by systematics, require particularly careful treatment. 

Quasars, being highly biased tracers of LSS, are excellent candidates to study the scale- and redshift dependence of galaxy bias, for example to constrain PNG. However, since current spectroscopic samples have low number densities and cannot compete with PNG constraints derived from the CMB, one must resort to photometric quasar catalogues. The largest catalogues currently available are extracted from the Sloan Digital Sky Survey (SDSS), and were used to study PNG and the integrated Sachs-Wolfe effect (ISW) \citep{SlosarHirata2008, Xia2010sdssqsoctheta, Xia2011sdssqsocell, Giannantonio2006isw, Giannantonio2008isw}. However, these studies demonstrated the high sensitivity of the correlation functions to the sky masks under consideration, indicating the presence of significant levels of contamination by stars and calibration-related systematics. In particular, recent work by \cite{PullenHirata2012}, corroborated by \cite{Giannantonio2013png}, confirmed the high levels of contamination in the \cite{Richards2008rqcat} catalogue of SDSS photometric quasars, leading to concerns about the use of this sample for clustering measurements. In this work, we use sample reduction, masking and mode projection to identify a subset of objects in this catalogue that can be used for cosmological analyses. We concentrate on the main systematics found by previous studies, and analyse their impact on the clustering measurements through auto- and cross-correlation of redshift subsamples with each other, and with templates of the systematics. \bl{Note that cross-correlating with external data can also prove useful in identifying and mitigating the systematics in quasar samples.}

When analysing photometric catalogues in redshift bins, the theory power spectrum predictions require precise estimates of the redshift distributions, which are compromised by the large uncertainties of the photometric redshifts. This issue is critical for photometric quasars, since their redshift estimates are significantly more uncertain than for other types of galaxies and include a significant fraction of catastrophic failures. We investigate the use of spectroscopic catalogues for calculating robust and unbiased redshift distribution estimates for the photometrically-selected quasar subsamples.

In addition to data quality and modelling issues, various methodological issues arise when estimating the power spectrum of a galaxy survey for comparison with theory. The pseudo-spectrum \citep{WHG00, HGN02} and quadratic maximum likelihood \citep{Teg97, BJK98} estimators were developed to measure the power spectrum in the presence of sky cuts. However, numerous technical subtleties and constraints due to pixelisation or limited computer resources are implicit in these estimators, and can create significant biases if not handled carefully, as concluded by several studies on the CMB (see e.g., \citealt{Efsta2003, Eriksen2007wmap3reanalysis, PP10, Copi2011smoothingbias}). This paper aims to clarify these technicalities in the context of galaxy surveys.
 
This article is organised as follows. In Sec.~2 we define and illustrate the properties of quadratic power spectrum estimators, and demonstrate their validity by applying them to a set of mock catalogues.  In Sec.~3 we turn to the \cite{Richards2008rqcat} catalogue of SDSS photometric quasars. We present our data samples, redshift distribution estimates, masks, and power spectrum measurements, and discuss the impact of the main systematics on these measurements. Our conclusions are presented in Sec.~4. Further technical details on smoothing and masking rules, Karhunen-Lo\`eve compression and $\chi^2$ measures are contained in appendices.

\section{Theory and methods}

\subsection{Background}\label{sec:background}

We consider a real signal $x(\ang)$ on the unit sphere $S^2$, equivalently described in terms of its spherical harmonic coefficients $\{x_{\ell m}\}$ with $\ell \in \mathbb{N}$ and $m \in \{-\ell,\dots, \ell\}$. The angular power spectrum $\{ \mCl \}$ of $x$ is defined as 
\equ{
	\mCl  \ = \   \sum_{m = -\ell}^{\ell} \frac{ |x_{\ell m}|^2 }{2\ell + 1} \label{harmps},
}	
and corresponds to the average power in fluctuations on scales of order $180/\ell$ degrees on the sphere. Assuming that $x$ is a realisation of an underlying random field denoted by $X$, the power spectrum of $x$ can be viewed as a compression technique, and used to perform statistical inference on physical models of $X$. In particular, this compression is lossless if $X$ is an isotropic Gaussian random field, and the power spectrum is then a sufficient statistic containing all the relevant information in the realisation $x$. Moreover, the `observed' power spectrum $ \mCl $ is a realisation of a `theory' power spectrum $ C_\ell$ that fully characterises the field of interest $X$. The variance of the former, known as cosmic variance, depends on the number of modes on the sky and is given by
\equ{
	{ {\rm Var}(\mCl)  }{}= \frac{2C^2_\ell}{2\ell+1} \label{cosmicvar}.
}	
In practice, real data contain a finite amount of information, and the continuous signal $x$ is observed at finite resolution on the sphere. In the context of LSS surveys, galaxy catalogues are usually constructed from raw imaging data and then reduced into pixelised overdensity maps $\vec{x} = (x_0, \dots, x_{\npix-1})$ where $x_i = x(\ang_i)$ and $\ang_i$ is the centre of the $i$th pixel on the sphere. More details on the construction of such maps from the source number counts will be given in Sec.~\ref{sec:modeproj}. The average correlation between the pixels, namely the pixel-pixel covariance matrix, depends on the theory power spectrum through a Legendre expansion, i.e.,
\equ{
	\mat{S} = \bra \vec{x}\vec{x}^t \ket = \sum_\ell  C_\ell \Pl, \label{signalcovar}
}
where $(\Pl)_{ij} =   ({2\ell+1})/{4\pi}\ P_\ell(\ang_i\cdot\ang_j)$ is a useful matrix notation \bl{\citep{Teg97}}. A quadratic estimator for the power spectrum of full-sky pixelised data is given by the projection of the data onto the Legendre matrices, i.e.,
\equ{
	\mCl = \vec{x}^t \Pl \vec{x}, \label{obsps}
}	
which is the pixel-space equivalent of Eq.~(\ref{harmps}) \bl{(see discussion and references in \citealt{PP10})}.

In this section we have used an arbitrary equal-area pixelisation scheme, but henceforth we will adopt the \textsc{healpix} conventions \citep{healpix1}. In defining Eq.~(\ref{obsps}), we only considered full sky coverage. This assumption will be relaxed in the next section. We also implicitly assumed that the power spectrum of the pixelised map $\vec{x}$ was equal to that of the continuous signal $x$. This approximation is only true at high resolution when the pixel size is small compared with $180/\ell$, and the integrals in the spherical harmonics and Legendre transforms are correctly approximated by matrix multiplications through quadrature, as in Eq.~(\ref{obsps}). The bias induced by pixelisation as a function of $\ell$ is critical for low-resolution power spectrum estimation, and needs to be corrected.  This issue is investigated in Appendix~\ref{app:bandlimitandsmoothing}, and the following sections will assume that the relevant corrections have been applied.

\subsection{Partial sky coverage and quadratic estimators}\label{sec:estimators}
  
Due to contamination or inaccessibility of certain regions of the sky, most cosmological applications involve signals that only cover a portion of the sphere. The power spectrum must then be calculated from a cut-sky map $\tilde{\vec{x}}$. From a theoretical perspective, the latter can be viewed as the restriction of the full sky map $\vec{x}$ using a binary mask $\vec{m} = (m_0, \dots, m_{\npix - 1})$, such that $m_i=0$ for masked pixels and $m_i=1$ elsewhere. Masked/unmasked vectors or matrices are related to each other through an operator $\mat{K}$, a diagonal matrix such that $(\mat{K})_{ij}=m_i\delta_{ij}$ (with $\delta_{ij}$ the Kronecker delta) removing pixels that lie inside the mask\footnote{Hence $\tilde{\vec{b}} = \mat{K}\vec{b}$ for any data vector $b$, while for any matrix $\mat{B}$ we write $\tilde{\mat{B}}=\mat{K}\mat{B}\mat{K}$ implicitly taking advantage of the  property $\mat{K}^{t}=\mat{K}$. }. In what follows, the addition of a tilde will represent cut-sky quantities. 

In the presence of partial sky coverage, applying Eq.~(\ref{obsps}) on the cut-sky map $\tilde{\vec{x}}$ leads to a cut-sky power spectrum $\{ \tilde{\mathcal{C}}_\ell \}$ that considers the zones inside the mask as \textit{data}, i.e., as pixels with $x(\ang) = 0$. Consequently $\tilde{\mathcal{C}}_\ell$ differs from the quantity of interest $\mCl$ and is not a realisation of the underlying theory spectrum $C_\ell$. Inverting this effect involves deconvolving the effect of the mask from the observed power spectrum $\{ \tilde{\mathcal{C}}_\ell \}$, leading to the definition of the `pseudo-spectrum' (\textrm{PCL}) estimator (\citealt{WHG00, HGN02, Efsta2004, BCT05}),
\equ{
	\hat{C}_\ell^{\textrm{PCL}} = \sum_{\ell'} (\mat{M}^{-1})_{\ell \ell'} \tilde{\mathcal{C}}_{\ell'},
}
where $\tilde{\mathcal{C}}_{\ell'} = \vec{\tilde{x}}^t \tilde{\mat{P}}^\ell  \vec{\tilde{x}}$ are the cut-sky estimates. The coupling matrix is defined as
\equ{
	(\mat{M})_{\ell \ell'}  =  \tr \tilde{\mat{P}}^\ell  \tilde{\mat{P}}^{\ell'},
}
and is a function of the mask only.
Since the variance of each $\ell$-mode depends on number of times it is observed, the minimum variance, namely the cosmic variance presented in Eq.~(\ref{cosmicvar}), is achieved on the full sky only. Partial sky coverage decreases the number of observed modes, and the variance of the PCL estimates in the absence of noise is approximately 
\equ{
	{\rm Var}(\hat{C}_\ell^{\textrm{PCL}}) \approx  \frac{1}{\fsky}{\rm Var}(\mCl)    \label{varapprox}.
}
Here, $\fsky = \sum_i m_i / \npix$ is the fraction of the sky covered by the mask, with $\fsky=1$ corresponding to full sky coverage. Equation~(\ref{varapprox}) is a good approximation for small scale modes, which remain numerous after masking.  The exact expression for the variance in the Gaussian framework is given in Eq.~(\ref{varianceestimates}), and must be used for low-$\ell$ modes since they are sensitive to the shape of the mask.

The PCL approach is simply an inversion of the mask and does not attempt to minimise the loss of information caused by the decrease in the number of observed modes. In fact, it is well known that the PCL estimates are only optimal (i.e., unbiased, minimum variance estimates) for a flat power spectrum (see e.g., \citealt{Efsta2004} and \citealt{PP10}). This equivalence will prove useful in the context of galaxy surveys, as we shall see in the next sections. To recall the definition of the optimal estimator, we consider the generic class of quadratic estimators of the form
\equ{
	\hat{C}_\ell = \vec{\tilde{x}}^t \mat{E}^\ell \vec{\tilde{x}}.
}	 
In this formalism, the \textrm{PCL} estimator reads
\equ{
	\mat{E}^\ell_{\textrm{PCL}} =  \sum_{\ell'} \ (\mat{M}^{-1})_{\ell \ell'}^{\textrm{PCL}} \ \tilde{\mat{P}}^{\ell'} , \label{pclel}
} 
with the coupling matrix $ (\mat{M})^{\textrm{PCL}}_{\ell \ell'} = \tr \tilde{\mat{P}}^{\ell}\tilde{\mat{P}}^{\ell'}$\hspace*{-2mm}. In the Gaussian case, the expected value of the generic quadratic estimator is  given by 
\equ{
	\bra \hat{C}_\ell \ket =\tr \tilde{\mat{C}} \mat{E}^\ell,
}
 and its variance by
\equ{
	\mat{V}_{\ell \ell'} = \bra \hat{C}_\ell \hat{C}_{\ell'} \ket - \bra \hat{C}_\ell \ket \bra \hat{C}_{\ell'} \ket = 2\tr \tilde{\mat{C}} {\mat{E}}^\ell \tilde{\mat{C}} {\mat{E}}^{\ell'}.\label{varianceestimates}
}
As a result, in the presence of sky cuts the uncertainties on the power spectrum estimates are typically significantly correlated. Uncorrelated error bars can be obtained by diagonalising the covariance matrix and using the resulting rotation matrix to transform the power spectrum estimates and the theory predictions \citep{Teg97, Tegmark2002earlysdss}. 

In the previous equations, $\tilde{\mat{C}}$ denotes the cut-sky pixel-pixel covariance matrix, which can be modelled as the superposition of a signal part $\tilde{\mat{S}}$ calculated with a theory prior $\{ C_\ell \}$ and noise, i.e.,
\equ{
	\tilde{\mat{C}} = \bra \tilde{\vec{x}}\tilde{\vec{x}}^t \ket =\tilde{\mat{S}} + \tilde{\mat{N}}. \label{tildecovarmatrix}
} 
The pixel-pixel covariance matrix must also incorporate any additional signal present in the data, such as the systematics, as detailed in Sec.~\ref{sec:modeproj}.

The minimum variance estimator in the Gaussian framework, first introduced in \cite{Teg97}, is the so-called quadratic maximum likelihood (QML) estimator. The latter reads
\equ{
	\mat{E}^\ell_{\textrm{QML}} =   \sum_{\ell'} \ (\mat{M}^{-1})^{\textrm{QML}}_{\ell \ell'} \  \frac{1}{2} \tilde{\mat{C}}^{-1} \tilde{\mat{P}}^{\ell'} \tilde{\mat{C}}^{-1} \label{qmlel},
}
and uses the deconvolution matrix
\equ{
	 (\mat{M})^{\textrm{QML}}_{\ell \ell'} =\frac{1}{2} \tr  \tilde{\mat{C}}^{-1} \tilde{\mat{P}}^{\ell} \tilde{\mat{C}}^{-1}  \tilde{\mat{P}}^{\ell'}  \label{fisherqml} .
}
In the Gaussian, isotropic case, QML is a lossless estimator that recovers all the relevant information contained in the data. The deconvolution and covariance matrices of the estimates are then equal to the inverse of the Fisher information matrix. While being no longer theoretically optimal for anisotropic theories, \cite{PP10} showed that QML remained superior to PCL.

PCL and QML can be contrasted in terms of the computational complexity and quality of the final power spectrum estimates. They both depend on the mask, but QML additionally requires an accurate model of the pixel-pixel covariance matrix. This model requires priors on the fiducial theory $\{ C_\ell \}$ and on the additional correlations present in the data, such as noise and systematics.  Although the pseudo-spectrum estimator does not explicitly use such priors, it is equivalent to a maximum likelihood analysis when a flat spectrum is assumed in place of a more motivated choice for the pixel-pixel covariance matrix \citep{Efsta2004, PP10}. As a result, PCL yields nearly optimal estimates when the power spectrum is close to flat and with no anisotropic contributions. Moreover, the sensitivity to the shape of the spectrum decreases at small scales as the number of observed modes increases. A simple inversion of the mask then maximises the likelihood function and the variance of PCL reaches its minimum, namely the inverse of the Fisher matrix \citep{Efsta2004, Efsta2006}, i.e.,
\equ{
	\mat{V}_{\ell \ell'}^{\textrm{PCL}} \approx \frac{2\mathcal{C}_\ell \mathcal{C}_{\ell'}}{2\ell + 1} (\mat{M}^{-1})^{\textrm{PCL}}_{\ell \ell'} \approx   (\mat{M}^{-1})^{\textrm{QML}}_{\ell \ell'} = \mat{V}_{\ell \ell'}^{\textrm{QML}}.
} 
Note that this result is only valid in the regime where the signal dominates in the covariance matrix. 

\begin{figure}\centering
\subfloat[CMB mask]{\includegraphics[trim = 0.1cm 2cm 0.1cm 1.8cm, clip, width=4.cm]{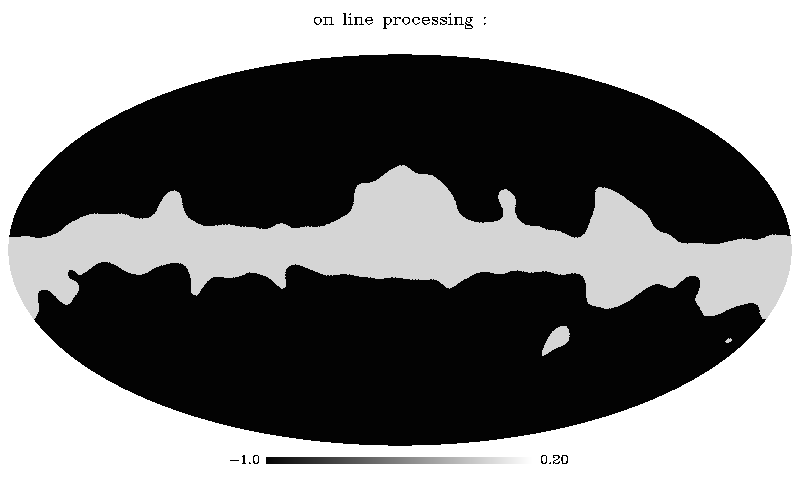}}
\subfloat[LSS mask]{\includegraphics[trim = 0.1cm 2cm 0.1cm 1.8cm, clip, width=4.cm]{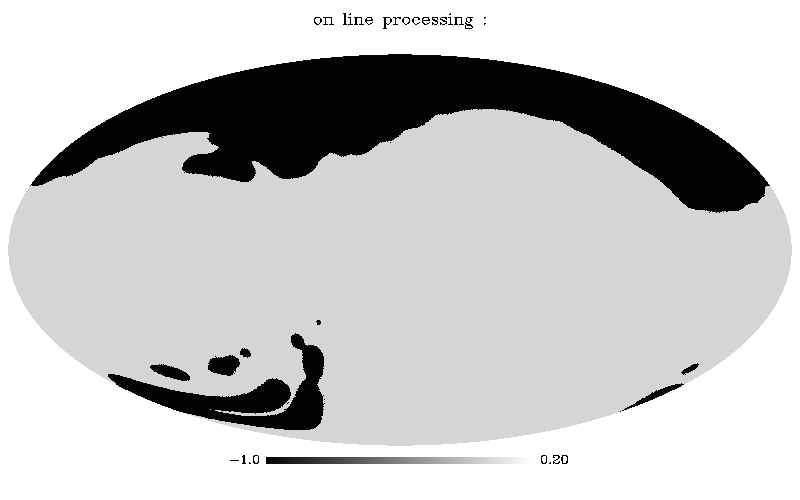}}
\caption{Fiducial CMB and galaxy survey masks used to calculate the covariance matrices of Fig.~\ref{fig:covmatrices}. These masks are in Galactic coordinates and approximate the WMAP KQ85 mask and the SDSS DR6 sky coverage respectively.}
\label{fig:masks} 
\end{figure}

\begin{figure}\centering
\setlength{\unitlength}{.5in}
\begin{picture}(9,2.5)(0,0)
\put(0.1,0.2){\subfloat[CMB spectrum]{\includegraphics[trim = 3.4cm 13.2cm 3.2cm 4.1cm, clip, width=4.1cm]{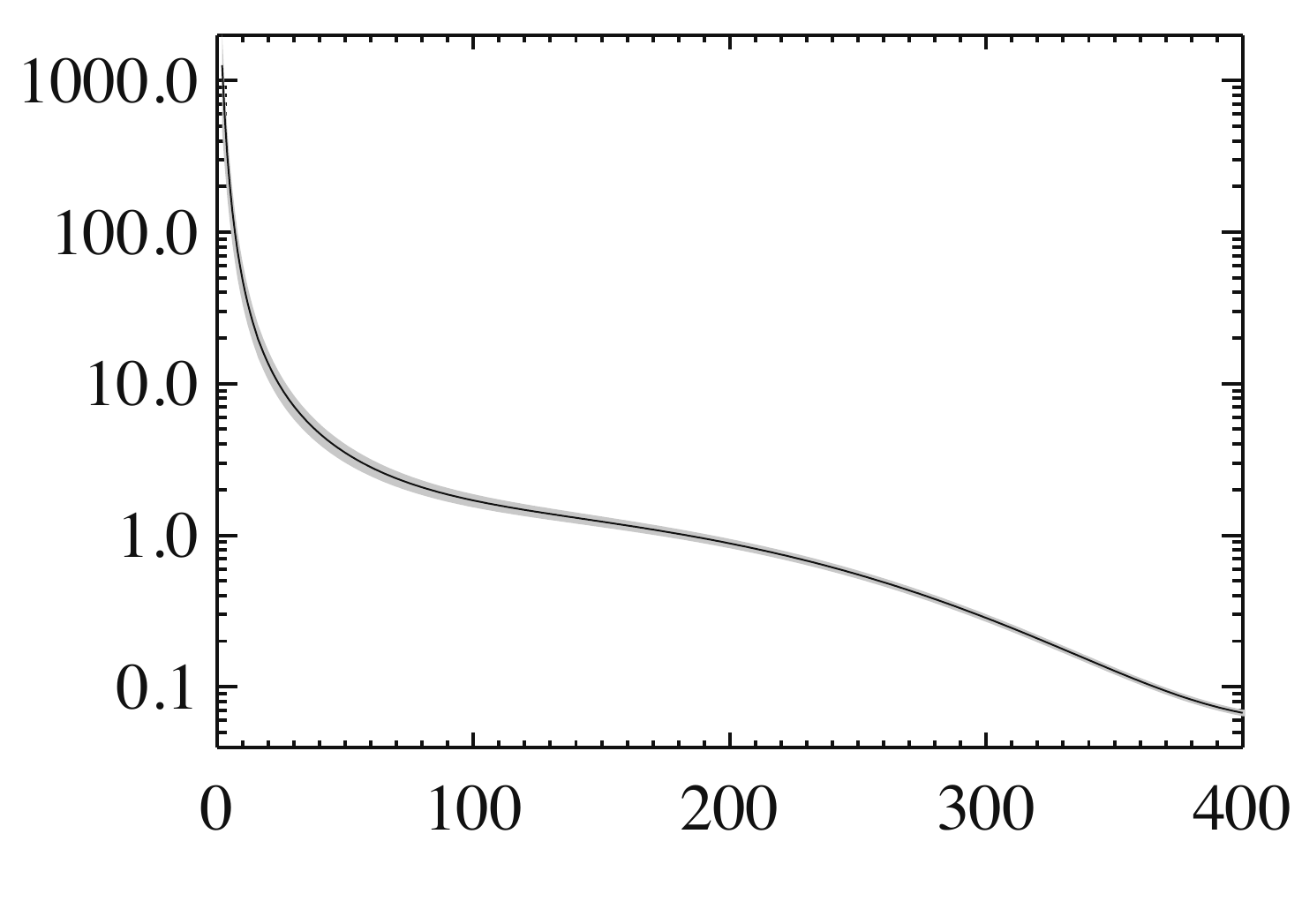}}}
\put(3.6,0.2){\subfloat[CMASS spectrum]{\includegraphics[trim = 4.5cm 13.2cm 3.1cm 4.1cm, clip, width=3.8cm]{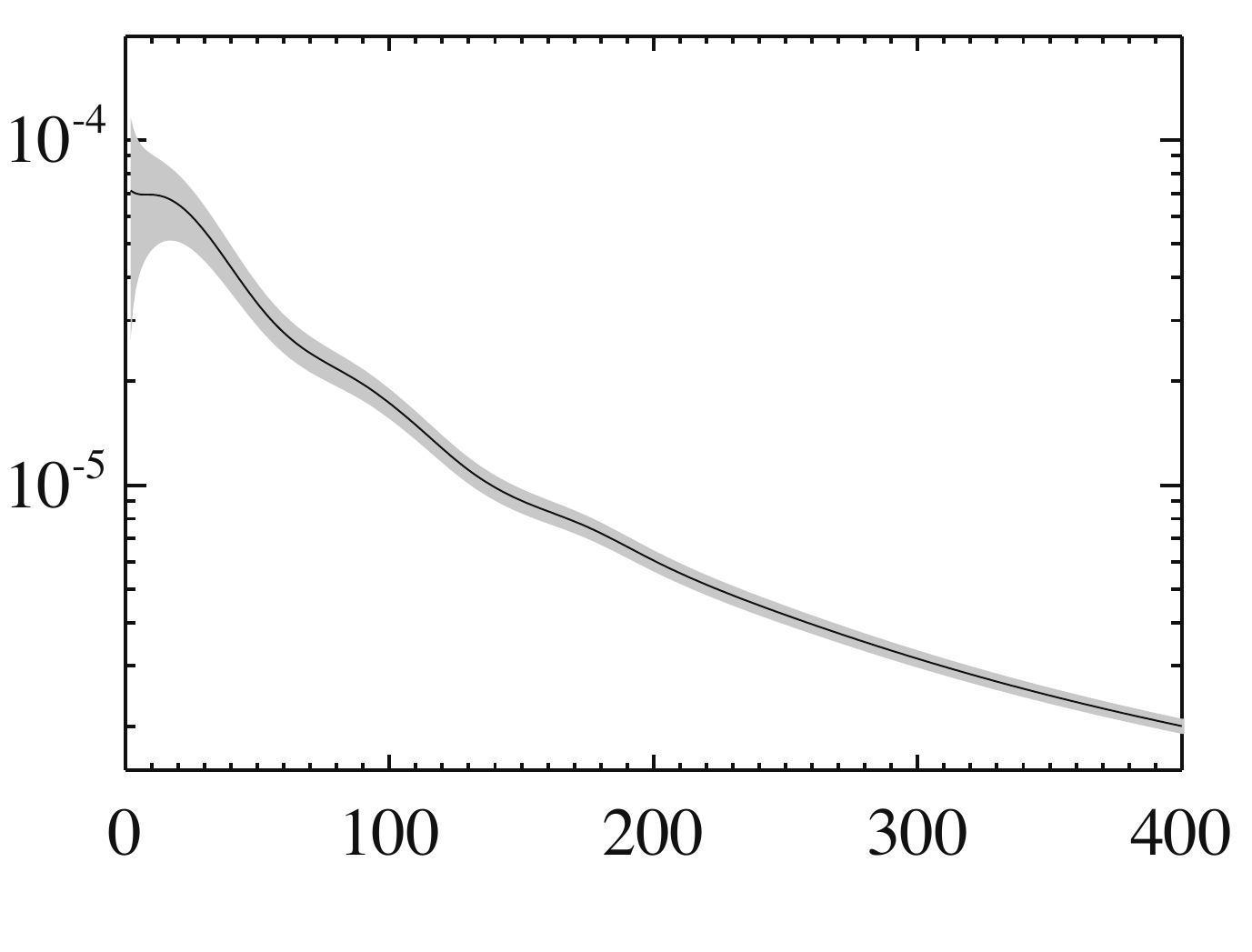}}}
\put(1.4,0.2){ Multipole $\ell$}
\put(4.7,0.2){ Multipole $\ell$}
\put(-0.05,1.0){\rotatebox{90}{ $C_\ell\ [\mu {\rm K}^2]$ }}
\put(3.4,1.3){\rotatebox{90}{ $C_\ell$}}
\end{picture}
\caption{Fiducial CMB and CMASS angular power spectra used to calculate the covariance matrices of Fig.~\ref{fig:covmatrices}. The grey bands show the cosmic variance.} 
\label{fig:theospectra}
\end{figure}
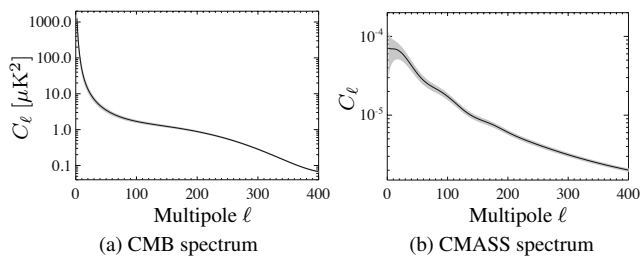

\begin{figure}\centering
\setlength{\unitlength}{.5in}
\begin{picture}(9,9.5)(0,0)
\put(1.9,9.3){ CMB mask \& CMB spectrum}
\put(0.8,9.1){ PCL estimates}
\put(4.5,9.1){ QML estimates}
\put(0.0,6.4){\includegraphics[trim = 2.4cm 13.5cm 5.2cm 1cm, clip, height=3.4cm]{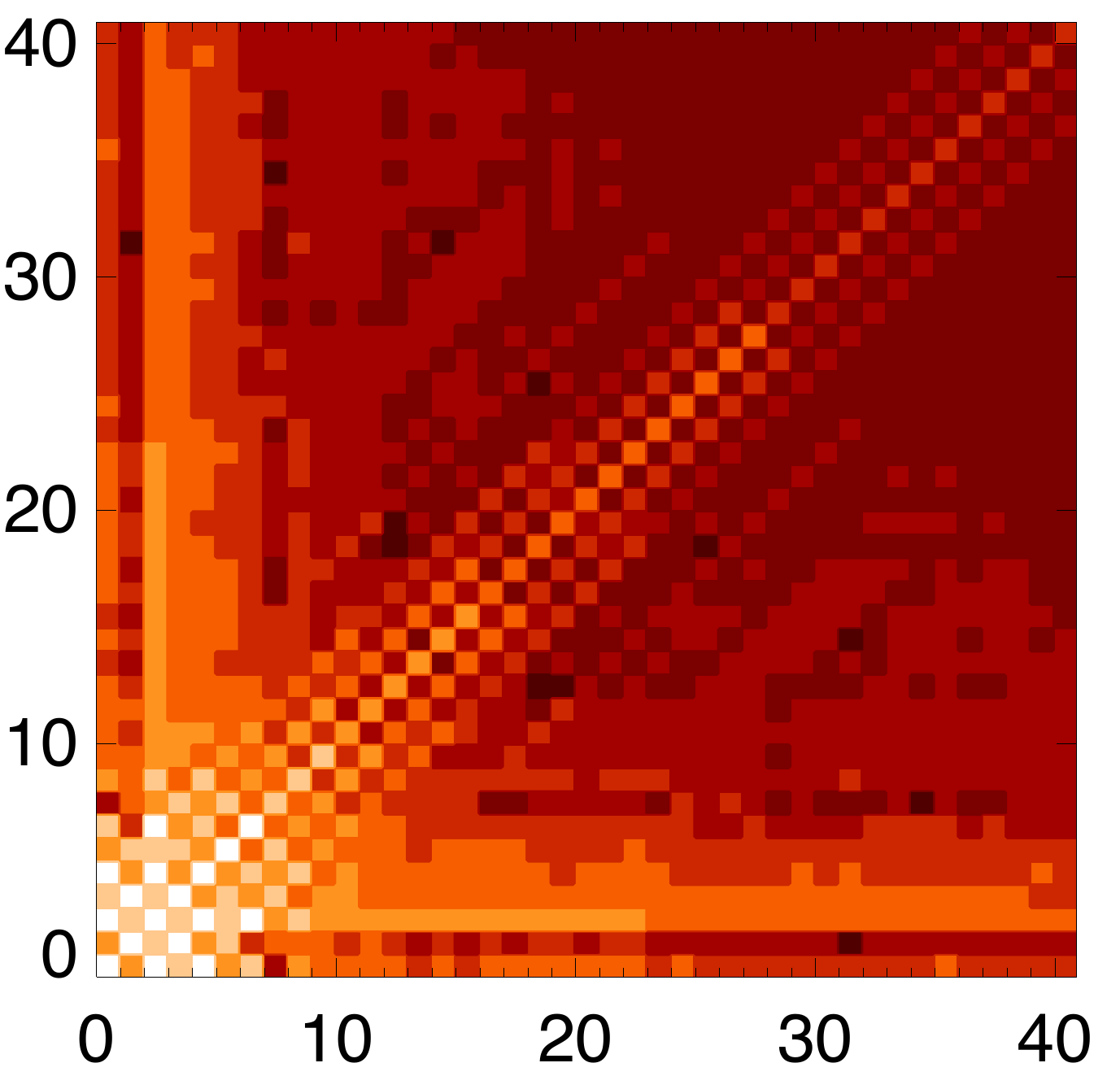} \
\includegraphics[trim = 16.5cm 13.5cm 1.6cm 1cm, clip, height=3.4cm]{pics/cmb_kq85_sm_covariance_16_1_47_1_PCL.pdf}
\includegraphics[trim = 2.3cm 13.5cm 5.5cm 1cm, clip, height=3.4cm]{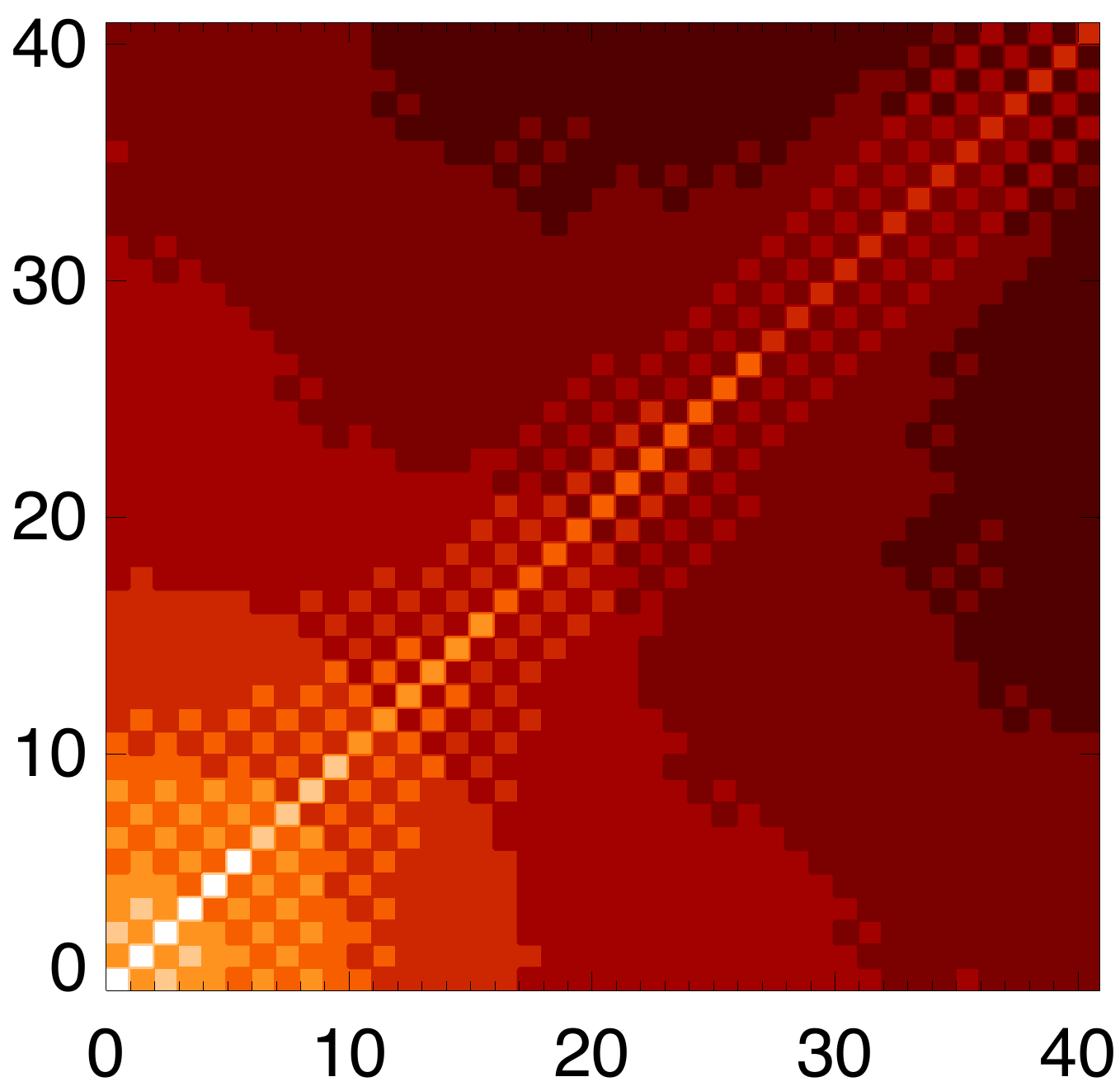} }
\put(2.0,6.1){ LSS mask \& CMB spectrum}
\put(0.32,6.35){$\ell$}
\put(4.00,6.35){$\ell$}
\put(0.05,6.85){$\ell$}
\put(3.73,6.85){$\ell$}
\put(0.8,5.9){ PCL estimates}
\put(4.5,5.9){ QML estimates}
\put(0.0,3.2){\includegraphics[trim = 2.4cm 13.5cm 5.2cm 1cm, clip, height=3.4cm]{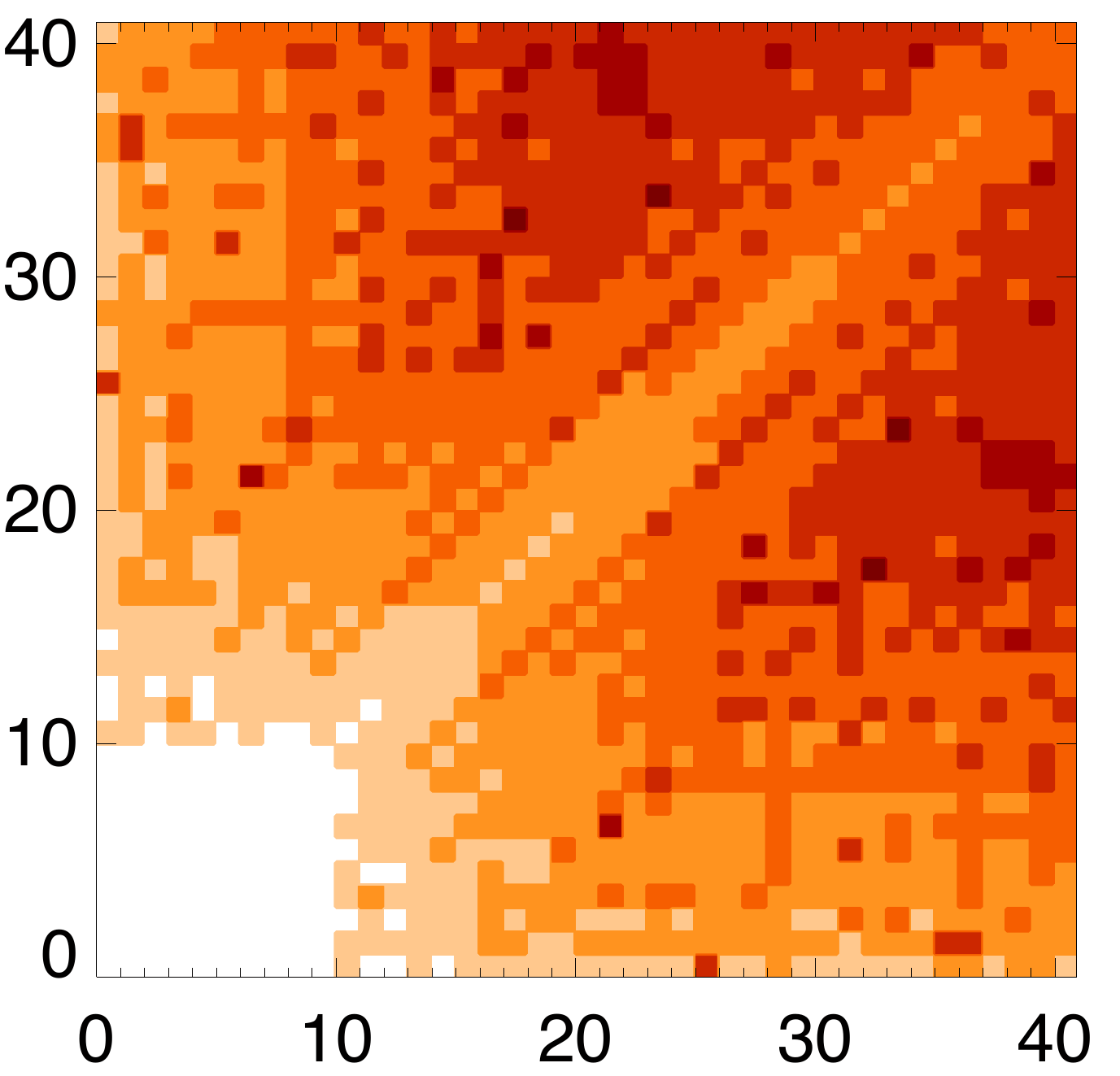} \
\includegraphics[trim = 16.5cm 13.5cm 1.6cm 1cm, clip, height=3.4cm]{pics/cmb_qso_enl_covariance_32_1_95_1_PCL.pdf}
\includegraphics[trim = 2.3cm 13.5cm 5.5cm 1cm, clip, height=3.4cm]{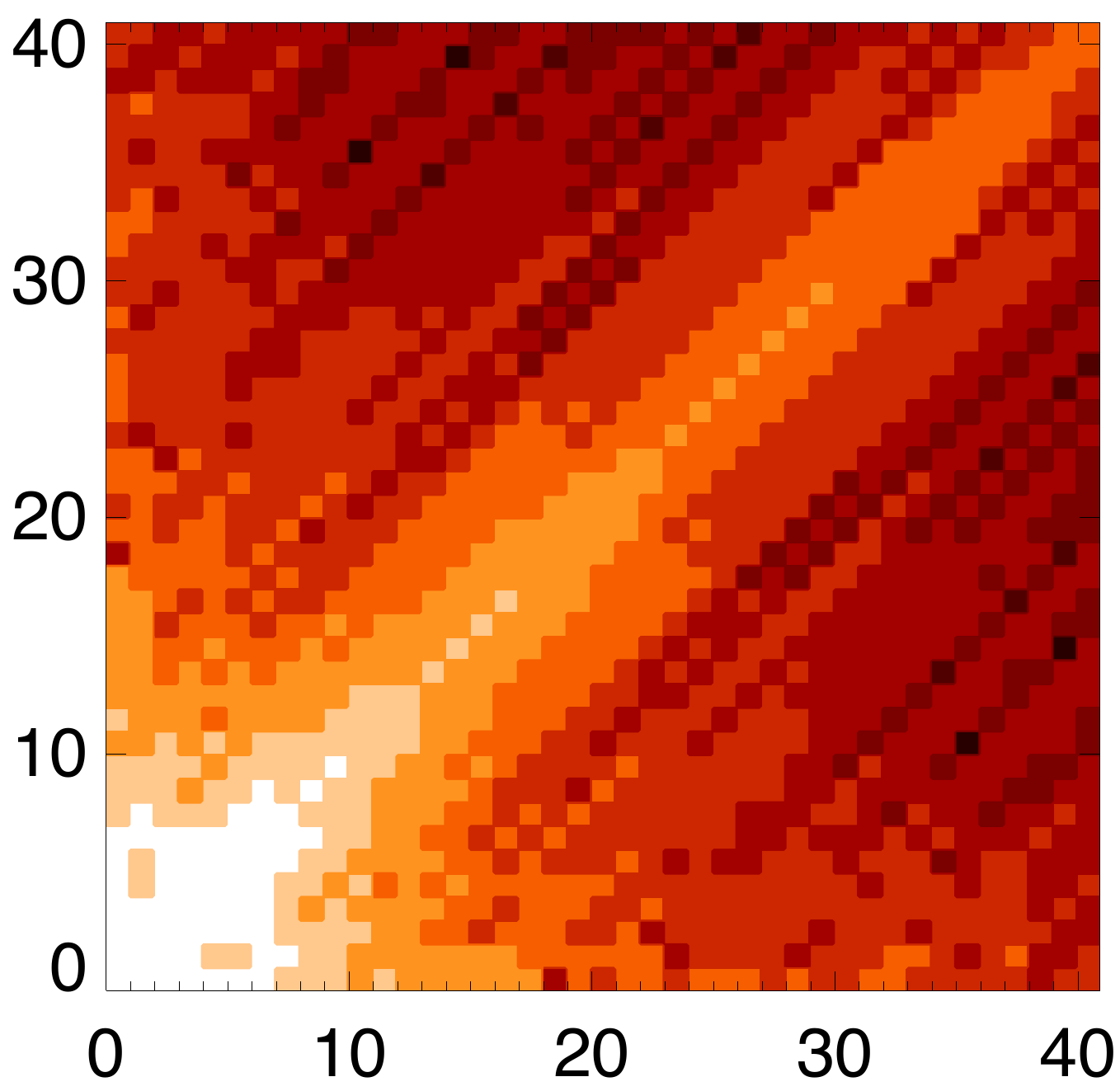}	}
\put(2.1,2.9){ LSS mask \& LSS spectrum}
\put(0.32,3.15){$\ell$}
\put(4.00,3.15){$\ell$}
\put(0.05,3.65){$\ell$}
\put(3.73,3.65){$\ell$}
\put(0.8,2.7){ PCL estimates}
\put(4.5,2.7){ QML estimates}
\put(0.0,0.0){\includegraphics[trim = 2.4cm 13.5cm 5.2cm 1cm, clip, height=3.4cm]{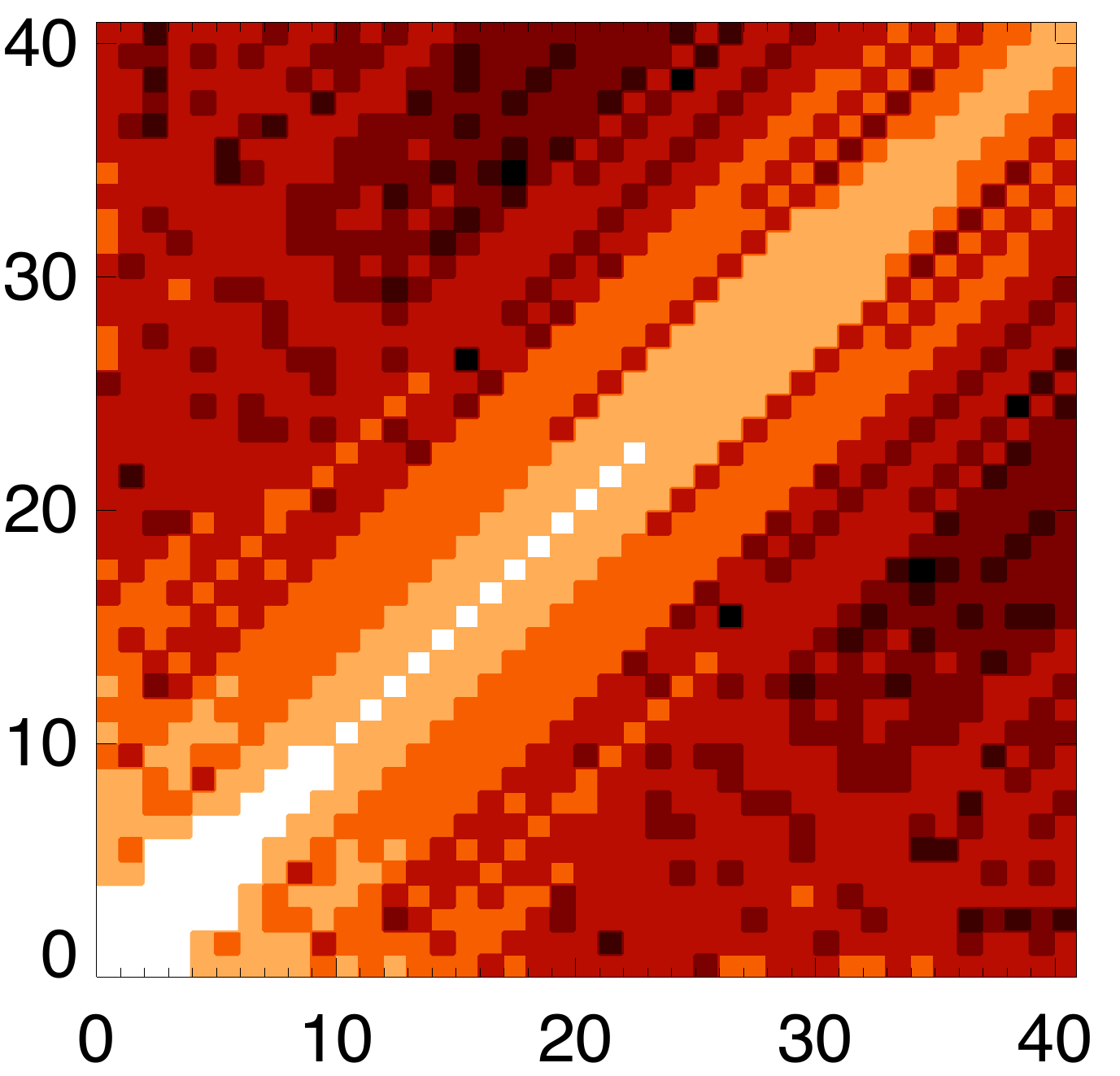} \
\includegraphics[trim = 16.5cm 13.5cm 1.6cm 1cm, clip, height=3.4cm]{pics/cmassbias2_qso_enl_covariance_32_1_95_1_PCL.pdf}
\includegraphics[trim = 2.3cm 13.5cm 5.5cm 1cm, clip, height=3.4cm]{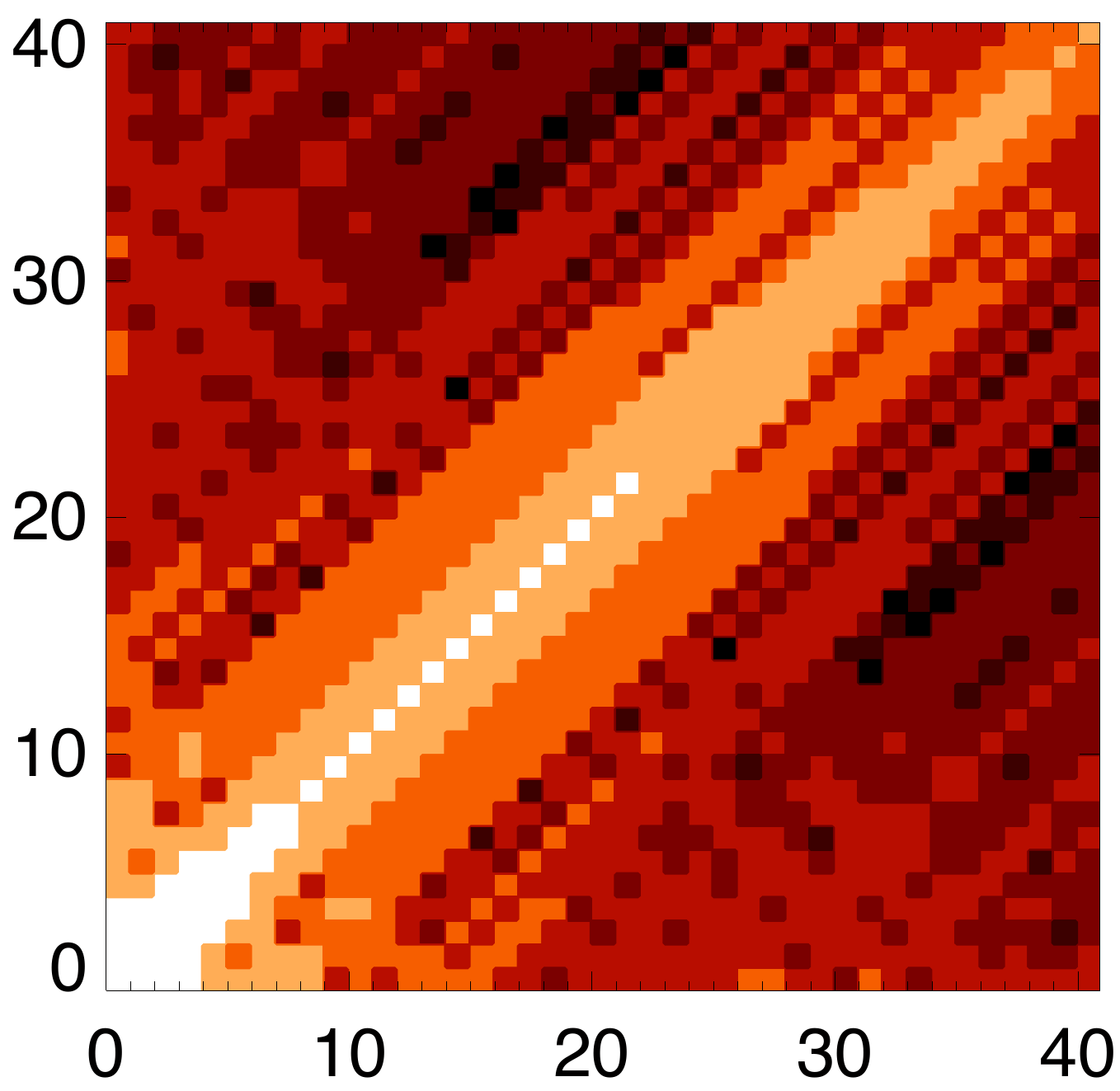}}
\put(0.32,-0.05){$\ell$}
\put(4.00,-0.05){$\ell$}
\put(0.05,0.45){$\ell$}
\put(3.73,0.45){$\ell$}
\end{picture}
\caption{Covariance matrices $|V_{\ell \ell'}|$ of the PCL and QML estimates calculated for the CMB and LSS spectra and masks in Figs.~\ref{fig:masks} and \ref{fig:theospectra} in the absence of noise and systematics. For a CMB spectrum, the QML variances (top and middle right) are smaller and less correlated than PCL variances (top and middle left), and the resulting PCL estimates are significantly suboptimal compared to QML. For the LSS spectrum, which is considerably flatter, this distinction is much less pronounced (bottom left and right panels), as expected by the reduction of QML to PCL in the case of a flat spectrum.}
\label{fig:covmatrices} 
\end{figure}

To illustrate the contrast between the PCL and QML estimates, we calculated and compared their covariance matrices in three realistic settings of interest, involving typical masks and spectra of CMB and galaxy catalogues. The masks are shown in Fig.~\ref{fig:masks}, the theory power spectra in Fig.~\ref{fig:theospectra}, and the resulting covariance matrices in Fig.~\ref{fig:covmatrices}. For the CMB, we considered the KQ85 mask (smoothed, galaxy part only) and the best-fit theory angular power spectrum of the Wilkinson Microwave Anisotropy Probe 9-year data release (WMAP, see e.g., \citealt{Bennett2012wmap9, Hinshaw2012wmap9}). As a realistic setting for a galaxy survey, we considered a mask created from the sky coverage of the SDSS DR6, enlarged and smoothed for stability of the estimates, shown in Fig.~\ref{fig:masks} (the variance rapidly becomes unstable for complex mask shapes and requires binning, as detailed in Sec.~\ref{sec:likelihood}). We used \textsc{camb\_sources}\footnote{\url{http://camb.info/sources/}} \citep{challinorlewis2011cambsources} to project the matter power spectrum $P(k)$ (corresponding to the WMAP 9-year cosmology) into an angular power spectrum using the redshift distribution of the CMASS sample, using a fixed astrophysical bias $b_g=2$. The CMASS sample will be described in further detail in Sec.~\ref{sec:cmassmock}. 

The CMB spectrum varies across three orders of magnitude in the range $0 < \ell < 50$ and, as expected, QML performs better than PCL in this range. In particular, the variance of the largest scale modes is up to $20\%$ smaller compared with PCL. For this mask, PCL is a good estimator for $\ell > 50$; its variance is typically within $\sim 10\%$ of that expected for a maximum likelihood estimator, although the degree of suboptimality will depend on the noise and the geometry of sky cut under consideration \citep{Efsta2003, Efsta2004, Efsta2006, HamimecheAndLewis2009}. The variance of the estimates is much larger for the galaxy survey mask, but since the CMASS spectrum varies by only one order of magnitude in $0 < \ell < 200$, the variance of the PCL and QML estimates in fact only differ by a few percent for $\ell > 15$. In conclusion, since LSS  power spectra are close to flat, PCL yields nearly optimal estimates, when no motivated prior for $\tilde{\mat{C}}$ is available. However, the degree of suboptimality increases when considering masks with complex geometries, and in the presence of non-isotropic contributions (e.g., spatial fluctuations due to calibration errors). In this case, the implicit assumptions in the PCL estimator are poor priors for a maximum likelihood analysis, and a better model of the pixel-pixel covariance matrix must be used in the QML estimator to obtain optimal estimates.

The complexity and resources involved in the PCL and QML algorithms also differ considerably. PCL benefits from fast, low-memory algorithms, and therefore can be applied to resolutions and multipole ranges which are amply sufficient for galaxy survey analyses. By contrast, QML involves the inversion of large covariance matrices and the execution of non-symmetric matrix multiplications. Although our optimised QML algorithm advantageously balances the work-load across processors and minimises memory use, typical galaxy surveys such as the SDSS ($\fsky \sim 1/6$) can only be analysed at \textsc{healpix} $\nside=64$ on a personal computer. To ensure that the estimates are minimally affected by the resolution, accurate smoothing and masking rules are detailed in Appendix~\ref{app:bandlimitandsmoothing}. Neglecting these considerations can lead to significant biases if the estimation is performed at low-resolution. Since high-$\ell$ maximum likelihood estimates might be desirable at high-resolution in some cases, \cite{GruetjenAndShellard2012} recently proposed a iterative algorithm (similar to the Newton-Raphson iterative scheme presented in the next section) to converge to the maximum likelihood solution using an augmented PCL basis. This approach will prove useful for obtaining optimal estimates of the damping tail of the CMB spectrum, where a full QML analysis is intractable, and PCL estimates are sub-optimal due to the exponential decay and anisotropic contributions to these modes \citep{HamimecheAndLewis2009}. However, this improved estimator is unnecessary for the study of galaxy surveys, since PCL is nearly optimal at the smallest scales where shot noise dominates. Alternatively, the complexity of the QML estimator can be reduced by adopting Karhunen-Lo\`eve compression \citep{VS96klcomp, T97largeklcomp, THS1998future, T01df}; for further details see Appendix~\ref{app:klcomp}. The results of this paper were obtained without this technique since we were able to run the estimator at $\nside=64$ without any other approximation, thus covering the scales which are not dominated by shot noise.

\subsection{Likelihood analysis and band-powers}\label{sec:likelihood}

The size and the shape of the mask strongly influence the variance of power spectrum estimates, and thus, the quality of any subsequent analysis. Typical CMB masks only cause a modest increase in the variance (e.g., $\sim 30\%$ for a WMAP-style mask with $\fsky = 75\%$) compared with the full sky case. By contrast, galaxy survey masks typically have $\fsky \la 20\%$ and can increase the variance of the individual multipoles by factors of 5-10. 

This problem is usually addressed by estimating the power spectrum in multipole bands, which smoothes the power spectrum, Gaussianises the likelihood function, reduces the sensitivity to the input prior, and decreases the variance of the estimates. In particular, this approach proves useful for PCL since its prior can be significantly suboptimal when using complex masks and in the presence of anisotropic contributions. The PCL and QML estimators in Eqs.~(\ref{pclel}) and (\ref{qmlel}) are straightforwardly transformed into band-power estimators by considering binned Legendre matrices $\tilde{\mat{P}}^b = \sum_{\ell \in L_b} \tilde{\mat{P}}^\ell$, where $L_b$ denotes the range of multipoles included in the $b$-th bin. The pixel-pixel covariance matrix $\tilde{\mat{C}}$ is unchanged and calculated as before, but the coupling matrices in the binned formulation are given by
\equ{
	(\mat{M})_{bb'} = \sum_{\ell\in L_b}\sum_{\ell'\in L_{b'}} (\mat{M})_{\ell \ell'}.
}
The expectation value of each band-power estimate reads
\equ{
	\bra \hat{C}_b \ket = \sum_{\ell} W_{b\ell} C_\ell  \label{binnedexpectedvalue}, 
}
where the window function $W_{b\ell}$ is constructed with the coupling matrix through
\equ{
	 W_{b\ell}  = \sum_{b'} (\mat{M})^{-1}_{bb'}  (\mat{M})_{b'\ell}. \label{windowfcts} 
} 
Hence, to be confronted with band-power estimates, the theory power spectrum must be transformed into band-powers using these window functions.

Using large band-powers guarantees that the likelihood function is sufficiently close to Gaussian for QML to deliver maximum-likelihood estimates. However, intermediate situations with smaller bins and complex masks might not fulfil this condition, leading to sub-optimal estimates. This issue can be addressed by iteratively converging to the maximum-likelihood solution using a Newton-Raphson scheme \citep{BJK98b,BJK98, KBJ98}. With the so-called Newton-Raphson maximum likelihood (NRML)  estimator, each iteration improves the previous one using 
\bl{\begin{equation}
	\delta C_\ell = \sum_{\ell'} (\mat{M}^{-1})^{\textrm{QML}}_{\ell \ell'} \frac{1}{2} \textrm{Tr}\left[ \left( \vec{x}\vec{x}^t - \tilde{\mat{C}} \right) \left( \tilde{\mat{C}}^{-1}  \tilde{\mat{P}}^\ell \tilde{\mat{C}}^{-1} \right)\right].
\end{equation}}
This approach is in fact equivalent to calculating the $i$-th estimate by feeding \textrm{QML} with the previous iteration.  The formulation of the NRML estimator then simplifies to
\equ{
	\hat{C}_\ell^{i+1} = \vec{\tilde{x}}^t (\mat{E}^\ell)^{i} \vec{\tilde{x}},
}
where $(\mat{E}^\ell)^{i}$ explicitly makes use of the covariance matrix calculated from $\hat{C}_\ell^{i}$, namely $\tilde{\mat{C}} = \sum_\ell  \tilde{\mat{P}}^{\ell'} \hat{C}_\ell^{i} + \tilde{\mat{N}}$. Hence, a critical issue is to construct a reliable covariance matrix at each iteration.

Both PCL and QML estimates can be extended to estimate the angular cross-power spectra (and hence cross band-powers) between two maps denoted by $\tilde{\vec{x}}_1$ and $\tilde{\vec{x}}_2$. The PCL estimator reads
\equ{
	\hat{C}_\ell^{\rm PCL, cross} =  \tilde{\vec{x}}^t_1 \mat{E}^\ell_{\textrm{PCL}} \tilde{\vec{x}}_2,
}
and implicitly assumes flat auto-power spectra and zero cross-power spectrum as priors. On the contrary, since QML derives from a likelihood function, the estimator must be adapted (following e.g., \citealt{tegmark2001, Padmanabhan_isw_2005}) by considering the input data vector as a concatenation of the two maps,
\equ{
	\tilde{\vec{x}} = \left( \begin{array}{c} \tilde{\vec{x}}_1 \\ \tilde{\vec{x}}_2 \end{array} \right),
}
and by using a pixel-pixel covariance matrix which incorporates all the information about the maps and their cross-correlation,
\equ{
	\tilde{\mat{C}} =  \left( \begin{array}{cc} \tilde{\mat{C}}_{11} & \tilde{\mat{C}}_{12} \\ \tilde{\mat{C}}^{\dagger}_{12} & \tilde{\mat{C}}_{22}  \end{array} \right). \label{crosscov}
}	
This formulation makes use of priors for the three power spectra, and must include models of the additional correlations and noise present in the data. The auto- and cross-spectra can be simultaneously estimated from $\tilde{\vec{x}}$ and $\tilde{\mat{C}}$ using the usual formulation of QML, i.e., Eqs.~(\ref{qmlel}) and (\ref{fisherqml}), \bl{In particular, the matrices to be used in the estimator with the covariance matrix of Eq.~(\ref{crosscov}) in order to calculate $\hat{C}_\ell^{11}$, $\hat{C}_\ell^{22}$ and $\hat{C}_\ell^{12}$ read
\equ{
	\tilde{\mat{P}}^\ell_{11} \mpt = \mpt  \left( \begin{array}{cc} \tilde{\mat{P}}^\ell & 0 \\ 0 & 0 \end{array} \right), \tilde{\mat{P}}^\ell_{22} \mpt=\mpt  \left( \begin{array}{cc} 0 & 0 \\ 0 & \tilde{\mat{P}}^\ell \end{array} \right), \tilde{\mat{P}}^\ell_{12} \mpt= \mpt \left( \begin{array}{cc} 0 & \tilde{\mat{P}}^\ell \\ \tilde{\mat{P}}^\ell & 0 \end{array} \right). 
}	
		}

\subsection{Galaxy surveys, shot noise and systematics}\label{sec:modeproj}

Galaxy catalogues are usually provided as lists of objects whose positions and properties, such as photometric colours, were measured by an instrument in the context of a sky survey. To relate to the dark matter distribution, a catalogue must be pixelised into a number count map $\tilde{\vec{G}}$, and then transformed into an overdensity map $\tilde{\vec{x}}$.  Given a pixelisation scheme, if $G_i$ denotes the number of objects in the $i$-th pixel (described by its size $\Omega_i$ and the position of its centre $\ang_i$), the overdensities are constructed as
\equ{
	\tilde{x}_i =  \frac{\tilde{G}_i}{\Omega_i \bar{G}} - 1,
}
where $\bar{G} = {\rm N_{\rm obj}}/{\Delta\Omega}$ is the average number of objects per steradian. ${\rm N}_{\rm obj}$ is the total number of objects in the catalogue, and $\Delta\Omega$ is the total surface outside the mask. The Poisson sampling of the observed tracers naturally gives rise to shot noise, characterised by a diagonal noise matrix $\tilde{\mat{N}}$ \citep{Huterer2001eds} such that $(\tilde{\mat{N}})_{ij}  = {\bar{G}^{-1}}\delta_{ij}$, where $\delta$ is the Kronecker delta.  Moreover, the power spectrum estimates of galaxy survey overdensity maps  always include a constant bias term due to shot noise,
\equ{
	\bra \hat{C}_\ell	\ket = C_\ell + \frac{1}{\bar{G}}.
}
Since the cosmological contribution $C_\ell$ steadily decreases with $\ell$ in the linear regime, shot noise usually dominates on small scales for catalogues with small number densities (i.e., for $\ell$ higher than a limit determined by $\bar{G}$).

So far, we assumed that the map $x$ was the result of cosmological clustering encapsulated by a theory power spectrum $\{ C_\ell \}$. In other words, the relevant correlations in the map were due to $\{ C_\ell \}$ and the shot noise ${1}/{\bar{G}}$. However, observations $\tilde{\vec{x}}^{\rm obs}$ are often contaminated by various signals introducing spurious correlations and requiring appropriate modelling to avoid suboptimal estimates. In particular, $\tilde{\vec{x}}^{\rm obs}$ can always be described as the superposition of the true cosmological signal $\tilde{\vec{x}}^{\rm true}$ and a contamination part due to systematics. We will assume that templates of the systematics are available, namely $n_{\rm sys}$ maps denoted by $\vec{c}^k$ with $k=1, \dots, n_{\rm sys}$. In this case, one can adopt a model for the contamination signal, and estimate its parameters from the data. The best-fit contamination model is then subtracted from the measured power spectra. This approach was recently used to correct the angular power spectra and 2-pt correlation functions of the CMASS sample with a best-fit linear contamination model \citep{ross2011weights, ho2012cosmoweights}. 

A more robust approach to mitigate the influence of the systematics is to incorporate their contribution in the pixel-pixel covariance matrix with large coefficients $\xi_k$, i.e.,
\equ{
	\tilde{\mat{C}}_{ij} = \tilde{\mat{S}}_{ij}   +   \tilde{\mat{N}}_{ij} + \sum_{k} \xi_k \tilde{c}_i^k \tilde{c}^k_j.
} 	 	
This technique, known as mode projection \citep{SlosarSeljak2004modeproj, ho_isw_2008}, assigns a very large variance to the modes corresponding to the systematics in pixel space, such that they do not influence the power spectrum estimates\footnote{\bl{This approach does not prevent the extraction of cosmological signals that happen to have the same power spectra as the systematics. In particular, projecting out a systematic is equivalent to ignoring one mode defined in pixel space.}}.  As a result, the QML estimates are unbiased in the power spectrum of $\tilde{\vec{x}}^{\rm true}$ as $\xi_i \rightarrow \infty$, provided that the contamination signal can be described as a linear combination of the templates. In the Bayesian perspective, mode projection is equivalent to marginalising over the parameters of a linear model of the contamination.  Hence, non-linear contamination or neglected systematics leave residual biases that must be eliminated by other means, for example through masking or modelling of the contamination signal.

\subsection{Illustration: recovering the power spectrum of the CMASS sample}\label{sec:cmassmock}

The ongoing Baryon Oscillation Spectroscopic Survey (BOSS) is part of SDSS-III \citep{Eisenstein2011} and aims to measure the spectroscopic redshifts of 1.5 million galaxies, \mbox{160,000} quasars and various ancillary targets from SDSS photometry \citep{Gunn1998, Gunn2006}. The CMASS \bl{spectroscopic} sample includes extended sources selected using colour magnitude cuts to produce a roughly volume limited sample in the redshift range $0.4 < z < 0.7$. In the DR9 release of BOSS, CMASS uses data taken up to the end of July 2011 and covers 3344 deg$^2$ in the Northern and Southern Galactic caps.

\cite{manera2012} presented a set of 600 mock catalogues for the CMASS \bl{DR9}  sample, constructed based on a $\Lambda$CDM cosmology defined by \{\mbox{$\Omega_m=0.274, \Omega_bh^2 = 0.0224, h=0.70, n_s=0.95, \sigma_8=0.8$}\}, and evolved using 2nd-order Lagrangian perturbation theory \citep{Scoccimarro19972lpt, Crocce20062lpt}.  These mock catalogues were used to compute accurate covariance matrices for CMASS and constrain cosmological parameters  \citep{sanchez2012bao},  test deviations from General Relativity \citep{Samushia2013grdeviation} and measure the scale-dependent halo bias \citep{ross2012png}. 

\begin{figure}\centering
\includegraphics[trim = 0.1cm 1.9cm 0.1cm 1.8cm, clip, width=7.2cm]{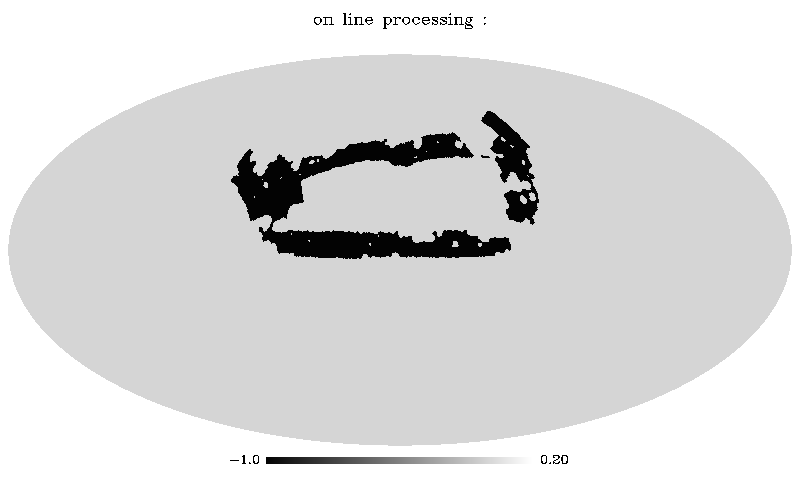}
\caption{Mask of the CMASS \bl{DR9} sample in the Northern Galactic Cap, in equatorial coordinates. } 
\label{fig:cmassmasks}
\end{figure}

To evaluate the performance of the previously described angular power spectrum estimators,  we compared the mean and standard deviation of the power spectrum estimates of the mock catalogues with the theoretical expectations. We only considered the CMASS \bl{DR9} mock catalogues in the Northern Galactic cap (NGC),  covering 2635 deg$^2$ as shown in Fig.~\ref{fig:cmassmasks}. We computed a theory prediction for the angular power spectrum of the underlying dark matter with \textsc{camb\_sources} \citep{challinorlewis2011cambsources} using a redshift distribution parametrised by an Edgeworth expansion of the redshift histogram of CMASS objects.

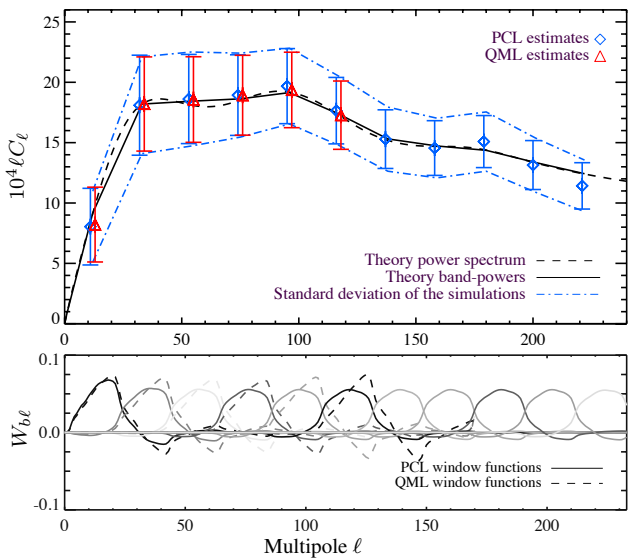
\begin{figure}\centering
\setlength{\unitlength}{.5in}
\begin{picture}(9,5.8)(0,0)
\put(0.2,2.15){\includegraphics[trim = 2.9cm 13.5cm 3.2cm 5.9cm, clip, width=8.2cm]{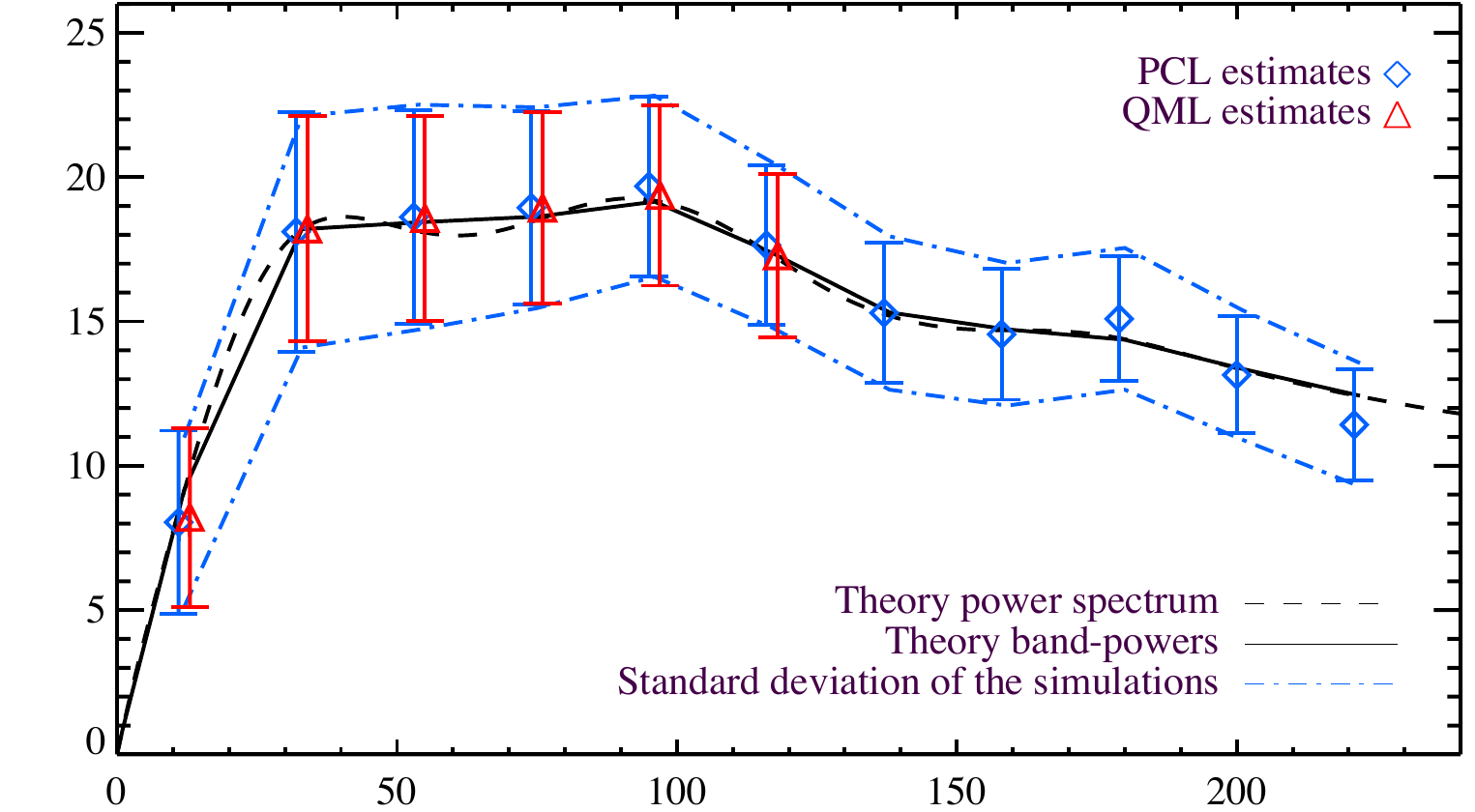}}
\put(0.2,0.2){\includegraphics[trim = 2.9cm 13.5cm 3.2cm 5.9cm, clip, width=8.2cm]{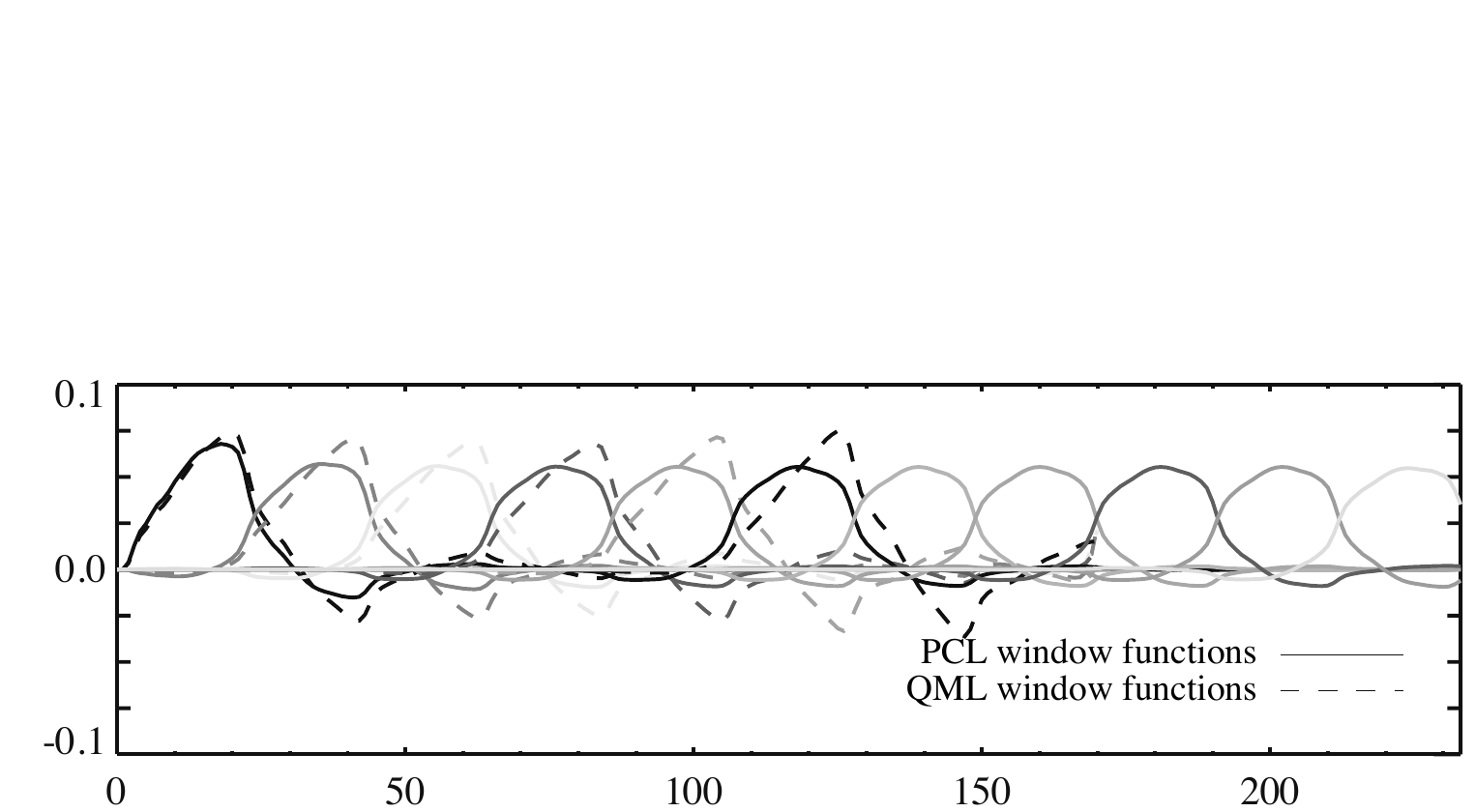}}
\put(2.8,0.0){\footnotesize Multipole $\ell$}
\put(0.1,3.7){\rotatebox{90}{\footnotesize $10^4 \ell C_\ell$}}
\put(0.16,1.1){\rotatebox{90}{\footnotesize $W_{b\ell}$}}
\end{picture}
\caption{Average PCL and QML estimates of the 600 CMASS mock catalogues \citep{manera2012}. Shot noise was subtracted from the estimates, and the angular power spectrum was calculated with \textsc{camb\_sources} (black dashed line) using a scale-independent astrophysical bias $b_g=1.9$. The theory band-powers (black solid line) were obtained by applying the exact PCL and QML binning window functions of Eqs.~(\ref{windowfcts}) and (\ref{binnedexpectedvalue}) (bottom panel). We were able to recover the theory power spectrum and the covariance of the estimates with good precision using the smoothing, masking and band-limit rules defined in Appendix~\ref{app:bandlimitandsmoothing}. } 
\label{fig:cmassspectra}
\end{figure}

We calculated the PCL and QML estimates of the individual mock catalogues at \textsc{healpix} resolutions ${\rm N_{side}^{ PCL}}$=128 and  ${\rm N_{side}^{QML}}$=64. The mean and variance of the estimates compared with theoretical expectations are shown in Fig.~\ref{fig:cmassspectra}. Note that the error bars of the power spectrum estimates are correlated; see Eq.~(\ref{varianceestimates}). We subtracted the shot noise and also multiplied the theory prediction by a scale-independent bias of $1.9$, the value used to construct these mock catalogues \citep{manera2012}. We were able to recover the theory power spectrum and the covariance of the estimates with good precision using the smoothing, masking and band-limit rules defined in Appendix~\ref{app:bandlimitandsmoothing}. 
The estimates and the theory predictions from \textsc{camb\_sources} were insensitive to small changes in the redshift distribution. However, we observed that inconsistent smoothing, masking and band-limit rules led to biases in the recovered power spectra, resulting from a mismatch between the model pixel-pixel covariance matrices and the information content in the data.
As expected, due to the flatness of the power spectrum and the simple geometry of the mask, QML performed only marginally better than PCL, which yielded nearly-optimal estimates in the absence of systematics. The theory spectrum was converted into band-powers using the exact window functions in Eqs.~(\ref{binnedexpectedvalue}) and (\ref{windowfcts}), which are shown in the bottom panel of Fig.~\ref{fig:cmassspectra}.

\section{Application to SDSS photometric quasars}\label{sec:quasarsample}

\subsection{Data and subsamples}

We considered the \cite{Richards2008rqcat} catalogue of photometric quasars, which is based on the Sixth Data Release (DR6) of the SDSS \citep{AdelmanMcCarthy2008dr6}. The objects in this catalogue -- which we call RQCat as in \cite{PullenHirata2012} -- were photometrically selected by a binary Bayesian classifier trained in 4D colour space using several catalogues of spectroscopically-confirmed stars and quasars. The final version of the catalogue includes 1,172,157 objects with several quality and technical flags, which can be exploited to apply further systematics cuts and obtain cleaner samples. The Bayesian classifier was initially applied to all point sources in the DR6 release without restriction, and explicitly assumed that 95\% of the input objects were stars (i.e., constant star and quasar priors $p_{\rm star}=0.95$ and $p_{\rm quasar}=0.05$). It was also applied with stronger priors ($p_{\rm star}=0.98$) to objects with photometric redshifts $\tilde{z}_p$ (estimated by the SDSS pipeline, as opposed to the true redshift $z$) in three redshift ranges $0 < \tilde{z}_p \le 2.2$, $2.2 < \tilde{z}_p \le  3.5$ and $\tilde{z}_p > 3.5$ to achieve higher efficiency\footnote{The efficiency, or purity, of a catalogue denotes the fraction of objects which are quasars. It characterises the ability of the classifier to separate quasars from stars.} and completeness. The efficiency is degraded in the two higher redshift ranges since the Bayesian classifier performs worse due to the overlap between the stellar and quasar loci in the colour space (at $z \sim 2.6$). More sophisticated algorithms such as XDQSO \citep{Bovy2010xdqso} were developed to specifically address this issue and identify higher-redshift quasars for spectroscopic follow-up in the context of BOSS \citep{Ross2012bosstarget}. However, in this work we focused on $z \leq 2.2$ objects, where the performance of the binary Bayesian classifier used by \cite{Richards2008rqcat} is satisfactory. 

Colour-based selection of quasars is difficult, and RQCat is expected to be significantly contaminated by stars, which are often misclassified as quasars due to similar colours. Hence, the full catalogue cannot be used as a statistical sample for direct power spectrum analysis due to its low efficiency (lower than $80\%$). In this work, we restricted ourselves to good UV-excess low-redshift objects, defined as $u - g < 1.0$ and $\tilde{z}_p<2.2$ (the corresponding flags are \textsc{good$>$0}, \textsc{uvx=1} and \textsc{lowz=1}). This sample, denoted by UVX-LOWZ, is the least contaminated by stars and achieves $96.3\% \pm 1.2$ efficiency \citep{Richards2008rqcat}. 

Previous studies considered the UVX sources in RQCat for cosmological analyses, e.g., \bl{studying the environment of quasars \citep{Myers2006first, Myers2007one}}, detecting the ISW effect \citep{Giannantonio2005de, Giannantonio2006isw, Giannantonio2008isw, Giannantonio01112012} and constraining PNG \citep{SlosarHirata2008, Xia2010sdssqsoctheta, Xia2011sdssqsocell}. Recent work by \cite{PullenHirata2012}, corroborated by \cite{Giannantonio2013png}, found that the UVX objects were significantly contaminated, as indicated by the cross-spectra of redshift bins which exhibited excess power at the largest scales. 

The two main sources of systematics in quasar photometric catalogues are contamination and calibration errors. The origin of contamination lies in the classification stage: selecting quasars based on photometric data is a complex task. Various objects can be misclassified as quasars, and are thus present in the final catalogues. Since the clustering properties of these contaminants differ from those of quasars, they affect the measured power spectra and can jeopardise the interpretation of the data. Calibration errors, on the other hand, are present in the catalogue regardless of the ability of the classifier to separate stars and quasars. In the ideal case, a perfect classifier applied to all point sources detected by a given instrument will lead to a sample with clustering properties purely due to cosmological physics. However, real instruments are not perfectly calibrated, and observing conditions also change with time, introducing spurious correlations due to variations in the number of detected sources on the sky. In addition, calibration errors impact the apparent magnitude estimates, which propagate through the Bayesian classifier (since the latter does not model or account for them), inducing a spatial dependence in its efficiency.

Contamination and calibration issues can be addressed in different ways. First, it is important to reduce the catalogue of interest by selecting the most reliable objects (here UVX sources) and also restricting the analysis to the most reliable areas of the sky. Secondly, corrections can be applied to the power spectrum estimates themselves to minimise the remaining spurious correlations. Alternatively, one can opt for a Bayesian analysis and marginalise over the systematics in the cosmological analysis. In this study, we focused on the sample reduction approach. We separated the UVX-LOWZ sample into four subsamples by selecting objects with photometric redshifts $\tilde{z}_p$ in bins with ranges $[0.5,1.3]$, $[1.3,1.8]$, $[1.8,2.2]$ and $[1.3,2.2]$. These samples are called Low-$z$, Mid-$z$, High-$z$  and Mid+High-$z$  respectively, as referred to as the RQCat subsamples. We rejected low redshift quasars ($\tilde{z}_p < 0.5$) because their power spectra were severely contaminated: this can be attributed to strong stellar contamination, and to the fraction of low-$z$  quasars which are extended sources and were therefore not processed by the Bayesian classifier. The corresponding incompleteness is non-trivial and likely to depend on observational effects such as dust absorption and seeing variations.

\subsection{Theory predictions}

\begin{figure}\centering
\setlength{\unitlength}{.5in}
\begin{picture}(8,6.5)(0,0)
\put(0.2,0.05){{\includegraphics[trim = 1.5cm 3.2cm 2.2cm 2.2cm, clip, width=8cm]{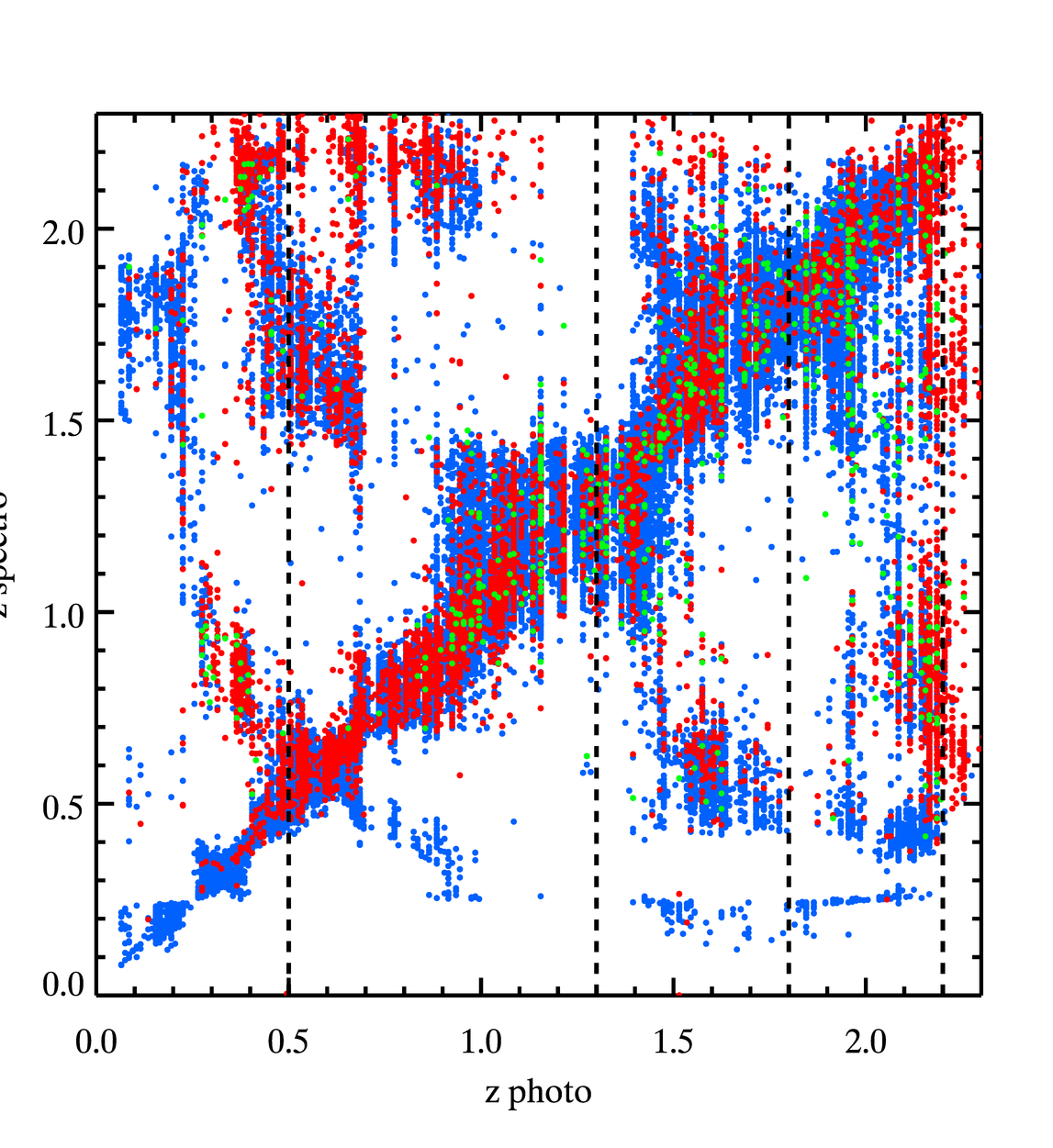}}}
\put(0.0,3.3){\footnotesize\rotatebox{90}{$\tilde{z}_s$}}
\put(3.4,-0.1){\footnotesize{$\tilde{z}_p$}}
\end{picture}
\caption{\bl{Distributions of the photometric and spectroscopic redshift estimates ($\tilde{z}_p$ and $\tilde{z}_s$) of the RQCat UVX-LOWZ sample, cross-matched with the SDSS-DR7, BOSS and 2SLAQ spectroscopic quasar catalogues (blue, red and green dots). The dashed lines indicate the photometric redshift cuts used to assemble the four RQCat subsamples.} The photometric redshift estimates are seen to be unreliable and cannot be used to estimate the redshift distributions of the photometrically-selected subsamples. On the contrary, the cross-matched samples have reliable spectroscopic redshifts, and can be used for this purpose. However, one must apply the relevant completeness corrections in order to account for the change of selection function between the photometric and cross-matched subsamples (due to e.g., different magnitude limits).} 
\label{fig:redshifterrors}
\end{figure}

In order to calculate theoretical predictions for the angular power spectra of the four RQCat subsamples, we used \textsc{camb\_sources} \citep{challinorlewis2011cambsources}, a high-precision code which projects the 3D matter power spectrum $P(k)$ into angular auto- and cross-power spectra. Since this study is focused on the impact of the systematics on the observed power spectra, we fixed the cosmological parameters to \textit{Planck} $\Lambda$CDM best-fit values\footnote{Fixed to $\Omega_ch^2=0.188, \Omega_m=0.315, \Omega_bh^2 = 0.02205 , H_0=67.3\ {\rm km s}^{-1} {\rm Mpc}^{-1}, \ln(10^{10}A_s)=3.089, n_s=0.9603, {\rm\ and \ } \tau = 0.089$.} \citep{Planck2013cosmologicalparams}.  We opted for a scale-independent linear bias to relate the observed galaxy clustering to dark matter.

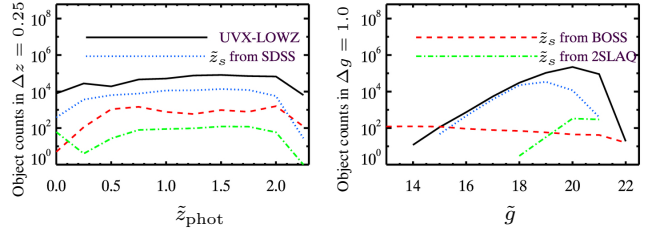
\begin{figure}\centering
\setlength{\unitlength}{.5in}
\begin{picture}(9,2.4)(0,0)
\put(0.2,0.0){{\includegraphics[trim = 1.5cm 1.5cm 2cm 2.7cm, clip, height=3.05cm]{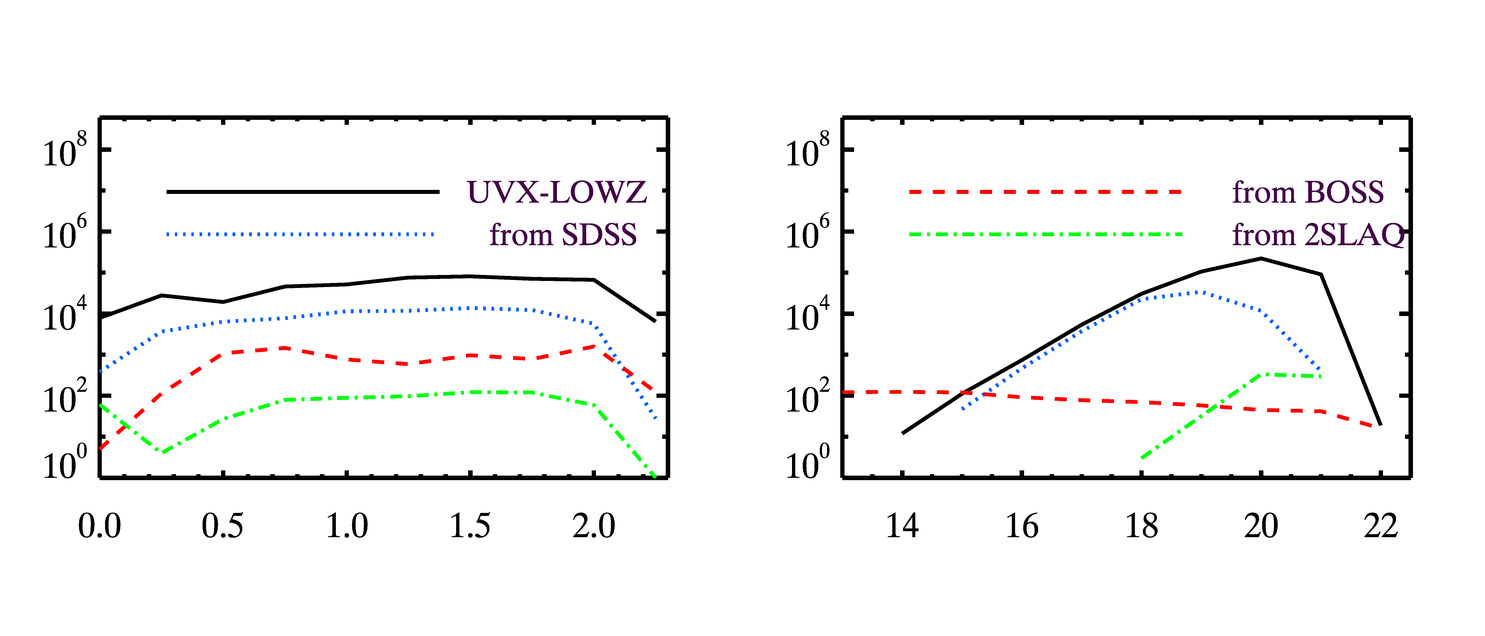}}}
\put(1.7,-0.05){\footnotesize{$\tilde{z}_{\rm phot}$}}
\put(0.0,0.2){\rotatebox{90}{\scriptsize{Object counts in $\Delta z= 0.25$}}}
\put(5.15,-0.05){\footnotesize{$\tilde{g}$}}
\put(3.4,0.25){\rotatebox{90}{\scriptsize{Object counts in $\Delta g= 1.0$}}}
\put(2.05,1.61){\scriptsize{$\tilde{z}_s$}}
\put(5.5,1.61){\scriptsize{$\tilde{z}_s$}}
\put(5.5,1.81){\scriptsize{$\tilde{z}_s$}}
\end{picture}
\caption{Histograms of the redshifts and apparent magnitudes of UVX-LOWZ objects in RQCat, and of objects with good spectra found in the SDSS-DR7, BOSS and 2SLAQ spectroscopic quasar catalogues. In addition to the higher number of objects, the SDSS-DR7 cross-matched sample has redshift and magnitude distributions close to that of RQCat, indicating similar selection functions. Hence, the redshift distributions of the RQCat subsamples can be estimated using the cross-matched sample, with only minor completeness corrections to account for the differences in magnitude limits. By contrast, BOSS and 2SLAQ target significantly different redshift and magnitude ranges, and the completeness corrections required to estimate the redshift distributions of RQCat are strongly redshift- and magnitude-dependent. In addition, the latter are limited by sample variance due to the smaller number of objects.} 
\label{fig:qsocompleteness1}
\end{figure}

\begin{figure}\centering
\setlength{\unitlength}{.5in}
\begin{picture}(9,2.2)(0,0)
\put(-0.2,0.0){{\includegraphics[trim = 1.6cm 1.3cm 0.cm 12.8cm, clip, width=9cm]{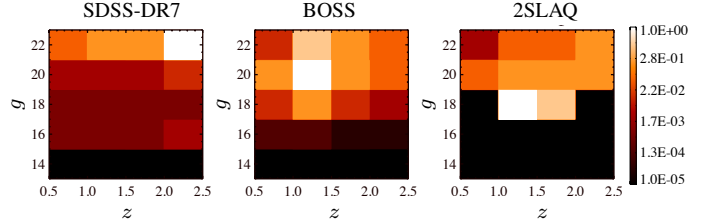}}}
\put(0.55,2.1){\footnotesize{SDSS-DR7}}
\put(2.85,2.1){\footnotesize{BOSS}}
\put(5.05,2.1){\footnotesize{2SLAQ}}
\put(0.95,0){\footnotesize{$z$}}
\put(-0.2,1.13){\footnotesize\rotatebox{90}{$g$}}
\put(3.12,0){\footnotesize{$z$}}
\put(1.95,1.13){\footnotesize\rotatebox{90}{$g$}}
\put(5.3,0){\footnotesize{$z$}}
\put(4.1,1.13){\footnotesize\rotatebox{90}{$g$}}
\end{picture}
\caption{Low-resolution redshift- and magnitude-dependent completeness corrections for estimating the redshift distributions of the RQCat subsamples through cross-matching with the SDSS-DR7, BOSS and 2SLAQ spectroscopic quasar catalogues. As expected from Fig.~\ref{fig:qsocompleteness1}, the redshift distribution of the cross-matched sample with SDSS-DR7 only requires weak magnitude corrections to relate to that of RQCat. On the contrary, BOSS and 2SLAQ require significant corrections, since they target different redshift and magnitude ranges than RQCat. } 
\label{fig:qsocompleteness2}
\end{figure}

Although the matter power spectrum $P(k)$ only depends on cosmological parameters, computing angular power spectrum predictions requires additional knowledge about the distributions and properties of the samples under consideration. In particular, the auto-angular power spectrum of a sample of interest reads
\equ{
	C_\ell = \frac{2}{\pi} \int dk k^2 P(k) [W_\ell(k)]^2,
	}
where the window function $W_\ell(k)$ includes several cosmological effects, as detailed in \cite{challinorlewis2011cambsources}. In fact, $W_\ell(k)$ requires two quantities in addition to the standard cosmological parameters: the unit-normalised redshift distribution of tracers, denoted by $n(z)$, and the logarithmic slope of the number counts $s$, which accounts for the effect of magnification due to lensing. A simplified formula for $W_\ell(k)$, which we use for illustration purposes only (see \cite{challinorlewis2011cambsources} for the full formalism), reads
\equ{
	W_\ell(k) = \int dz \left[ b_g n(z) + 2(2.5 s -1) f(z) \right] D(z) j_\ell(kr), \hspace*{-1mm} \label{simplewindow}
	}
where $b_g$ is the linear galaxy bias, $D(z)$ is the growth factor, $j_\ell$ is the spherical Bessel function, $r(z)$ is the comoving distance, and $f(z)$ is the lensing window function, giving rise to magnification. In this work, we assumed that the logarithmic slope of the number counts, defined as
\equ{
	s = \frac{ d \log N(m) }{dm},
}
was constant in each redshift range. We estimated its value for the four RQCat subsamples by calculating the slope of the histogram of number counts in terms of the $g$-band PSF magnitude at $g = 21$. We found $s=0.18, 0.89, 0.89$ and $0.87$ respectively for these subsamples, consistent with previous studies in similar redshift ranges \citep{Xia2009highzisw, PullenHirata2012}. \bl{The choice of a scale- and redshift-independent linear bias is motivated by previous studies of RQCat and of low-redshift quasars in general (e.g., \citealt{Myers2007one, PullenHirata2012, Sherwin2012qsolensing, Giannantonio2013png}). The redshift evolution of the bias in $0.5 < z < 2.2$ proves to be smaller than the uncertainty on the completeness corrections, and thus marginally affects the predicted angular power spectra.}

\subsection{Redshift distributions estimates}\label{sec:redshiftdistrib}

\begin{figure*}\centering
\setlength{\unitlength}{1cm}
\begin{picture}(18,8.7)(0,0)
\put(0,0){\includegraphics[trim = 3.6cm 2.3cm 4.3cm 8.0cm, clip, width=15.4cm]{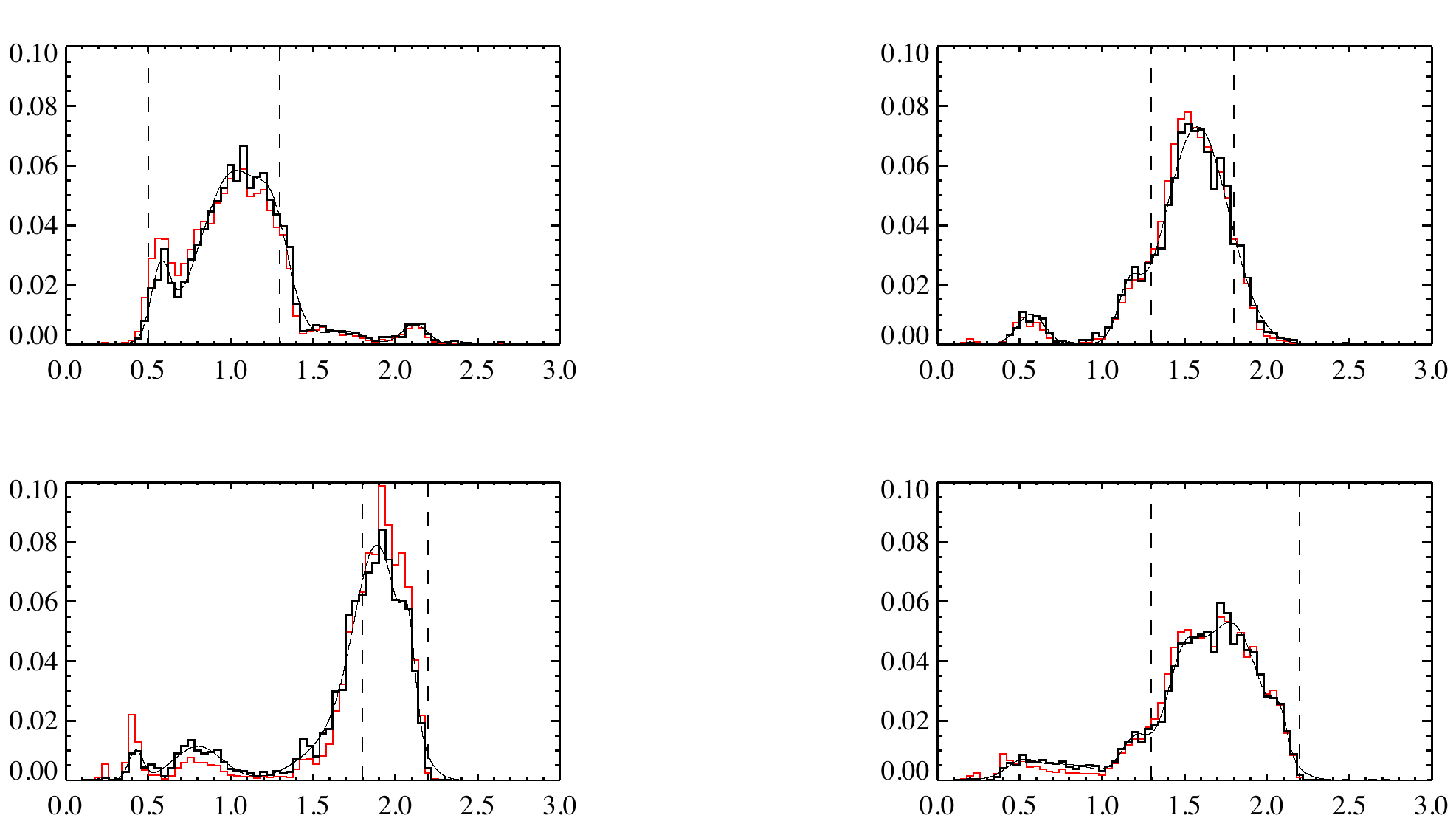}}
\put(6.2,4.7){\includegraphics[trim = 6cm 0.5cm 7.5cm 15.7cm, clip, width=2.2cm]{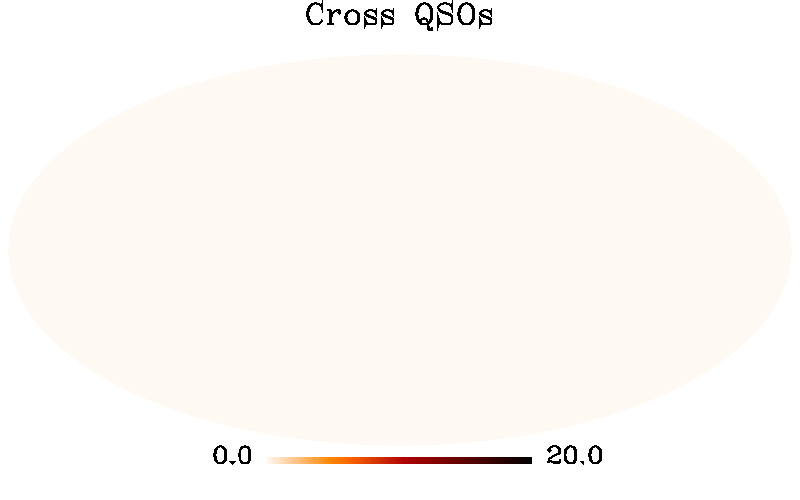}}
\put(6,5.1){\includegraphics[trim = 7.7cm 8.3cm 9.2cm 1.9cm, clip, width=2.5cm]{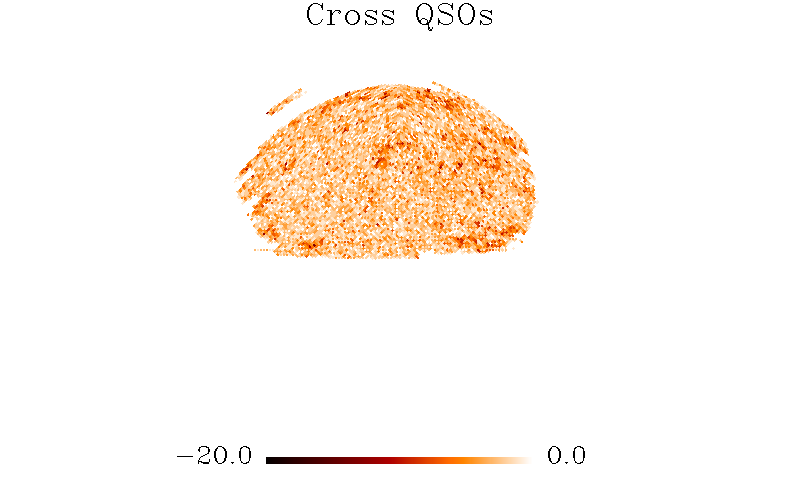}}
\put(6,6.7){\includegraphics[trim = 7.7cm 8.3cm 9.2cm 1.9cm, clip, width=2.5cm]{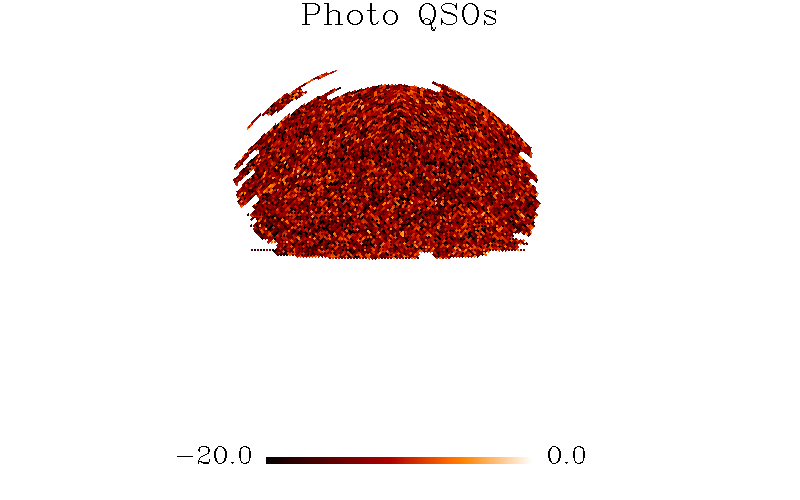}}
\put(1.9,8.4){{Low-$z$ : $0.5 < \tilde{z}_{\rm p} \leq 1.3$}}
\put(3.25,4.4){{$z$}}
\put(-0.30,6.4){\rotatebox{90}{{$\tilde{n}(z)$}}}
\put(6.2,0.1){\includegraphics[trim = 6cm 0.5cm 7.5cm 15.7cm, clip, width=2.2cm]{pics/qso_crosscat_bar1.png}}
\put(6,0.5){\includegraphics[trim = 7.7cm 8.3cm 9.2cm 1.9cm, clip, width=2.5cm]{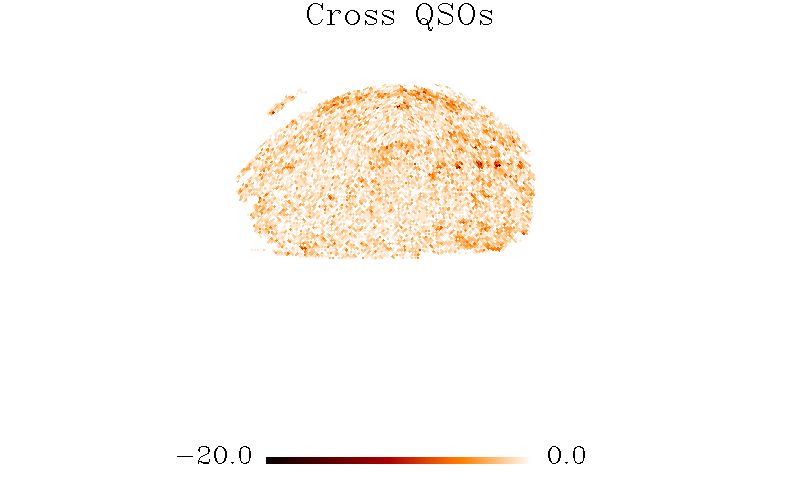}}
\put(6,2.1){\includegraphics[trim = 7.7cm 8.3cm 9.2cm 1.9cm, clip, width=2.5cm]{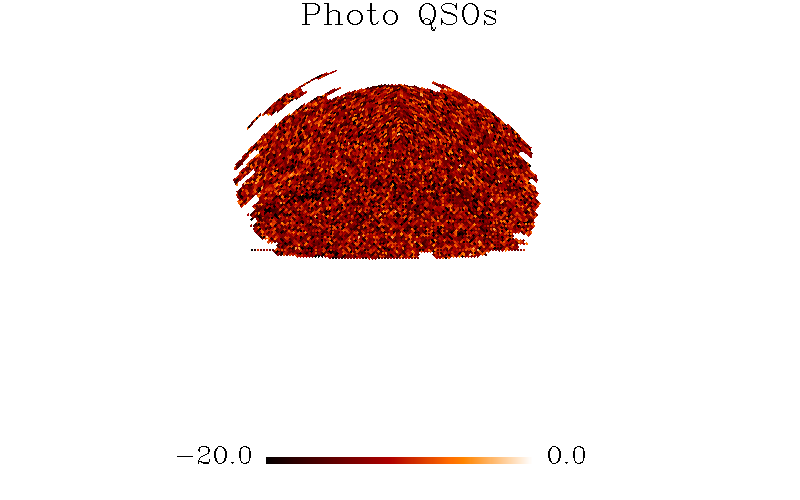}}
\put(1.9,3.8){{High-$z$ : $1.8 < \tilde{z}_{\rm p} \leq 2.2$}}
\put(3.25,-0.2){{$z$}}
\put(-0.30,1.8){\rotatebox{90}{{$\tilde{n}(z)$}}}
\put(15.45,4.7){\includegraphics[trim = 6cm 0.5cm 7.5cm 15.7cm, clip, width=2.2cm]{pics/qso_crosscat_bar1.png}}
\put(15.25,5.1){\includegraphics[trim = 7.7cm 8.3cm 9.2cm 1.9cm, clip, width=2.5cm]{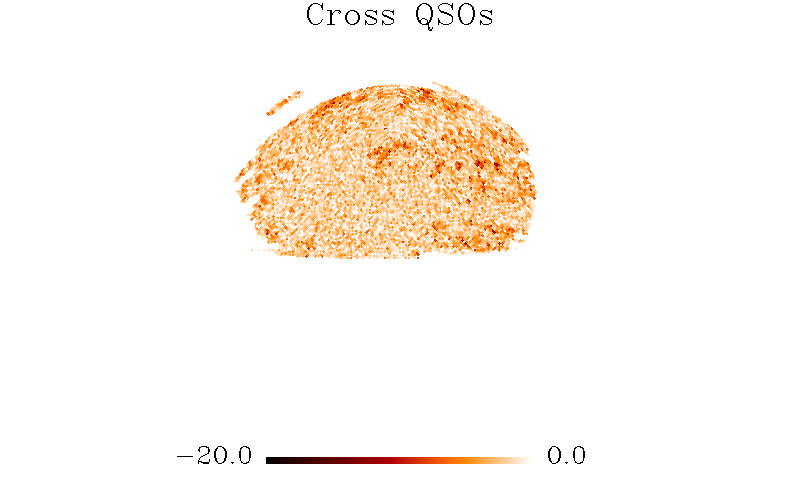}}
\put(15.25,6.7){\includegraphics[trim = 7.7cm 8.3cm 9.2cm 1.9cm, clip, width=2.5cm]{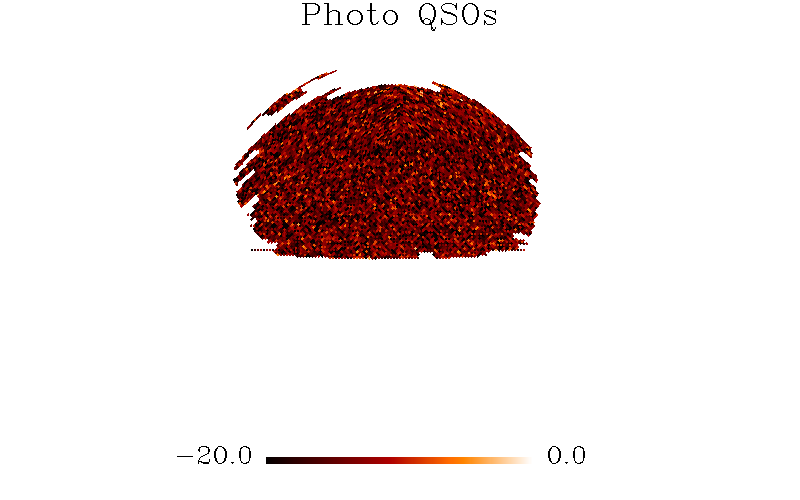}}
\put(11.15,8.4){{Mid-$z$ : $1.3 < \tilde{z}_{\rm p} \leq 1.8$}}
\put(12.5,4.4){{$z$}}
\put(8.9,6.4){\rotatebox{90}{{$\tilde{n}(z)$}}}
\put(15.45,0.1){\includegraphics[trim = 6cm 0.5cm 7.5cm 15.7cm, clip, width=2.2cm]{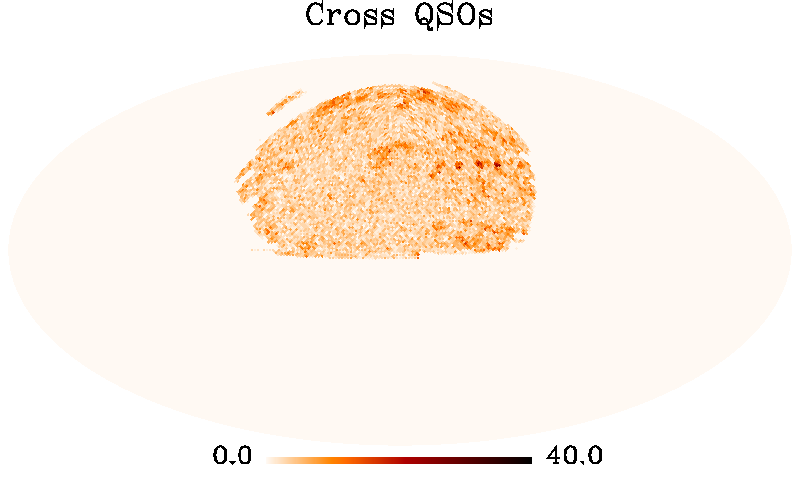}}
\put(15.25,0.5){\includegraphics[trim = 7.7cm 8.3cm 9.2cm 1.9cm, clip, width=2.5cm]{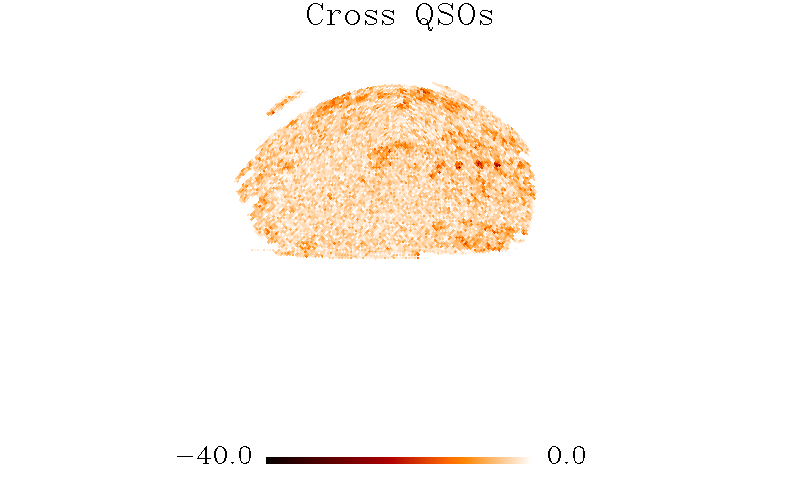}}
\put(15.25,2.1){\includegraphics[trim = 7.7cm 8.3cm 9.2cm 1.9cm, clip, width=2.5cm]{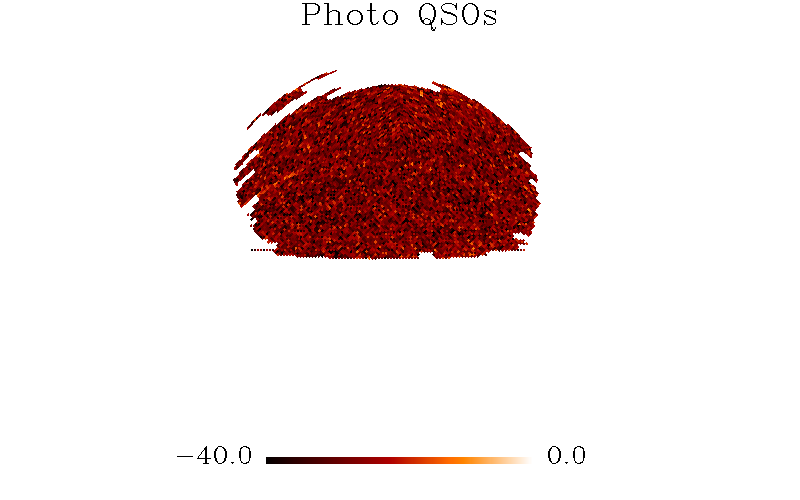}}
\put(10.75,3.8){{Mid+High-$z$ : $1.3 < \tilde{z}_{\rm p} \leq 2.2$}}
\put(12.5,-0.2){{$z$}}
\put(8.9,1.8){\rotatebox{90}{{$\tilde{n}(z)$}}}
\end{picture}
\caption{Final estimates of the redshift distributions of the four RQCat photometric subsamples, calculated by cross-matching with the SDSS-DR7 spectroscopic quasar catalogue \citep{Schneider2010qsodr7cat}. The accuracy of the redshift distribution estimates is essential for obtaining robust angular power spectrum estimates. Maps of the photometric and cross-matched samples are shown on the right subpanels at \textsc{healpix} resolution $\nside=64$. The dashed vertical lines in the main subpanels indicate the photometric redshift cuts used to construct the RQCat subsamples. 
Since the cross-matched samples have different selection functions, their redshift distributions (red thin histograms) must be corrected in order to accurately estimate the redshift distributions of the RQCat samples.
The thick black histograms show the final estimates obtained by applying magnitude- and pixel-dependent completeness corrections, and were fitted with a superposition of Gaussian distributions (solid lines) for use in \textsc{camb\_sources}.} 
\label{fig:qsoredshiftdistribution}
\end{figure*}

The quasars in each RQCat subsample are characterised by a normalised redshift distribution $n(z)$, with $n(z)dz$ corresponding to the probability of finding a quasar with redshift between $z$ and $z+dz$. Consequently, $n(z)$ incorporates the physical distribution (i.e., originating from the quasar luminosity function), the survey characteristics (such as the magnitude limits), and the photometric redshift cuts used to construct the subsample under consideration. The simplest estimator for $n(z)$ is a normalised histogram of the photometric redshifts of all objects in each subsample of interest. However, in practice, this approach does not yield good estimates due to the large uncertainties in quasar photometric redshift estimates. To illustrate this issue, we used the SDSS-DR7, BOSS and 2SLAQ spectroscopic quasar catalogues \citep{Schneider2010qsodr7cat, paris2012bossqsodr9, Croom20092slaqcat} to find objects in the four RQCat subsamples for which reliable spectra, and thus good spectroscopic redshifts, were available. Figure~\ref{fig:redshifterrors} shows the photometric and spectroscopic redshifts ($\tilde{z}_p$ and $\tilde{z}_s$, respectively) of the cross-matched objects. The dispersion of the points $(\tilde{z}_p, \tilde{z}_s)$ around $\tilde{z}_p=\tilde{z}_s$ demonstrates that the photometric redshift estimates of quasars suffer from large uncertainties and catastrophic failures, yielding a large fraction of spectroscopic redshifts outside the photometric windows used to construct the RQCat subsamples, indicated by the dashed lines. The photometric redshift estimates do not accurately follow the underlying redshift distributions $n(z)$, and cannot be used to compute accurate angular power spectrum predictions.

Nevertheless, one can use the redshift distributions of the cross-matched samples, which can be calculated with great accuracy using the spectroscopic redshift estimates, whose uncertainties are negligible compared to the precision required for $n(z)$.  However, the redshift, magnitude and spatial distributions of the cross-matched samples may deviate from those of the photometric samples due to differences in their selection functions, caused, e.g., by different magnitude limits, redshift ranges, or sky coverage. In order to avoid biases in the redshift distribution estimates, one must include a completeness correction factor, denoted by $f_c$, which is a function of redshift $z$, magnitude $g$ and position on the sky $\ang$.

In practice, we divide the redshift and magnitude domains into bins denoted by $[z_i^{\rm min}, z_i^{\rm max}]$ (centred at $z_i$) and $[g_j^{\rm min}, g_j^{\rm max}]$ (centred at $g_j$) respectively, and the sky into pixels denoted by $k$. For each RQCat subsample, an estimator of $n(z)$ at redshift $z=z_i$ is given by
\equ{
	 \est{n}(z_i) = C\ \sum_{jk} \frac{ {N_{\rm obj}}({\rm cross}, [z_i^{\rm min}, z_i^{\rm max}], [g_j^{\rm min}, g_j^{\rm max}], k) }{ \est{f}_c( z_i,  g_j, k)  } \label{nzestspec},
	 } 
where $C$ is a normalisation constant (such that $\sum_i \est{n}(z_i) = 1$) and ${N_{\rm obj}}({\rm cross}, [z_i^{\rm min}, z_i^{\rm max}], [g_j^{\rm min}, g_j^{\rm max}], k)$ is the number of objects in the \textit{cross-matched} sample in the $i$th redshift bin, $j$th magnitude bin and $k$th pixel.
 
With an estimator of $n(z)$ in hand, we now discuss how to construct the completeness correction. A simple estimator for the correction in the $(i,j,k)$th volume element reads
\equ{
	\est{f}_c( z_i,  g_j, k ) = \frac{  {N_{\rm obj}}({\rm cross}, [z_i^{\rm min}, z_i^{\rm max}], [g_j^{\rm min}, g_j^{\rm max}], k)  }{  {N_{\rm obj}}({\rm photo},  [z_i^{\rm min}, z_i^{\rm max}], [g_j^{\rm min}, g_j^{\rm max}], k) },
}
where ${N_{\rm obj}}({\rm photo}, [z_i^{\rm min}, z_i^{\rm max}], [g_j^{\rm min}, g_j^{\rm max}], k)$ is the number of objects in the \textit{photometric} sample (namely, one of the RQCat subsamples), thus constructed using the photometric redshifts.

The resolution of previous multidimensional histograms is limited by sample variance and by the uncertainties on (both photometric and spectroscopic) redshift and magnitude estimates\footnote{We will neglect the uncertainty in the sky position since it is negligible compared to the pixel size used for the completeness correction.}. To estimate the redshift distribution of the RQCat subsamples, we investigated the use of the SDSS-DR7, BOSS and 2SLAQ spectroscopic quasar catalogues. The redshift and magnitude distributions of the UVX-LOWZ objects in RQCat cross-matched with these catalogues are shown in Fig.~\ref{fig:qsocompleteness1}. We found that the BOSS- and 2SLAQ-based cross-matched samples were small and had selection functions quite different from RQCat. As a result, the estimator required significant completeness corrections, and the multidimensional histograms had to be constructed with large bins to reduce sample variance and biases due to uncertainties on the redshift and magnitude estimates.  On the other hand, cross-matching with SDSS-DR7 led to large samples which had selection functions similar to that of the RQCat subsamples.  Fig.~\ref{fig:qsocompleteness2} shows low-resolution estimates of completeness corrections arising when cross-matching the whole UVX-LOWZ sample (union of the RQCat subsamples) with SDSS-DR7, BOSS and 2SLAQ, where spatial dependence was neglected for purposes of illustration. Even at such low resolution, the completeness corrections for BOSS and 2SLAQ exhibit strong redshift and magnitude dependences (and are limited by sample variance due to the low number of cross-matched objects). For these reasons, the final redshift distribution estimates of the four RQCat samples were calculated using the SDSS-DR7 catalogue. The right panels of Fig.~\ref{fig:qsoredshiftdistribution} show number count maps of the RQCat subsamples and the SDSS-DR7-based cross-matched samples. The numbers of objects in the respective subsamples are summarised in Table~\ref{nobjtable}.

\begin{table}
\centering
\begin{minipage}{7.3cm}
\centering
	\caption{Number of objects in the four RQCat redshift subsamples (N$_{\rm obj}$ photo), and for which good spectra (and thus good spectroscopic redshifts) were found in the SDSS-DR7 quasar catalogue (N$_{\rm obj}$ cross). These cross-matched samples were used to estimate the redshift distributions shown in Fig~\ref{fig:qsoredshiftdistribution}. }
	\label{nobjtable}
	\hspace*{-4mm}	
	\begin{tabular}{l|r|r|r|r|r}
	\hline & Low-$z$ & Mid-$z$ & High-$z$ & Mid+High-$z$\\ \hline
	N$_{\rm obj}$ photo& 95,185 & 109,713 & 92,740 & 202,453 \\
	N$_{\rm obj}$ cross & 19,328  & 17,589 & 10,654 & 28,243 \\ \hline 
	\end{tabular}
\end{minipage}
\end{table}

With the high number of cross-matched objects, we were able to test various assumptions for the completeness corrections, such as the weakness of the redshift-dependence, and choose resolutions that yielded the best estimates. Since SDSS-DR7  spans the same redshift range as RQCat, no redshift-dependent corrections were required (Fig.~\ref{fig:qsocompleteness2} shows that the correction is only weakly redshift dependent, apart from a mild transition at $z=2$. Neglecting this transition did not significantly impact the final estimates). We used magnitude-dependent completeness corrections at resolution $\Delta g = 1.0$, due to the different magnitude limits in the SDSS photometric and spectroscopic data. Finally, the SDSS-DR7 spectroscopic quasar catalogue was assembled from different data releases of SDSS and is known to be non-uniform on the sky \citep{Schneider2010qsodr7cat}. Indeed, Fig.~\ref{fig:qsoredshiftdistribution} shows that the cross-matched samples contain regions with a greater number of objects, in some cases exploring fainter magnitudes. To address this spatial dependence, we calculated the magnitude-dependent completeness corrections in individual pixels at \textsc{healpix} resolution $\nside = 16$. The final estimates $\est{n}(z_i)$ are shown in Fig.~\ref{fig:qsoredshiftdistribution}, and were fitted by superpositions of Gaussian distributions for use in \textsc{camb\_sources}. 

\begin{figure}\centering
\setlength{\unitlength}{.5in}
\begin{picture}(8,3.5)(0,0)
\put(0.3,0.2){{\includegraphics[trim = 1.4cm 1.1cm 7cm 10.5cm, clip, width=8cm]{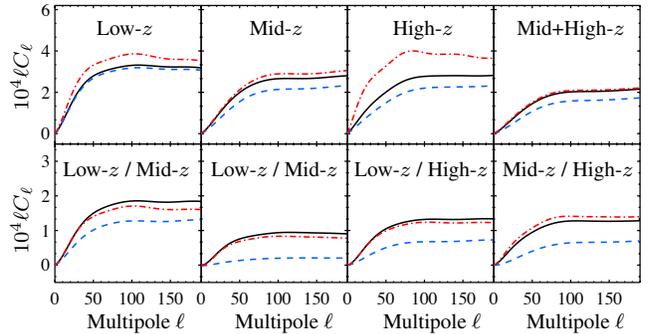}}}
\put(0.9,3.05){{Low-$z$}}
\put(2.5,3.05){{Mid-$z$}}
\put(4.0,3.05){{High-$z$}}
\put(5.3,3.05){{Mid+High-$z$}}
\put(0.55,1.57){{Low-$z$ / Mid-$z$}}
\put(2.1,1.57){{Low-$z$ / Mid-$z$}}
\put(3.6,1.57){{Low-$z$ / High-$z$}}
\put(5.15,1.57){{Mid-$z$ / High-$z$}}
\put(0.0,0.7){\rotatebox{90}{ $10^4 \ell C_\ell$}}
\put(0.0,2.2){\rotatebox{90}{ $10^4 \ell C_\ell$}}
\put(0.8,-0.02){{Multipole $\ell$}}
\put(2.35,-0.02){{Multipole $\ell$}}
\put(3.9,-0.02){{Multipole $\ell$}}
\put(5.4,-0.02){{Multipole $\ell$}}
\end{picture}
\caption{Theory predictions for the four RQCat subsamples computed with \textsc{camb\_sources}. The black lines show the final theory predictions, calculated with the best redshift distribution estimates presented in Sec.~\ref{sec:redshiftdistrib}, for a \textit{Planck} cosmology and a fixed galaxy bias $b_g=2.3$. The blue dashed lines show the predictions obtained by neglecting the effect of magnification, and the red dot-dashed lines by neglecting the completeness correction in the estimated redshift distributions (i.e., using the redshift histograms of the cross-matched samples without accounting for the differences in magnitude limits).} 
\label{fig:cltheo}
\end{figure}

Figure~\ref{fig:cltheo} shows the estimates obtained with  \textsc{camb\_sources} (thus using the full formalism from \citealt{challinorlewis2011cambsources}) with our best estimates for $s$ and $n(z)$, compared with the same estimates where magnification and completeness corrections were neglected. The large differences between the resulting angular power spectra demonstrate that these effects must be accounted for, and carefully estimated from the data in order to avoid significant biases in the theory predictions.

\subsection{Masks and systematics}\label{sec:masks}

\begin{figure*}
	\vspace{-2mm}
		\subfloat[Stellar density]{\includegraphics[trim = 7.5cm 8.2cm 9.2cm 1.8cm, clip, width=3.5cm]{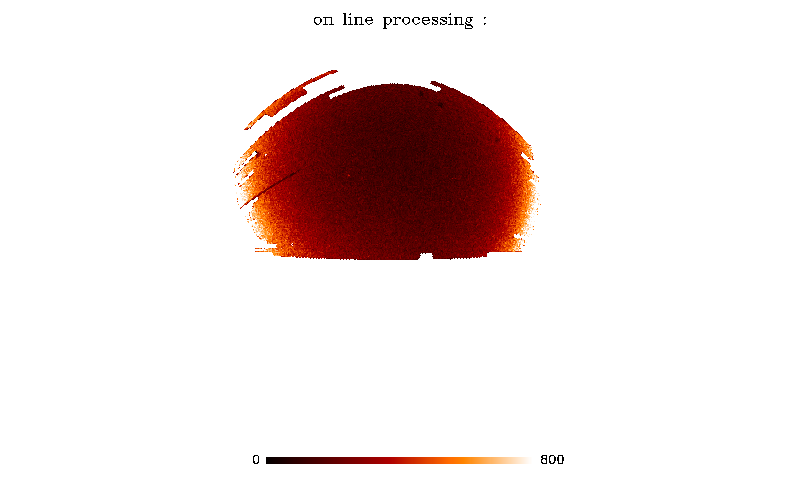}}
		\subfloat[Extinction]{\includegraphics[trim = 7.5cm 8.2cm 9.2cm 1.8cm, clip, width=3.5cm]{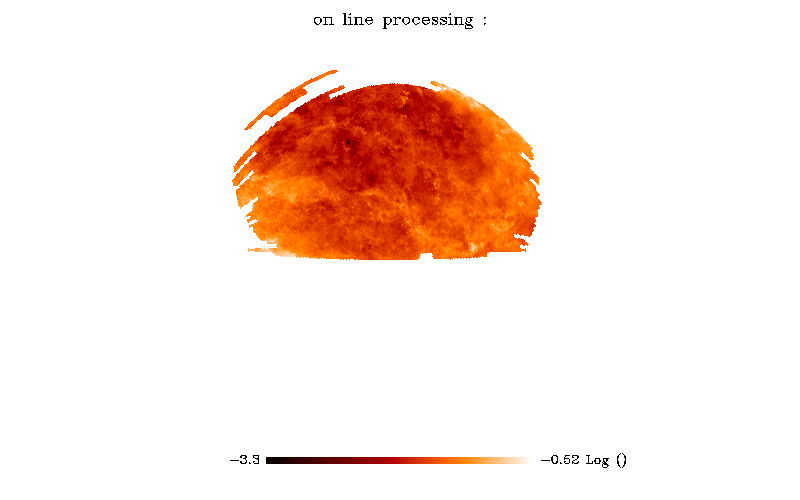}}
		\subfloat[Airmass]{\includegraphics[trim = 7.5cm 8.2cm 9.2cm 1.8cm, clip, width=3.5cm]{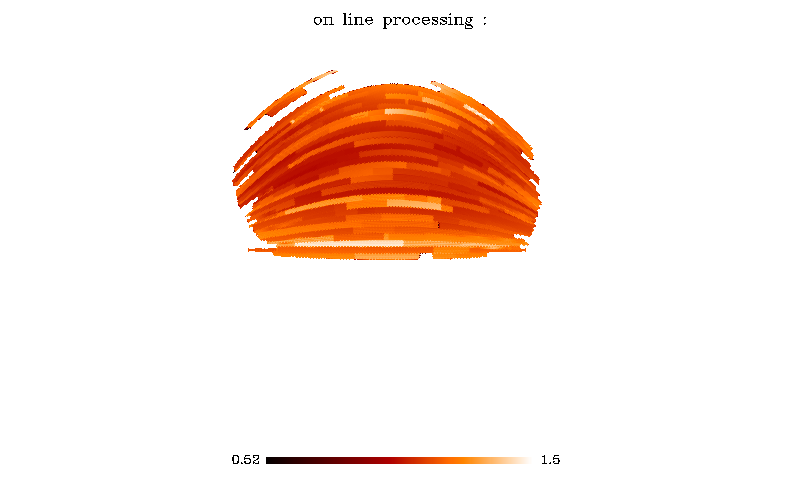}} 
		\subfloat[Seeing]{\includegraphics[trim = 7.5cm 8.2cm 9.2cm 1.8cm, clip, width=3.5cm]{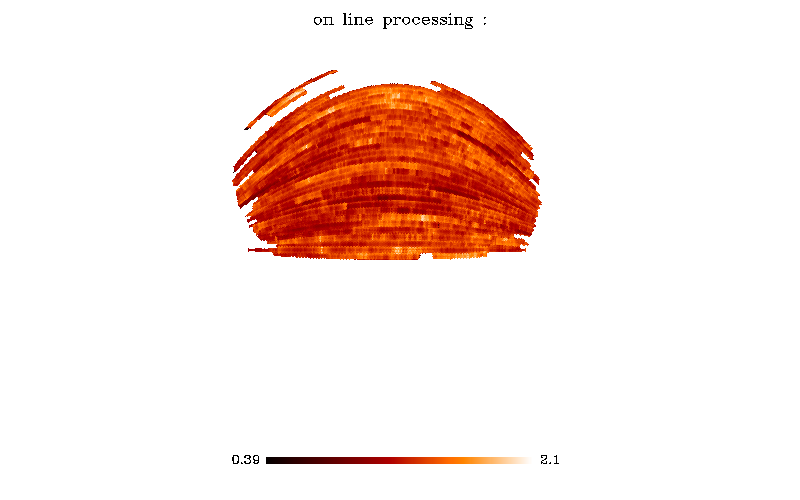}}
		\subfloat[Sky brightness]{\includegraphics[trim = 7.5cm 8.2cm 9.2cm 1.8cm, clip, width=3.5cm]{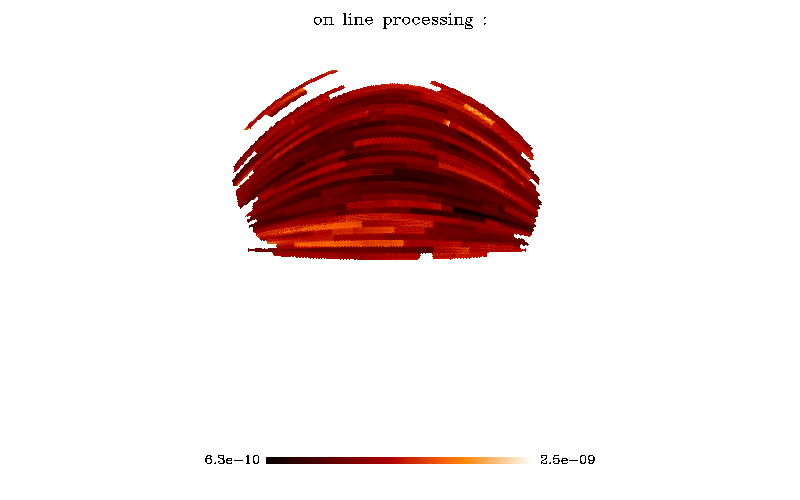}} \\[0.5\baselineskip]
\setlength{\unitlength}{.5in}
\begin{picture}(18,1.4)(0,0)
\put(0,0.1){{\includegraphics[trim = 1.6cm 14.7cm 2.5cm 4.65cm, clip, width=18cm]{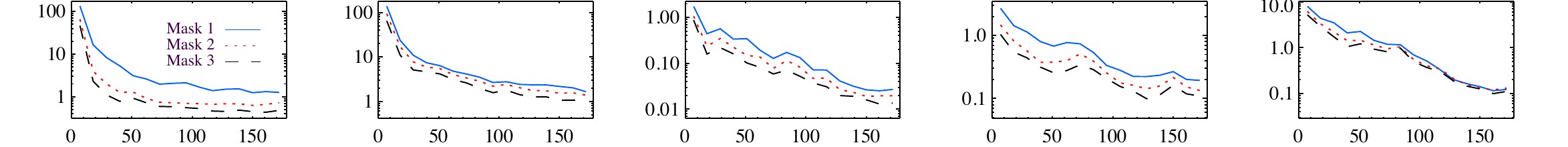}}}
\put(1.15,-0.1){{Multipole $\ell$}}
\put(3.95,-0.1){{Multipole $\ell$}}
\put(6.7,-0.1){{Multipole $\ell$}}
\put(9.5,-0.1){{Multipole $\ell$}}
\put(12.3,-0.1){{Multipole $\ell$}}
\put(0.05,0.45){\rotatebox{90}{ $10^4 \ell C_\ell$}}
\put(2.8,0.45){\rotatebox{90}{ $10^4 \ell C_\ell$}}
\put(5.55,0.45){\rotatebox{90}{ $10^4 \ell C_\ell$}}
\put(8.35,0.45){\rotatebox{90}{ $10^4 \ell C_\ell$}}
\put(11.15,0.45){\rotatebox{90}{ $10^4 \ell C_\ell$}}
\end{picture}
\caption{Systematics templates used in this analysis, and the (dimensionless) angular power spectra $\tilde{\mathcal{C}}_\ell$ of their overdensity maps. }
\label{fig:qsosys}
\end{figure*}
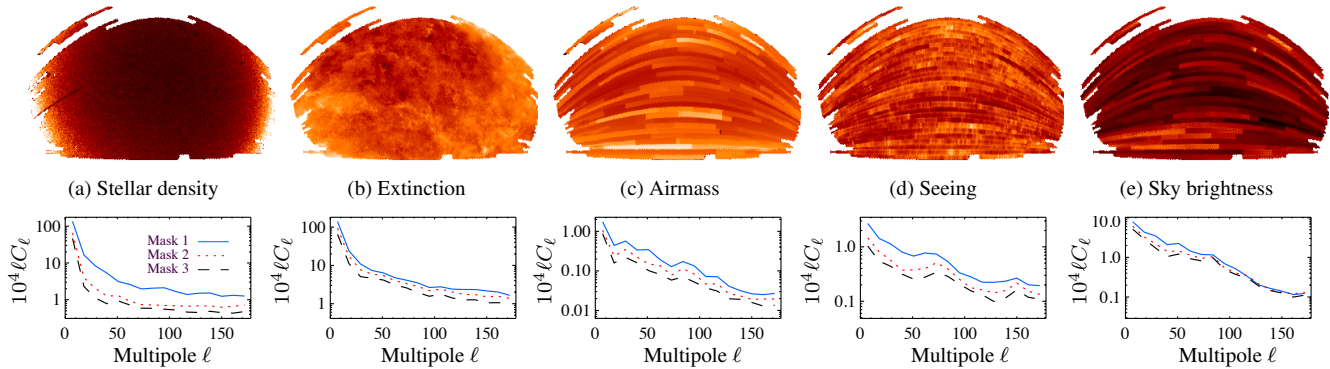

We considered five sources of systematics: stellar contamination, dust absorption, seeing, airmass and sky brightness. Following \cite{PullenHirata2012}, we constructed the stellar density map from SDSS DR6 point sources with $18.0 < r <18.5$ and $i < 21.3$. For the extinction, we used the dust maps from \cite{Schlegel1998dust} with the corrections by \cite{PeekGraves2010dust}. Templates for seeing, airmass and sky brightness were constructed with the \textsc{mangle} software \citep{Hamilton2004mangle, Swanson2008mangle} using data retrieved from the \textsc{Fields} table in the SDSS CAS server. All maps were binned onto the \textsc{healpix} grid at resolution $\nside=128$, and are shown in Fig.~\ref{fig:qsosys}.

\begin{figure}
	\vspace{-4mm}
		\subfloat[Mask 1]{\includegraphics[trim = 8.5cm 8.2cm 9.9cm 2.5cm, clip, width=2.8cm]{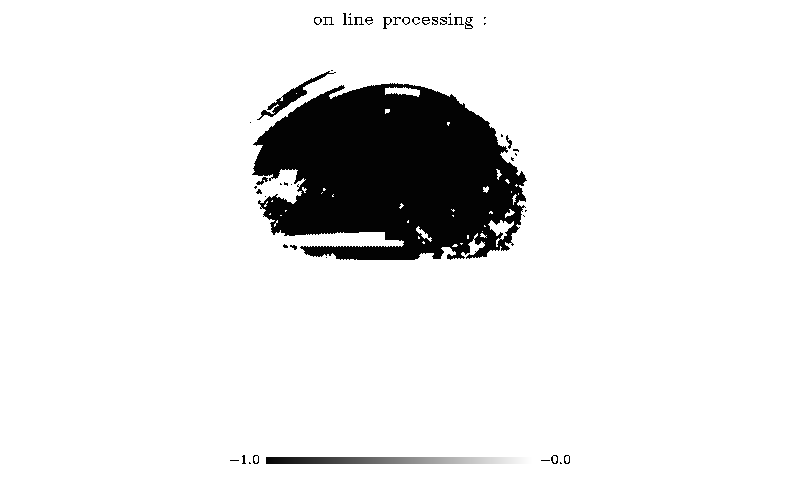}}
		\subfloat[Mask 2]{\includegraphics[trim = 8.5cm 8.2cm 9.9cm 2.5cm, clip, width=2.8cm]{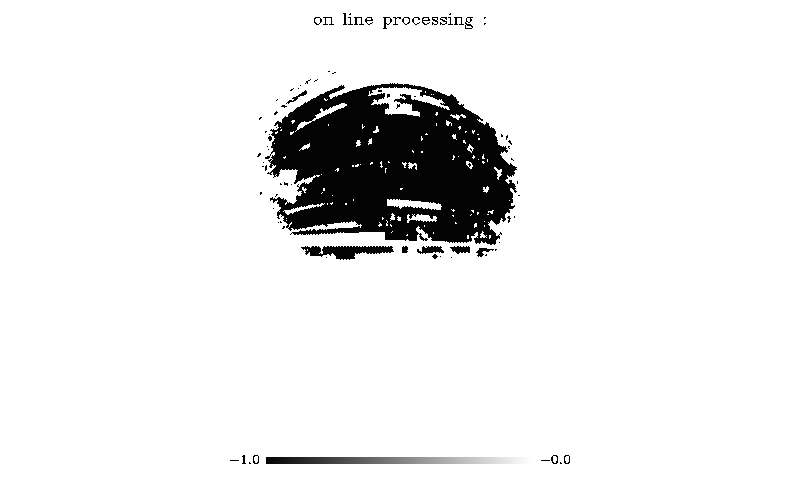}}
		\subfloat[Mask 3]{\includegraphics[trim = 8.5cm 8.2cm 9.9cm 2.5cm, clip, width=2.8cm]{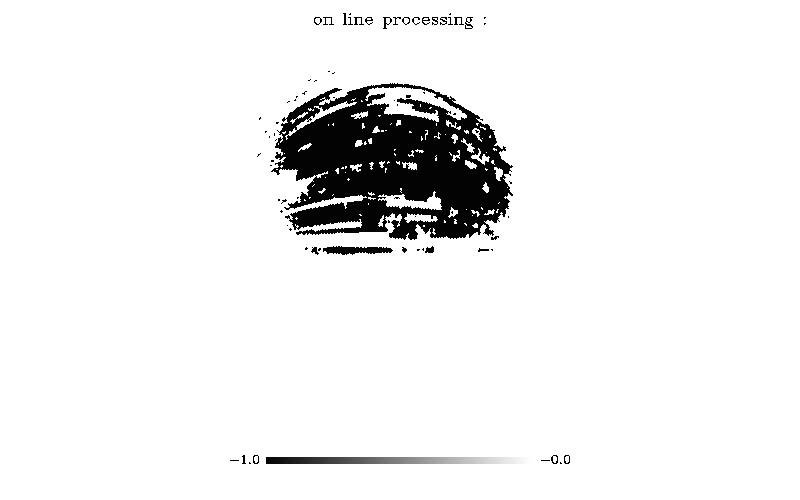}}
\caption{Masks used for the power spectrum analysis of RQCat, in Equatorial coordinates. Retained regions are based on thresholds summarised in Table \ref{systematiccuts} and the systematics templates of Fig.~\ref{fig:qsosys}. Additional excised rectangles follow Pullen \& Hirata (2012). The three masks respectively have $\fsky = 0.148, 0.121, {\rm and\ }0.101$.}
\label{fig:qsomasks}
\end{figure}

We designed three sky masks by excluding pixels based on their values in the systematics maps. The thresholds are summarised in Table \ref{systematiccuts}, and the resulting masks are presented in Fig.~\ref{fig:qsomasks}. Following \cite{PullenHirata2012}, we also excised rectangular regions with missing data\footnote{In equatorial (J2000) coordinates, the discarded angular rectangles are $(\alpha, \delta) = (122\degree-139\degree,-1.5-(-0.5)\degree), (121\degree-126\degree,0\degree-4\degree), (119\degree-128\degree,4\degree-6\degree), (111\degree-119\degree,6\degree-25\degree), (111.5\degree-117.5\degree,25\degree-30\degree), (110\degree-116\degree,32\degree-35\degree), (246\degree-251\degree,8.5\degree,13.5\degree), (255\degree-270\degree,20\degree-40\degree), (268\degree-271\degree,46\degree-49\degree), (232\degree-240\degree,26\degree-30\degree).$}. Our first mask is therefore very similar to those used in previous studies of RQCat and constitutes the reference mask. The two other masks use more aggressive systematics cuts.

The cut-sky angular power spectra, $\tilde{\mathcal{C}}_\ell$, of the systematics maps for the three masks are shown in Fig.~\ref{fig:qsosys}. Interestingly, stellar density and dust absorption templates display strong large-scale power ($\ell <30$), and calibration error templates (seeing, airmass, sky brightness) have notable features at $\ell \sim 70, 110 {\rm\ and\ } 150$. The masked template maps have these features reduced, but not eliminated. Since the data are known to be affected by these systematics, the measured spectra are likely to be contaminated at these multipoles. 

\begin{table}
\begin{minipage}{8.4cm}
\centering
	\caption{Systematics thresholds used to restrict the power spectrum analysis of RQCat to the most reliable regions of the sky to minimise contamination from calibration errors. The maps of the systematics are shown in Fig.~\ref{fig:qsosys}, and the resulting masks in Fig.~\ref{fig:qsomasks}.}
	\label{systematiccuts}
	\begin{tabular}{l|r|r|r|r}
		\hline Systematic (unit) & Mask 1 & Mask 2 & Mask 3 \\ \hline
		Seeing (arcsec) & 2.0 & 1.6 & 1.55 \\
		Reddening (mag) & 0.05 & 0.05 & 0.045 \\
		Stellar density (stars/deg$^2$)  &  562 & 400 & 350 \\
		Airmass (mag) & 1.4 & 1.3 & 1.25 \\
		Sky brightness  ({nmgy/arcsec$^2$}) \mpt\mpt & $2\mpt\times\mpt10^{-9}$\mpt\mpt & $1.8\mpt\times\mpt10^{-9}$ \mpt\mpt\mpt & $1.75\mpt\times\mpt10^{-9}$ \mpt\mpt \\ \hline
	\end{tabular}
\end{minipage}
\end{table}

\begin{figure}\centering
\setlength{\unitlength}{.5in}
\begin{picture}(16,9.3)(0,0)
\put(0.3,2.9){\includegraphics[trim = 0.8cm 3.4cm 1.5cm 3.cm, clip, width=8.1cm]{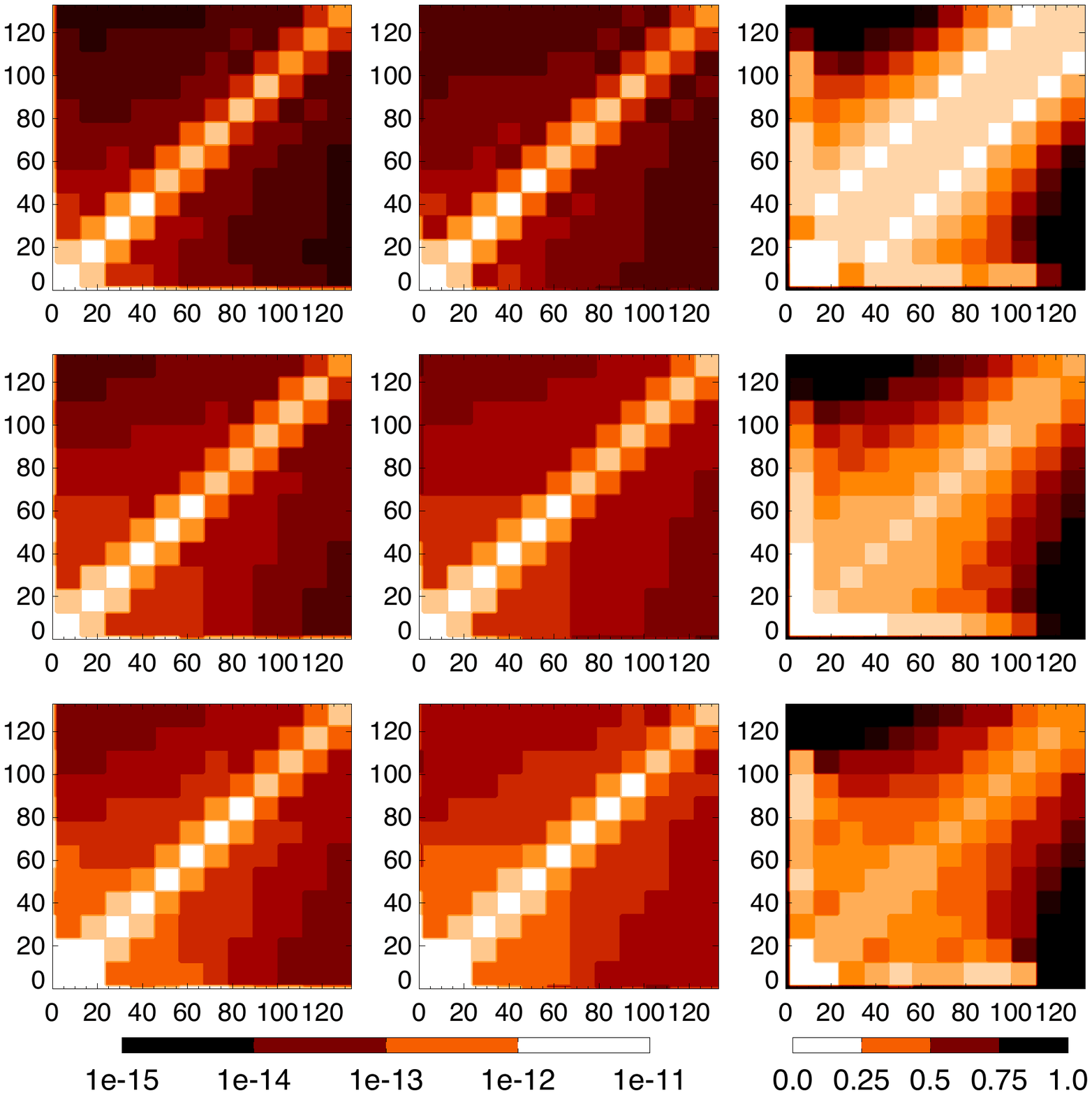}}
\put(0.3,2.5){\includegraphics[trim = 0.8cm 2.3cm 1.5cm 24.6cm, clip, width=8.1cm]{pics/covars.pdf}}
\put(0.1,7.6){\rotatebox{90}{Mask 1}}
\put(0.1,5.6){\rotatebox{90}{Mask 2}}
\put(0.1,3.6){\rotatebox{90}{Mask 3}}
\put(1,8.95){\footnotesize $|V_{bb'}^{\rm QML}|$}
\put(3.1,8.95){\footnotesize $|V_{bb'}^{\rm PCL}|$}
\put(4.7,8.95){\footnotesize 1 - $|V_{bb'}^{\rm PCL}/V_{bb'}^{\rm QML}|$}
\put(0.1,0.1){\rotatebox{90}{ $10^4 \ell (\hat{C}^{\rm PCL}_b-\hat{C}^{\rm QML}_b)$}}
\put(0.4,0.1){\includegraphics[trim = 2.0cm 12.9cm 10cm 11.5cm, clip, width=7.9cm]{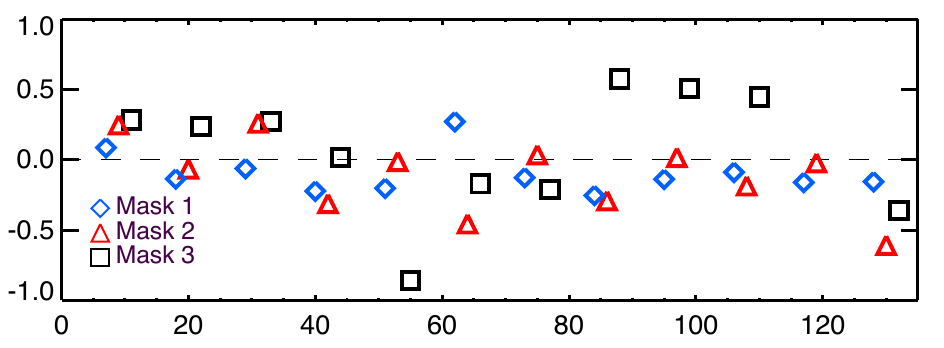}}
\put(3,-0.1){ Multipole $\ell$}
\put(0.9,1.95){ Mid+High-$z$}
\end{picture}
\caption{{Illustration of the suboptimality of the PCL estimator in the case of the Mid+High-$z$  subsample. The top panels show the covariance matrices of the PCL and QML estimates with $\Delta \ell = 11$ for the three masks of Fig.~\ref{fig:qsomasks}. The suboptimality of the PCL prior is measured by the fractional increase of variance compared to QML, shown in the right column. The bottom panel shows the resulting effects on the power spectrum estimates, which are more pronounced for the second and third masks due to their complex geometry.}}
\label{fig:qsopclqml}
\end{figure}

\subsection{Power spectrum results}

We obtained angular band-power estimates with the QML estimator and multipole bins of size $\Delta \ell = 11$, which led to a good balance in terms of multipole resolution and variance of the estimates. We did not use the PCL estimator for the final results because the geometry of the second and third masks, in addition to the presence of systematics, yielded significantly suboptimal estimates. To illustrate this point, Fig.~\ref{fig:qsopclqml} shows a comparison of the PCL and QML covariance matrices and the band-power estimates of the Mid+High-$z$ subsample for the three masks. Any significant increase of the PCL variance compared to that of QML, especially on diagonal- and nearly-diagonal elements which contain the most significant contributions, demonstrates the suboptimality of the PCL prior. For the first mask, the PCL variance of these elements is at most $\sim 20\%$ greater than the QML variance, indicating that the resulting estimates are nearly optimal. However, for the second and third masks, these elements have a PCL variance up to $\sim 50\%$ greater than that of QML, and the resulting PCL estimates significantly differ from the optimal QML estimates, as shown in the bottom panel of Fig.~\ref{fig:qsopclqml}. This effect is less pronounced for larger multipole bins (e.g., $\Delta \ell = 31$), as the likelihood becomes less sensitive to the priors on the pixel-pixel covariance matrix. However, the resulting loss of resolution prevents the study of localised multipole ranges affected by systematics. For these reasons we opted for the QML estimator with $\Delta \ell = 11$ in the final analysis. \bl{We systematically marginalised over the values the monopole and the dipole by projecting them out. They are poorly constrained from cut-sky data, and may affect the power spectrum estimates over a wide range of multipoles when deconvoling the cut-sky power spectrum into full sky estimates\footnote{\bl{In standard $P(k)$ analyses, this issue is resolved by applying an integral constraint to the power spectrum estimates (see, .e.g, \citealt{Tegmark2002earlysdss}).}}.}  We used the values $\bar{G}^{-1} = 1.95\cdot 10^{-5}, 1.55\cdot 10^{-5}, 1.85\cdot 10^{-5}$ and $8.15\cdot 10^{-6}$ respectively for the shot noise of the four RQCat subsamples, calculated from the average number count per steradian assuming $5\%$ stellar contamination. 

\begin{figure*}\centering
\setlength{\unitlength}{.5in}
\begin{picture}(16,8)(0,0)
\put(0.0,0.45){\includegraphics[trim = 0.9cm 4.0cm 1.2cm 3.5cm, clip, width=17.7cm]{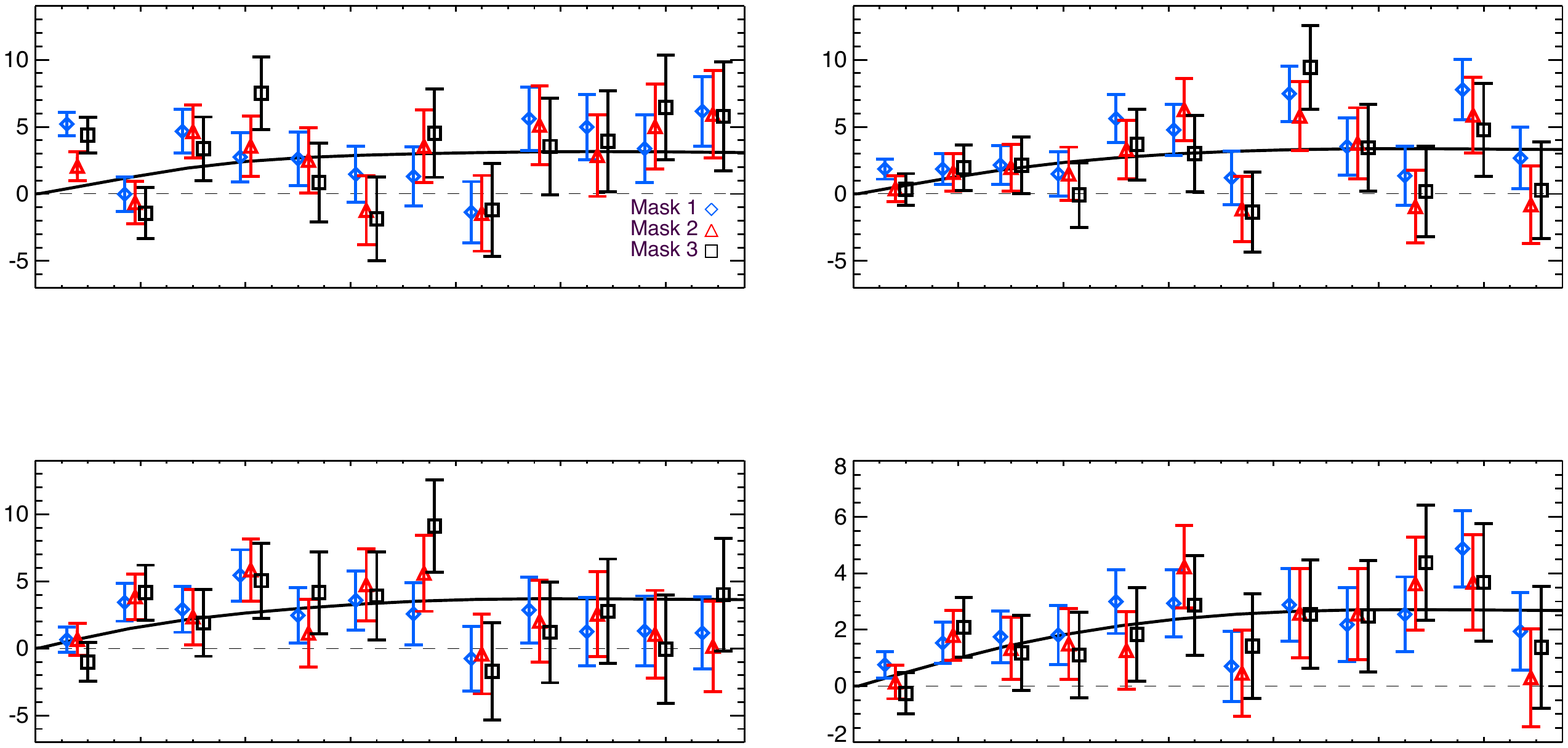}}
\put(0.0,-0.05){\includegraphics[trim = 0.9cm 4cm 1.2cm 15cm, clip, width=17.7cm]{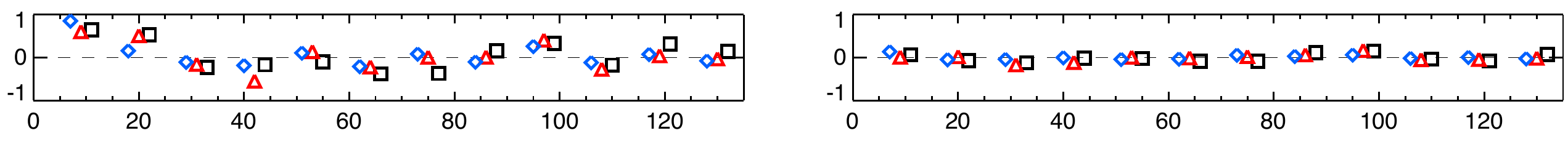}}
\put(0.0,4.){\includegraphics[trim = 0.9cm 7cm 1.2cm 5cm, clip, width=17.7cm]{pics/qso_autoqml_temp.pdf}}
\put(2.85,4){ Multipole $\ell$}
\put(2.85,-0.05){ Multipole $\ell$}
\put(10.15,4){ Multipole $\ell$}
\put(10.15,-0.05){ Multipole $\ell$}
\put(-0.2,2.1){\rotatebox{90}{ $10^4 \ell C_\ell$}}
\put(-0.2,6.1){\rotatebox{90}{ $10^4 \ell C_\ell$}}
\put(7.08,2.1){\rotatebox{90}{ $10^4 \ell C_\ell$}}
\put(7.08,6.1){\rotatebox{90}{ $10^4 \ell C_\ell$}}
\put(7.08,4.35){\rotatebox{90}{ $10^4 \ell \Delta C_\ell$}}
\put(7.08,0.3){\rotatebox{90}{ $10^4 \ell \Delta C_\ell$}}
\put(-0.2,4.35){\rotatebox{90}{ $10^4 \ell \Delta C_\ell$}}
\put(-0.2,0.3){\rotatebox{90}{ $10^4 \ell \Delta C_\ell$}}
\put(0.6,7.52){\rotatebox{0}{Low-$z$ : $0.5 < \tilde{z}_p \leq 1.3$}}
\put(0.6,3.48){\rotatebox{0}{High-$z$ : $1.8 < \tilde{z}_p \leq 2.2$}}
\put(7.9,7.52){\rotatebox{0}{Mid-$z$ : $1.3 < \tilde{z}_p \leq 1.8$}}
\put(7.9,3.48){\rotatebox{0}{Mid+High-$z$ : $1.3 < \tilde{z}_p \leq 2.2$}}
\end{picture}
\caption{QML estimates of the (dimensionless) auto-power spectra of the overdensity maps of the four RQCat samples presented in Fig.~\ref{fig:qsoredshiftdistribution}. The estimates were calculated for the three masks in Fig.~\ref{fig:qsomasks} (blue diamond, red triangles and black squares) for multipole bins of size $\Delta \ell = 11$, and compared with the theory prediction for a \textit{Planck} cosmology (solid line). The top panel shows the estimates obtaining by mode-projecting the five systematics of Fig~\ref{fig:qsosys}. The bottom panel indicates the difference in the estimates when mode projection is not used, i.e., $ \hat{C}_b^{\rm no\ mp}-\hat{C}_b^{\rm mp}$, and can be added to the top panel to recover the estimates without mode projection (the change in the covariance of the estimates due to mode projection is negligible).}
\label{fig:autoqml}
\end{figure*}

\begin{figure*}\centering
\setlength{\unitlength}{.5in}
\begin{picture}(16,8)(0,0)
\put(0.0,0.45){\includegraphics[trim = 0.9cm 4cm 1.2cm 3.5cm, clip, width=17.7cm]{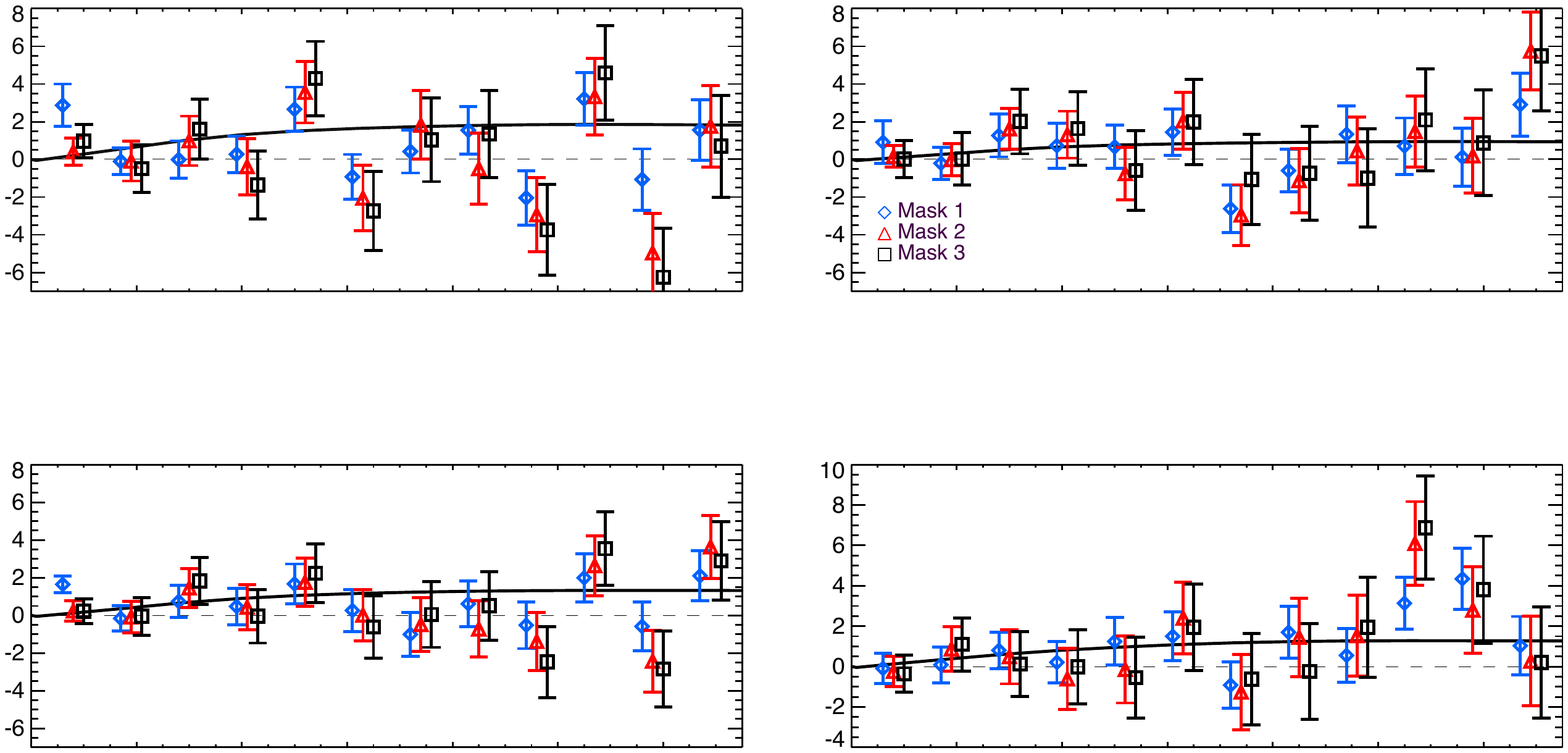}}
\put(0.0,-0.05){\includegraphics[trim = 0.9cm 4cm 1.2cm 15cm, clip, width=17.7cm]{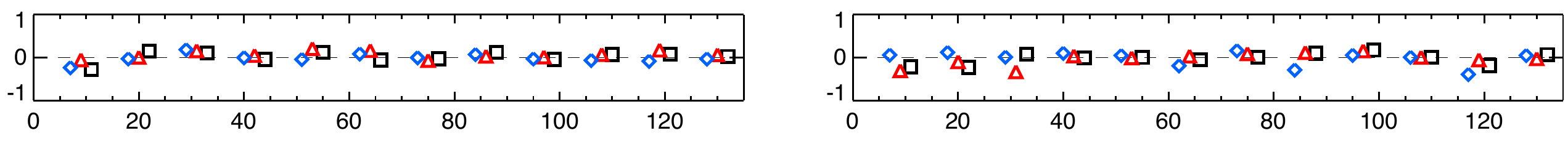}}
\put(0.0,4.){\includegraphics[trim = 0.9cm 7cm 1.2cm 5cm, clip, width=17.7cm]{pics/qso_crossqml_temp.pdf}}
\put(2.85,4){ Multipole $\ell$}
\put(2.85,-0.05){ Multipole $\ell$}
\put(10.15,4){ Multipole $\ell$}
\put(10.15,-0.05){ Multipole $\ell$}
\put(-0.2,2.1){\rotatebox{90}{ $10^4 \ell C_\ell$}}
\put(-0.2, 6.1){\rotatebox{90}{ $10^4 \ell C_\ell$}}
\put(7.08,2.1){\rotatebox{90}{ $10^4 \ell C_\ell$}}
\put(7.08,6.1){\rotatebox{90}{ $10^4 \ell C_\ell$}}
\put(7.08,4.35){\rotatebox{90}{ $10^4 \ell \Delta C_\ell$}}
\put(7.08,0.3){\rotatebox{90}{ $10^4 \ell \Delta C_\ell$}}
\put(-0.2,4.35){\rotatebox{90}{ $10^4 \ell \Delta C_\ell$}}
\put(-0.2,0.3){\rotatebox{90}{ $10^4 \ell \Delta C_\ell$}}
\put(0.6,7.52){\rotatebox{0}{Low-$z$ / Mid-$z$}}
\put(7.9,7.52){\rotatebox{0}{Low-$z$ / High-$z$}}
\put(0.6,3.48){\rotatebox{0}{Low-$z$ / Mid+High-$z$}}
\put(7.9,3.48){\rotatebox{0}{Mid-$z$ / High-$z$}}
\end{picture}
\caption{QML estimates of the (dimensionless) cross-power spectra of the RQCat overdensity maps using the same conventions as Fig.~\ref{fig:autoqml}.}
\label{fig:crossqml}
\end{figure*}

The auto- and cross-spectra of the four RQCat samples are presented in Figs.~\ref{fig:autoqml} and \ref{fig:crossqml}, and the $\chi^2$ values of the theory prediction are listed in Table~\ref{chi2vals1}. We subtracted the shot noise from the auto-spectra, and used a constant bias, $b_g=2.3$, following previous studies of these data \citep{SlosarHirata2008, Giannantonio2006isw, Giannantonio2008isw, Xia2010sdssqsoctheta, PullenHirata2012}. The theory predictions are summarised in Fig.~\ref{fig:cltheo}. We also used the exact window functions $W_{b\ell}$ for converting the theory power spectra into band-powers; see Eq.~(\ref{binnedexpectedvalue}). Figure~\ref{fig:crosssys} shows the cross-correlation power spectra of the quasar samples with the systematics templates, and Table~\ref{chi2vals2} lists the corresponding $\chi^2$ values. Details of the $\chi^2$ computation are contained in Appendix~\ref{app:chi2}. 

In Figs.~\ref{fig:autoqml} and \ref{fig:crossqml}, the top panels show the final band-power estimates, where the \bl{pixel space} modes corresponding to the five systematics templates were projected out. The effect of mode projection on the estimates is illustrated in the bottom panels, showing the differences in the QML estimates. Hence, these values can be added to the estimates in the top panels to recover the results without mode projection. The change in the covariance of the estimates due to mode projection is negligible.

\subsubsection{Reference mask}

\begin{figure*}\centering
\setlength{\unitlength}{.5in}
\begin{picture}(13.5,7.4)(0,0)
\put(-0.2,0.1){\includegraphics[trim = 1.9cm 1.6cm 3cm 8.6cm, clip, height=8.8cm]{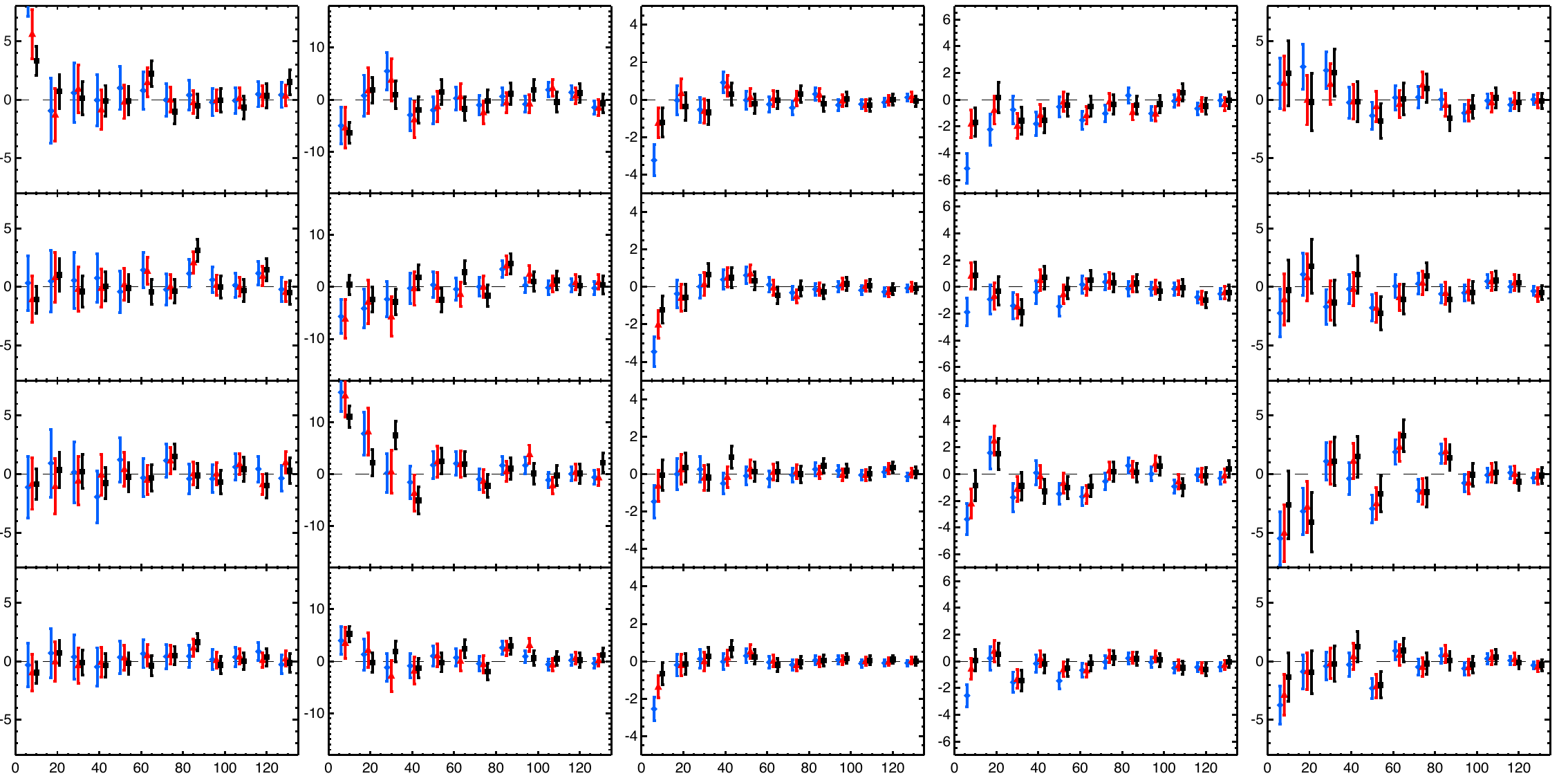}}
\put(-0.45,5.8){\rotatebox{90}{ $10^4 \ell C_\ell$}}
\put(-0.45,4.15){\rotatebox{90}{ $10^4 \ell C_\ell$}}
\put(-0.45,2.5){\rotatebox{90}{ $10^4 \ell C_\ell$}}
\put(-0.45,0.8){\rotatebox{90}{ $10^4 \ell C_\ell$}}
\put(0.9,-0.1){ Multipole $\ell$}
\put(3.6,-0.1){ Multipole $\ell$}
\put(6.4,-0.1){ Multipole $\ell$}
\put(9.1,-0.1){ Multipole $\ell$}
\put(11.9,-0.1){ Multipole $\ell$}
\put(0.15,6.71){ Low-$z$ / Stellar density}
\put(3.2,6.71){ Low-$z$ / Extinction}
\put(6.1,6.71){ Low-$z$ / Airmass}
\put(8.9,6.71){ Low-$z$ / Seeing}
\put(11.3,6.71){ Low-$z$ / Sky brightness}
\put(0.15,5){ Mid-$z$ / Stellar density}
\put(3.2,5){ Mid-$z$ / Extinction}
\put(6.1,5){ Mid-$z$ / Airmass}
\put(8.9,5){ Mid-$z$ / Seeing}
\put(11.3,5){ Mid-$z$ / Sky brightness}
\put(0.15,3.34){ High-$z$ / Stellar density}
\put(3.2,3.34){ High-$z$ / Extinction}
\put(6.1,3.34){ High-$z$ / Airmass}
\put(8.9,3.34){ High-$z$ / Seeing}
\put(11.3,3.34){ High-$z$ / Sky brightness}
\put(0.3,1.66){ Mid+High-$z$}
\put(0.7,1.45){  / Stellar density}
\put(3.25,1.66){ Mid+High-$z$}
\put(3.65,1.45){  / Extinction}
\put(6.15,1.66){ Mid+High-$z$}
\put(6.55,1.45){  / Airmass}
\put(8.95,1.66){ Mid+High-$z$ }
\put(9.35,1.45){  / Seeing}
\put(11.35,1.66){ Mid+High-$z$ }
\put(11.75,1.45){  / Sky brightness}
\end{picture}
\caption{QML estimates for the (dimensionless) cross-power spectra of the four RQCat overdensity maps with the systematics templates using the same conventions as Fig.~\ref{fig:autoqml}.}
\label{fig:crosssys}
\end{figure*}

Our first mask, which is similar to that used in previous studies of RQCat \citep{SlosarHirata2008, Giannantonio2006isw, Giannantonio2008isw, Xia2010sdssqsoctheta, PullenHirata2012}, is mostly based on extinction, stellar density and seeing cuts, and also excises a few pixels with extreme values of airmass and sky brightness. When using this reference mask, the auto-spectrum estimates of the four RQCat subsamples exhibit significant excess power in the first multipole bin. In particular, the cross-correlation of the Low-$z$  sample with the other samples confirm the presence of systematics in common. The cross-spectra of the quasar subsamples with the systematics templates, shown in Fig.~\ref{fig:crosssys}, enable us to identify the main sources of contamination responsible for this excess power. In addition to seeing and airmass, which are the main contaminants in the four samples, stellar contamination affects the Low-$z$  sample, and dust extinction and sky brightness contaminate the Mid-$z$  and High-$z$  samples. 

The auto- and cross-spectra are marginally improved by projecting out the modes corresponding to the systematics templates, as shown by the small decrease in the $\chi^2$ values, summarised in Tables~\ref{chi2vals1} and~\ref{chi2vals2}. In particular, the large-scale power excess persists, confirming the conclusions by \cite{PullenHirata2012} that the contamination must involve non-linear combinations of systematics, or else systematics which have not been accounted for.

\subsubsection{Improved masks}

Our second mask is based on more restrictive cuts on the systematics, the most important of which are seeing and stellar density cuts. Using this mask not only improves the overall quality of the estimates, as measured by the $\chi^2$, but also eliminates the excess power at low $\ell$ in all subsamples except the Low-$z$  one.  Interestingly, the cross-spectra of the Low-$z$  sample with the others exhibit no excess power. This indicates that the systematics responsible for the excess in the Low-$z$  sample are successfully mitigated in the Mid-$z$  and High-$z$  samples, and thus in the Mid+High-$z$  sample. The cross-spectra with the systematics templates are significantly decreased, although dust extinction and seeing still affect the Mid-$z$  and High-$z$  samples. Mode-projection further improves the quality of the estimates, but does not eliminate the excess power in the Low-$z$  sample.

The third mask is based even more stringent cuts on the systematics. The cross-spectra with the systematics templates show that using this mask further decreases the influence of extinction, airmass and seeing on the Mid-$z$  and High-$z$  samples. However, it fails to remove the excess power in the Low-$z$  sample, and dust extinction continues to impact the Mid-$z$  sample. Yet, no statistical anomalies are observed in the auto-spectra of Mid-$z$ , High-$z$  and Mid+High-$z$  samples. Mode-projection further improves the $\chi^2$ values of all auto- and cross-spectra. In particular, the $\chi^2$ values for the cross-spectra between the Mid-$z$  and the High-$z$  samples significantly improve, indicating a successful mitigation of the remaining levels of extinction. 

In summary, when using the third mask and mode projection, the auto-spectra of the Mid-$z$ , the High-$z$  and Mid+High-$z$  samples exhibit no evidence of systematics, and are well-fitted by the theory prediction. The cross-spectra of these samples with the Low-$z$  sample are also not anomalous, indicating the absence of systematics in common. Mode-projection eliminates the contributions (to linear order) of the five systematics we have considered, so that any remaining contamination can only involve non-linear combinations or unidentified systematics. Their presence is confirmed in the Low-$z$  subsample, but the other subsamples exhibit no evidence of contamination, indicating that any residual systematics are well within the sample variance.

\begin{table*}
\begin{minipage}{14.0cm}
\centering
	\caption{The chi square values for the auto- and cross-power spectra of the four RQCat samples presented in Figs.~\ref{fig:autoqml} and \ref{fig:crossqml}, with and without mode projection (mp). The number of degrees of freedom is $\nu-p=13$, and the probability to exceed (PTE) the observed chi squares are shown in parentheses. The results are shown in bold when PTE $<1\%$ (corresponding to $\chi^2_{13} = 27.7$). }
	\label{chi2vals1}
	\begin{tabular}{l|r|r|r|r|r|r}
	\hline $\chi_{\nu-p}^2$ & \multicolumn{2}{|c|}{Mask 1} & \multicolumn{2}{|c|}{Mask 2} & \multicolumn{2}{|c|}{Mask 3} \\ 
	 & \multicolumn{1}{c}{no mp} & \multicolumn{1}{c}{mp} & \multicolumn{1}{c}{no mp} & \multicolumn{1}{c}{mp} & \multicolumn{1}{c}{no mp} & \multicolumn{1}{c}{mp} \\ \hline
	  Low-$z$ &  {\bf 56.60} (2e-7) & {\bf 44.40} (3e-5) &  16.43 (0.23) & 12.40 (0.49) & { 25.60} (0.02)  & { 20.24} (0.09) \\
	  Mid-$z$ & 23.90 (0.03) & 22.99 (0.04) & 11.20 (0.59)  & 11.18 (0.60) & 8.86 (0.78) & 8.72 (0.79) \\
	  High-$z$ & 16.47 (0.22)  & 14.46 (0.34) & 12.98 (0.45) & 11.84 (0.54) & 11.05 (0.61) & 11.51 (0.57) \\
	  Mid+High-$z$ &  13.67 (0.40) & 13.11 (0.44) & 8.25 (0.83) & 8.30 (0.82) &  4.24 (0.99) & 4.54 (0.98) \\ \hline
	  Low-$z$ / Mid-$z$ &  {\bf 33.26} (2e-4) & { 23.76} (5e-3) &  {\bf 27.72} (1e-4) & 15.82 (0.07) & {\bf 27.93} (1e-4) & 15.35 (0.08) \\
	  Low-$z$ / High-$z$ & 10.22 (0.68) & 11.54 (0.24) & 12.13  (0.52) & 7.34 (0.60) & 6.20 (0.94) & 3.52 (0.94) \\
	  Low-$z$ / Mid+High-$z$ & { 26.30} (0.02) & { 19.97} (0.02) & { 17.90} (0.16) & 9.76 (0.37) & 13.94 (0.38) & 8.36 (0.50) \\
	  Mid-$z$ / High-$z$ &  6.18 (0.94) & 4.33 (0.89) & 11.05 (0.61) & 3.92 (0.92) & 9.27 (0.75) & 2.78 (0.97) \\ \hline
	\end{tabular}
\end{minipage}
\end{table*}

\begin{table*}
\begin{minipage}{15.4cm}
\centering
	\caption{The chi square values for the cross-power spectra of the four RQCat samples with the systematics templates, presented in Fig.~\ref{fig:crosssys}, using the same conventions as Table~\ref{chi2vals1}.}
	\label{chi2vals2}
	\hspace*{-4mm}	\begin{tabular}{l|r|r|r|r|r|r}
	\hline $\chi_{\nu-p}^2$ & \multicolumn{1}{|c|}{Stellar density} & \multicolumn{1}{|c|}{Extinction} & \multicolumn{1}{|c|}{Airmass} & \multicolumn{1}{|c|}{Seeing} & \multicolumn{1}{|c|}{Sky brightness}  \\ 
	 & \multicolumn{1}{c}{Masks 1 / 2 / 3} & \multicolumn{1}{c}{Masks 1 / 2 / 3} & \multicolumn{1}{c}{Masks 1 / 2 / 3} & \multicolumn{1}{c}{Masks 1 / 2 / 3} & \multicolumn{1}{c}{Masks 1 / 2 / 3} \\ \hline
	 Low-$z$ & 17.74 / 11.25 / 15.54 & 10.58 / 9.14 / 14.39 & { 23.97} / 8.24 / 6.79 & {\bf 50.85} / {25.29} / 12.51 & 11.31 / 4.16 / 6.99 \\
	 Mid-$z$ & 4.06 / 9.54 / 15.35 & 12.39 / 17.72 / 13.19 & 26.97 / 16.14 / 8.21 & 19.90 / 11.17 / 11.37 & 8.45 / 6.41 / 7.73 \\
	 High-$z$ & 3.35 / 5.75 / 6.55 & {\bf 28.33} / { 26.91} / {\bf 47.14} & 5.87 / 3.08 / 6.82 & {\bf 32.06} / { 20.14} / 12.18 & {\bf 27.98} / 18.64 / 15.31  \\ 
	 Mid+High & 2.88 / 5.33 / 7.77 & 11.26 / 14.33 / { 23.17} & { 20.26} / 9.11 / 6.15 & {\bf 34.75} / 14.88 / 9.63 & { 19.80} / 10.84 / 6.65 \\ \hline
	\end{tabular}
\end{minipage}
\end{table*}

\section{Discussion}\label{sec:discussion}

We have investigated the problem of estimating angular auto- and cross-power spectra on the largest scales in the presence of systematics, and applied this framework to the UVX sources in the RQCat catalogue of SDSS photometric quasars. Previous studies \citep{Xia2011sdssqsocell, PullenHirata2012, Giannantonio2013png} indicated that this catalogue was not suitable for clustering analyses due to the high-levels of contamination by the systematics. We examined these conclusions by focusing on $0.5< \tilde{z}_p \leq 2.2$ objects, divided into four redshift bins, and attempted to remove the influence of the main systematics using improved sky masks and mode projection. We also improved the theoretical predictions by making use of refined redshift distributions estimated by cross-matching objects in RQCat with the SDSS-DR7 spectroscopic catalogue and applying robust completeness corrections.

In agreement with previous studies, we found that $\tilde{z}_p < 1.3$ objects exhibited significant levels of contamination by systematics, in particular dust absorption, airmass and seeing, which could not be eliminated by masking and mode projection. The remaining excess power on large scales points to the presence of unknown or non-linear combinations of the systematics in this redshift bin. 

The large-scale excess power observed in the auto- and cross-spectra of $1.3 < \tilde{z}_p \leq 2.2$ objects was eliminated by using improved sky masks based on templates of five of the main systematics. The agreement with the theory predictions further improved when projecting out the modes corresponding to these systematics. Within the statistical uncertainties, we found no evidence for remaining contamination in this sample. We conclude that photometric quasar samples can be made suitable for cosmological studies.

We did not attempt to model the contamination signal, but rather constructed a sample (through object selection and masking) which exhibited negligible levels of contamination. This approach relies on the ability to measure these levels of contamination, in our case through auto- and cross-spectra of the samples and cross-spectra with systematics templates, which are limited by the variance of the estimates due to the shot noise and sky coverage. Consequently, more contaminated samples (such as the whole RQCat) with higher number densities require a model of the systematics in order to obtain clustering measurements that are not dominated by spurious correlations. In this context, mode projection can be used to marginalise over the parameters of linear contamination models while estimating the power spectrum. However, future surveys may require more sophisticated algorithms, or even Bayesian component separation models similar to \textsc{Commander-Ruler} \mbox{\citep{Eriksen2006commander, Eriksen2008commander}} used in the context of \textit{Planck} \citep{Planck2013component}.

The clustering properties of the large galaxy and quasar catalogues produced by next generation photometric surveys such as DES will put tight constraints on models of galaxy bias and PNG. However, stellar contamination and calibration errors will always be present in galaxy survey data, and may compromise our understanding of the observables if not correctly treated. Thus, as sample variance becomes steadily smaller with increasing catalogue sizes, efficient strategies for mitigating the systematics, such as those presented here, will become critical for the cosmological interpretation of these surveys.

\section*{Aknowledgements}
We thank Stephen Feeney, Filipe Abdalla and Ofer Lahav for useful discussions and comments. BL is supported by the Perren Fund and the IMPACT Fund. HVP is supported by STFC, the Leverhulme Trust, and the European Research Council under the European Community's Seventh Framework Programme (FP7/2007- 2013) / ERC grant agreement no 306478-CosmicDawn. ABL is supported by the Leverhulme Trust and STFC. AP is supported by the Oxford Martin School. We acknowledge use of the following public software packages: \textsc{healpix} \citep{healpix1}; \textsc{mangle} \citep{Hamilton2004mangle, Swanson2008mangle} and \textsc{camb\_sources} \citep{challinorlewis2011cambsources}. We acknowledge use of the Legacy Archive for Microwave Background Data Analysis (LAMBDA). Support for LAMBDA is provided by the NASA Office of Space Science.

\footnotesize{
  \bibliographystyle{mn2e}
\providecommand{\eprint}[1]{\href{http://arxiv.org/abs/#1}{arXiv:#1}}	
  \bibliography{bib}

\begin{thebibliography}{}

\bibitem[\protect\citeauthoryear{{Adelman-McCarthy} et~al.,}{{Adelman-McCarthy}
  et~al.}{2008}]{AdelmanMcCarthy2008dr6}
{Adelman-McCarthy} J.~K.  et~al., 2008, Astrophys.\ J.\ Supp., 175, 297,
  \eprint{0707.3413}

\bibitem[\protect\citeauthoryear{Albrecht \& Steinhardt}{Albrecht \&
  Steinhardt}{1982}]{AlbrechtSteinhardt1982inflation}
Albrecht A.,  Steinhardt P.~J.,  1982, Phys. Rev. Lett., 48, 1220

\bibitem[\protect\citeauthoryear{{Aurich} \& {Lustig}}{{Aurich} \&
  {Lustig}}{2011}]{Aurich2011smoothingbias}
{Aurich} R.,  {Lustig} S.,  2011, Mon.\ Not.\ Roy.\ Astron.\ Soc., 411, 124,
  \eprint{1005.5069}

\bibitem[\protect\citeauthoryear{{Bennett} et~al.,}{{Bennett}
  et~al.}{2012}]{Bennett2012wmap9}
{Bennett} C.~L.  et~al., 2012, ArXiv e-prints, \eprint{1212.5225}

\bibitem[\protect\citeauthoryear{{Bond}, {Jaffe} \& {Knox}}{{Bond}
  et~al.}{1998}]{BJK98b}
{Bond} J.~R.,  {Jaffe} A.~H.,    {Knox} L.,  1998, Phys.\ Rev.\ D., 57, 2117,
  \eprint{astro-ph/9708203}

\bibitem[\protect\citeauthoryear{{Bond}, {Jaffe} \& {Knox}}{{Bond}
  et~al.}{2000}]{BJK98}
{Bond} J.~R.,  {Jaffe} A.~H.,    {Knox} L.,  2000, Astrophys.\ J., 533, 19,
  \eprint{astro-ph/9808264}

\bibitem[\protect\citeauthoryear{{Bovy} et~al.,}{{Bovy}
  et~al.}{2011}]{Bovy2010xdqso}
{Bovy} J.  et~al., 2011, Astrophys.\ J., 729, 141, \eprint{1011.6392}

\bibitem[\protect\citeauthoryear{{Brown}, {Castro} \& {Taylor}}{{Brown}
  et~al.}{2005}]{BCT05}
{Brown} M.~L.,  {Castro} P.~G.,    {Taylor} A.~N.,  2005, Mon.\ Not.\ Roy.\
  Astron.\ Soc., 360, 1262, \eprint{astro-ph/0410394}

\bibitem[\protect\citeauthoryear{{Challinor} \& {Lewis}}{{Challinor} \&
  {Lewis}}{2011}]{challinorlewis2011cambsources}
{Challinor} A.,  {Lewis} A.,  2011, Phys.\ Rev.\ D., 84, 043516,
  \eprint{1105.5292}

\bibitem[\protect\citeauthoryear{{Copi}, {Huterer}, {Schwarz} \&
  {Starkman}}{{Copi} et~al.}{2011}]{Copi2011smoothingbias}
{Copi} C.~J.,  {Huterer} D.,  {Schwarz} D.~J.,    {Starkman} G.~D.,  2011,
  ArXiv e-prints, \eprint{1103.3505}

\bibitem[\protect\citeauthoryear{Corasaniti, Giannantonio \&
  Melchiorri}{Corasaniti et~al.}{2005}]{Giannantonio2005de}
Corasaniti P.-S.,  Giannantonio T.,    Melchiorri A.,  2005, Phys. Rev. D, 71,
  123521

\bibitem[\protect\citeauthoryear{{Crocce}, {Pueblas} \& {Scoccimarro}}{{Crocce}
  et~al.}{2006}]{Crocce20062lpt}
{Crocce} M.,  {Pueblas} S.,    {Scoccimarro} R.,  2006, Mon.\ Not.\ Roy.\
  Astron.\ Soc., 373, 369, \eprint{astro-ph/0606505}

\bibitem[\protect\citeauthoryear{{Croom} et~al.,}{{Croom}
  et~al.}{2009}]{Croom20092slaqcat}
{Croom} S.~M.  et~al., 2009, Mon.\ Not.\ Roy.\ Astron.\ Soc., 392, 19,
  \eprint{0810.4955}

\bibitem[\protect\citeauthoryear{{Dalal}, {Dor{\'e}}, {Huterer} \&
  {Shirokov}}{{Dalal} et~al.}{2008}]{Dalal2008png}
{Dalal} N.,  {Dor{\'e}} O.,  {Huterer} D.,    {Shirokov} A.,  2008, Phys.\
  Rev.\ D., 77, 123514, \eprint{0710.4560}

\bibitem[\protect\citeauthoryear{{Efstathiou}}{{Efstathiou}}{2004a}]{Efsta2003}
{Efstathiou} G.,  2004a, Mon.\ Not.\ Roy.\ Astron.\ Soc., 348, 885,
  \eprint{astro-ph/0310207}

\bibitem[\protect\citeauthoryear{{Efstathiou}}{{Efstathiou}}{2004b}]{Efsta2004}
{Efstathiou} G.,  2004b, Mon.\ Not.\ Roy.\ Astron.\ Soc., 349, 603,
  \eprint{astro-ph/0307515}

\bibitem[\protect\citeauthoryear{{Efstathiou}}{{Efstathiou}}{2006}]{Efsta2006}
{Efstathiou} G.,  2006, Mon.\ Not.\ Roy.\ Astron.\ Soc., 370, 343,
  \eprint{astro-ph/0601107}

\bibitem[\protect\citeauthoryear{{Eisenstein} et~al.,}{{Eisenstein}
  et~al.}{2011}]{Eisenstein2011}
{Eisenstein} D.~J.  et~al., 2011, Astrophys.\ J., 142, 72, \eprint{1101.1529}

\bibitem[\protect\citeauthoryear{{Eriksen} et~al.,}{{Eriksen}
  et~al.}{2006}]{Eriksen2006commander}
{Eriksen} H.~K.  et~al., 2006, Astrophys.\ J., 641, 665,
  \eprint{astro-ph/0508268}

\bibitem[\protect\citeauthoryear{{Eriksen} et~al.,}{{Eriksen}
  et~al.}{2007}]{Eriksen2007wmap3reanalysis}
{Eriksen} H.~K.  et~al., 2007, Astrophys.\ J., 656, 641,
  \eprint{astro-ph/0606088}

\bibitem[\protect\citeauthoryear{{Eriksen}, {Jewell}, {Dickinson}, {Banday},
  {G{\'o}rski} \& {Lawrence}}{{Eriksen} et~al.}{2008}]{Eriksen2008commander}
{Eriksen} H.~K.,  {Jewell} J.~B.,  {Dickinson} C.,  {Banday} A.~J.,
  {G{\'o}rski} K.~M.,    {Lawrence} C.~R.,  2008, Astrophys.\ J., 676, 10,
  \eprint{0709.1058}

\bibitem[\protect\citeauthoryear{{Feeney}, {Peiris} \& {Pontzen}}{{Feeney}
  et~al.}{2011}]{FPP11}
{Feeney} S.~M.,  {Peiris} H.~V.,    {Pontzen} A.,  2011, Phys.\ Rev.\ D., 84,
  103002, \eprint{1107.5466}

\bibitem[\protect\citeauthoryear{Giannantonio, Crittenden, Nichol \&
  Ross}{Giannantonio et~al.}{2012}]{Giannantonio01112012}
Giannantonio T.,  Crittenden R.,  Nichol R.,    Ross A.~J.,  2012, Mon.\ Not.\
  Roy.\ Astron.\ Soc., 426, 2581,
  \eprint{http://mnras.oxfordjournals.org/content/426/3/2581.full.pdf+html}

\bibitem[\protect\citeauthoryear{{Giannantonio} et~al.,}{{Giannantonio}
  et~al.}{2006}]{Giannantonio2006isw}
{Giannantonio} T.  et~al., 2006, Phys.\ Rev.\ D., 74, 063520

\bibitem[\protect\citeauthoryear{{Giannantonio}, {Ross}, {Percival},
  {Crittenden}, {Bacher}, {Kilbinger}, {Nichol} \& {Weller}}{{Giannantonio}
  et~al.}{2013}]{Giannantonio2013png}
{Giannantonio} T.,  {Ross} A.~J.,  {Percival} W.~J.,  {Crittenden} R.,
  {Bacher} D.,  {Kilbinger} M.,  {Nichol} R.,    {Weller} J.,  2013, ArXiv
  e-prints, \eprint{1303.1349}

\bibitem[\protect\citeauthoryear{{Giannantonio}, {Scranton}, {Crittenden},
  {Nichol}, {Boughn}, {Myers} \& {Richards}}{{Giannantonio}
  et~al.}{2008}]{Giannantonio2008isw}
{Giannantonio} T.,  {Scranton} R.,  {Crittenden} R.~G.,  {Nichol} R.~C.,
  {Boughn} S.~P.,  {Myers} A.~D.,    {Richards} G.~T.,  2008, Phys.\ Rev.\ D.,
  77, 123520, \eprint{0801.4380}

\bibitem[\protect\citeauthoryear{{G{\'o}rski}, {Hivon}, {Banday}, {Wandelt},
  {Hansen}, {Reinecke} \& {Bartelmann}}{{G{\'o}rski} et~al.}{2005}]{healpix1}
{G{\'o}rski} K.~M.,  {Hivon} E.,  {Banday} A.~J.,  {Wandelt} B.~D.,  {Hansen}
  F.~K.,  {Reinecke} M.,    {Bartelmann} M.,  2005, Astrophys.\ J., 622, 759,
  \eprint{astro-ph/0409513}

\bibitem[\protect\citeauthoryear{{Gruetjen} \& {Shellard}}{{Gruetjen} \&
  {Shellard}}{2012}]{GruetjenAndShellard2012}
{Gruetjen} H.~F.,  {Shellard} E.~P.~S.,  2012, ArXiv e-prints,
  \eprint{1212.6945}

\bibitem[\protect\citeauthoryear{{Gunn} et~al.,}{{Gunn}
  et~al.}{1998}]{Gunn1998}
{Gunn} J.~E.  et~al., 1998, Astrophys.\ J., 116, 3040,
  \eprint{astro-ph/9809085}

\bibitem[\protect\citeauthoryear{{Gunn} et~al.,}{{Gunn}
  et~al.}{2006}]{Gunn2006}
{Gunn} J.~E.  et~al., 2006, Astrophys.\ J., 131, 2332,
  \eprint{astro-ph/0602326}

\bibitem[\protect\citeauthoryear{{Guth}}{{Guth}}{1981}]{Guth1981inflation}
{Guth} A.~H.,  1981, Phys.\ Rev.\ D., 23, 347

\bibitem[\protect\citeauthoryear{{Hamilton} \& {Tegmark}}{{Hamilton} \&
  {Tegmark}}{2004}]{Hamilton2004mangle}
{Hamilton} A.~J.~S.,  {Tegmark} M.,  2004, Mon.\ Not.\ Roy.\ Astron.\ Soc.,
  349, 115, \eprint{astro-ph/0306324}

\bibitem[\protect\citeauthoryear{{Hamimeche} \& {Lewis}}{{Hamimeche} \&
  {Lewis}}{2009}]{HamimecheAndLewis2009}
{Hamimeche} S.,  {Lewis} A.,  2009, Phys.\ Rev.\ D., 79, 083012,
  \eprint{0902.0674}

\bibitem[\protect\citeauthoryear{{Hinshaw} et~al.,}{{Hinshaw}
  et~al.}{2012}]{Hinshaw2012wmap9}
{Hinshaw} G.  et~al., 2012, ArXiv e-prints, \eprint{1212.5226}

\bibitem[\protect\citeauthoryear{{Hivon}, {G{\'o}rski}, {Netterfield}, {Crill},
  {Prunet} \& {Hansen}}{{Hivon} et~al.}{2002}]{HGN02}
{Hivon} E.,  {G{\'o}rski} K.~M.,  {Netterfield} C.~B.,  {Crill} B.~P.,
  {Prunet} S.,    {Hansen} F.,  2002, Astrophys.\ J., 567, 2,
  \eprint{astro-ph/0105302}

\bibitem[\protect\citeauthoryear{{Ho} et~al.,}{{Ho}
  et~al.}{2012}]{ho2012cosmoweights}
{Ho} S.  et~al., 2012, Astrophys.\ J., 761, 14, \eprint{1201.2137}

\bibitem[\protect\citeauthoryear{{Ho}, {Hirata}, {Padmanabhan}, {Seljak} \&
  {Bahcall}}{{Ho} et~al.}{2008}]{ho_isw_2008}
{Ho} S.,  {Hirata} C.,  {Padmanabhan} N.,  {Seljak} U.,    {Bahcall} N.,  2008,
  Phys.\ Rev.\ D., 78, 043519, \eprint{0801.0642}

\bibitem[\protect\citeauthoryear{{Huterer}, {Cunha} \& {Fang}}{{Huterer}
  et~al.}{2013}]{Huterer2012calibrationerrors}
{Huterer} D.,  {Cunha} C.~E.,    {Fang} W.,  2013, Mon.\ Not.\ Roy.\ Astron.\
  Soc., \eprint{1211.1015}

\bibitem[\protect\citeauthoryear{{Huterer}, {Knox} \& {Nichol}}{{Huterer}
  et~al.}{2001}]{Huterer2001eds}
{Huterer} D.,  {Knox} L.,    {Nichol} R.~C.,  2001, Astrophys.\ J., 555, 547,
  \eprint{astro-ph/0011069}

\bibitem[\protect\citeauthoryear{{Jaffe}, {Bond}, {Ferreira} \& {Knox}}{{Jaffe}
  et~al.}{1999}]{Jaffe1999cmblikelihoods}
{Jaffe} A.~H.,  {Bond} J.~R.,  {Ferreira} P.~G.,    {Knox} L.~E.,  1999, in
  {Maiani} L.,  {Melchiorri} F.,   {Vittorio} N.,  eds,  American Institute of
  Physics Conference Series Vol. 476, 3K cosmology. pp 249--365

\bibitem[\protect\citeauthoryear{{Knox}, {Bond} \& {Jaffe}}{{Knox}
  et~al.}{1998}]{KBJ98}
{Knox} L.,  {Bond} J.~R.,    {Jaffe} A.~H.,  1998, in {A.~V.~Olinto,
  J.~A.~Frieman, \& D.~N.~Schramm} ed., Eighteenth Texas Symposium on
  Relativistic Astrophysics. p.~282

\bibitem[\protect\citeauthoryear{Linde}{Linde}{1982}]{Linde1982inflation}
Linde A.,  1982, Physics Letters B, 108, 389

\bibitem[\protect\citeauthoryear{{Manera} et~al.,}{{Manera}
  et~al.}{2013}]{manera2012}
{Manera} M.  et~al., 2013, Mon.\ Not.\ Roy.\ Astron.\ Soc., 428, 1036,
  \eprint{1203.6609}

\bibitem[\protect\citeauthoryear{{Matarrese} \& {Verde}}{{Matarrese} \&
  {Verde}}{2008}]{matarrese2008}
{Matarrese} S.,  {Verde} L.,  2008, Astrophys.\ J.\ Lett., 677, L77,
  \eprint{0801.4826}

\bibitem[\protect\citeauthoryear{{Myers}, {Brunner}, {Nichol}, {Richards},
  {Schneider} \& {Bahcall}}{{Myers} et~al.}{2007}]{Myers2007one}
{Myers} A.~D.,  {Brunner} R.~J.,  {Nichol} R.~C.,  {Richards} G.~T.,
  {Schneider} D.~P.,    {Bahcall} N.~A.,  2007, Astrophys.\ J., 658, 85,
  \eprint{arXiv:astro-ph/0612190}

\bibitem[\protect\citeauthoryear{{Myers} et~al.,}{{Myers}
  et~al.}{2006}]{Myers2006first}
{Myers} A.~D.  et~al., 2006, Astrophys.\ J., 638, 622,
  \eprint{arXiv:astro-ph/0510371}

\bibitem[\protect\citeauthoryear{{Padmanabhan}, {Hirata}, {Seljak}, {Schlegel},
  {Brinkmann} \& {Schneider}}{{Padmanabhan}
  et~al.}{2005}]{Padmanabhan_isw_2005}
{Padmanabhan} N.,  {Hirata} C.~M.,  {Seljak} U.,  {Schlegel} D.~J.,
  {Brinkmann} J.,    {Schneider} D.~P.,  2005, Phys.\ Rev.\ D., 72, 043525,
  \eprint{astro-ph/0410360}

\bibitem[\protect\citeauthoryear{{P{\^a}ris} et~al.,}{{P{\^a}ris}
  et~al.}{2012}]{paris2012bossqsodr9}
{P{\^a}ris} I.  et~al., 2012, Astron.\ \& Astrophys., 548, A66,
  \eprint{1210.5166}

\bibitem[\protect\citeauthoryear{{Peek} \& {Graves}}{{Peek} \&
  {Graves}}{2010}]{PeekGraves2010dust}
{Peek} J.~E.~G.,  {Graves} G.~J.,  2010, Astrophys.\ J., 719, 415,
  \eprint{1006.3310}

\bibitem[\protect\citeauthoryear{{Planck Collaboration}}{{Planck
  Collaboration}}{2013a}]{Planck2013component}
{Planck Collaboration} 2013a, ArXiv e-prints, \eprint{1303.5072}

\bibitem[\protect\citeauthoryear{{Planck Collaboration}}{{Planck
  Collaboration}}{2013b}]{Planck2013cosmologicalparams}
{Planck Collaboration} 2013b, ArXiv e-prints, \eprint{1303.5076}

\bibitem[\protect\citeauthoryear{{Pontzen} \& {Peiris}}{{Pontzen} \&
  {Peiris}}{2010}]{PP10}
{Pontzen} A.,  {Peiris} H.~V.,  2010, Phys.\ Rev.\ D., 81, 103008,
  \eprint{1004.2706}

\bibitem[\protect\citeauthoryear{{Pullen} \& {Hirata}}{{Pullen} \&
  {Hirata}}{2012}]{PullenHirata2012}
{Pullen} A.~R.,  {Hirata} C.~M.,  2012, ArXiv e-prints, \eprint{1212.4500}

\bibitem[\protect\citeauthoryear{{Richards} et~al.,}{{Richards}
  et~al.}{2009}]{Richards2008rqcat}
{Richards} G.~T.  et~al., 2009, Astrophys.\ J.\ Supp., 180, 67,
  \eprint{0809.3952}

\bibitem[\protect\citeauthoryear{{Ross} et~al.,}{{Ross}
  et~al.}{2011}]{ross2011weights}
{Ross} A.~J.  et~al., 2011, Mon.\ Not.\ Roy.\ Astron.\ Soc., 417, 1350,
  \eprint{1105.2320}

\bibitem[\protect\citeauthoryear{{Ross} et~al.,}{{Ross}
  et~al.}{2013}]{ross2012png}
{Ross} A.~J.  et~al., 2013, Mon.\ Not.\ Roy.\ Astron.\ Soc., 428, 1116,
  \eprint{1208.1491}

\bibitem[\protect\citeauthoryear{{Ross} et~al.,}{{Ross}
  et~al.}{2012a}]{ross2012systematics}
{Ross} A.~J.  et~al., 2012a, Mon.\ Not.\ Roy.\ Astron.\ Soc., 424, 564,
  \eprint{1203.6499}

\bibitem[\protect\citeauthoryear{{Ross} et~al.,}{{Ross}
  et~al.}{2012b}]{Ross2012bosstarget}
{Ross} N.~P.  et~al., 2012b, Astrophys.\ J.\ Supp., 199, 3, \eprint{1105.0606}

\bibitem[\protect\citeauthoryear{{Samushia} et~al.,}{{Samushia}
  et~al.}{2013}]{Samushia2013grdeviation}
{Samushia} L.  et~al., 2013, Mon.\ Not.\ Roy.\ Astron.\ Soc., 429, 1514,
  \eprint{1206.5309}

\bibitem[\protect\citeauthoryear{{S{\'a}nchez} et~al.,}{{S{\'a}nchez}
  et~al.}{2012}]{sanchez2012bao}
{S{\'a}nchez} A.~G.  et~al., 2012, Mon.\ Not.\ Roy.\ Astron.\ Soc., 425, 415,
  \eprint{1203.6616}

\bibitem[\protect\citeauthoryear{{Schlegel}, {Finkbeiner} \&
  {Davis}}{{Schlegel} et~al.}{1998}]{Schlegel1998dust}
{Schlegel} D.~J.,  {Finkbeiner} D.~P.,    {Davis} M.,  1998, Astrophys.\ J.,
  500, 525, \eprint{astro-ph/9710327}

\bibitem[\protect\citeauthoryear{{Schneider} et~al.,}{{Schneider}
  et~al.}{2010}]{Schneider2010qsodr7cat}
{Schneider} D.~P.  et~al., 2010, Astrophys.\ J., 139, 2360, \eprint{1004.1167}

\bibitem[\protect\citeauthoryear{{Scoccimarro}}{{Scoccimarro}}{1997}]{Scoccimarro19972lpt}
{Scoccimarro} R.,  1997, ArXiv Astrophysics e-prints, \eprint{astro-ph/9711187}

\bibitem[\protect\citeauthoryear{{Sherwin} et~al.,}{{Sherwin}
  et~al.}{2012}]{Sherwin2012qsolensing}
{Sherwin} B.~D.  et~al., 2012, Phys.\ Rev.\ D., 86, 083006, \eprint{1207.4543}

\bibitem[\protect\citeauthoryear{{Slosar}, {Hirata}, {Seljak}, {Ho} \&
  {Padmanabhan}}{{Slosar} et~al.}{2008}]{SlosarHirata2008}
{Slosar} A.,  {Hirata} C.,  {Seljak} U.,  {Ho} S.,    {Padmanabhan} N.,  2008,
  Journal of Cosmology and Astroparticle Physics, 8, 31, \eprint{0805.3580}

\bibitem[\protect\citeauthoryear{{Slosar}, {Seljak} \& {Makarov}}{{Slosar}
  et~al.}{2004}]{SlosarSeljak2004modeproj}
{Slosar} A.,  {Seljak} U.,    {Makarov} A.,  2004, Phys.\ Rev.\ D., 69, 123003,
  \eprint{astro-ph/0403073}

\bibitem[\protect\citeauthoryear{{Swanson}, {Tegmark}, {Hamilton} \&
  {Hill}}{{Swanson} et~al.}{2008}]{Swanson2008mangle}
{Swanson} M.~E.~C.,  {Tegmark} M.,  {Hamilton} A.~J.~S.,    {Hill} J.~C.,
  2008, Mon.\ Not.\ Roy.\ Astron.\ Soc., 387, 1391, \eprint{0711.4352}

\bibitem[\protect\citeauthoryear{{Tegmark}}{{Tegmark}}{1997}]{Teg97}
{Tegmark} M.,  1997, Phys.\ Rev.\ D., 55, 5895, \eprint{astro-ph/9611174}

\bibitem[\protect\citeauthoryear{{Tegmark} \& {de Oliveira-Costa}}{{Tegmark} \&
  {de Oliveira-Costa}}{2001}]{tegmark2001}
{Tegmark} M.,  {de Oliveira-Costa} A.,  2001, Phys.\ Rev.\ D., 64, 063001,
  \eprint{astro-ph/0012120}

\bibitem[\protect\citeauthoryear{{Tegmark} et~al.,}{{Tegmark}
  et~al.}{2002}]{Tegmark2002earlysdss}
{Tegmark} M.  et~al., 2002, Astrophys.\ J., 571, 191, \eprint{astro-ph/0107418}

\bibitem[\protect\citeauthoryear{{Tegmark}, {Hamilton}, {Strauss}, {Vogeley} \&
  {Szalay}}{{Tegmark} et~al.}{1998}]{THS1998future}
{Tegmark} M.,  {Hamilton} A.~J.~S.,  {Strauss} M.~A.,  {Vogeley} M.~S.,
  {Szalay} A.~S.,  1998, Astrophys.\ J., 499, 555, \eprint{astro-ph/9708020}

\bibitem[\protect\citeauthoryear{{Tegmark}, {Hamilton} \& {Xu}}{{Tegmark}
  et~al.}{2002}]{T01df}
{Tegmark} M.,  {Hamilton} A.~J.~S.,    {Xu} Y.,  2002, Mon.\ Not.\ Roy.\
  Astron.\ Soc., 335, 887, \eprint{astro-ph/0111575}

\bibitem[\protect\citeauthoryear{{Tegmark}, {Taylor} \& {Heavens}}{{Tegmark}
  et~al.}{1997}]{T97largeklcomp}
{Tegmark} M.,  {Taylor} A.~N.,    {Heavens} A.~F.,  1997, Astrophys.\ J., 480,
  22, \eprint{astro-ph/9603021}

\bibitem[\protect\citeauthoryear{{Thomas}, {Abdalla} \& {Lahav}}{{Thomas}
  et~al.}{2010}]{Thomas2010lrgs1}
{Thomas} S.~A.,  {Abdalla} F.~B.,    {Lahav} O.,  2010, Mon.\ Not.\ Roy.\
  Astron.\ Soc., \eprint{1011.2448}

\bibitem[\protect\citeauthoryear{{Thomas}, {Abdalla} \& {Lahav}}{{Thomas}
  et~al.}{2011}]{Thomas2010lrgs2}
{Thomas} S.~A.,  {Abdalla} F.~B.,    {Lahav} O.,  2011, Phys.\ Rev.\ Lett.,
  106, 241301, \eprint{1012.2272}

\bibitem[\protect\citeauthoryear{{Vogeley} \& {Szalay}}{{Vogeley} \&
  {Szalay}}{1996}]{VS96klcomp}
{Vogeley} M.~S.,  {Szalay} A.~S.,  1996, Astrophys.\ J., 465, 34,
  \eprint{astro-ph/9601185}

\bibitem[\protect\citeauthoryear{{Wandelt}, {Hivon} \& {G{\'o}rski}}{{Wandelt}
  et~al.}{2001}]{WHG00}
{Wandelt} B.~D.,  {Hivon} E.,    {G{\'o}rski} K.~M.,  2001, Phys.\ Rev.\ D.,
  64, 083003, \eprint{astro-ph/0008111}

\bibitem[\protect\citeauthoryear{{Xia}, {Baccigalupi}, {Matarrese}, {Verde} \&
  {Viel}}{{Xia} et~al.}{2011}]{Xia2011sdssqsocell}
{Xia} J.-Q.,  {Baccigalupi} C.,  {Matarrese} S.,  {Verde} L.,    {Viel} M.,
  2011, Journal of Cosmology and Astroparticle Physics, 8, 33,
  \eprint{1104.5015}

\bibitem[\protect\citeauthoryear{{Xia}, {Bonaldi}, {Baccigalupi}, {De Zotti},
  {Matarrese}, {Verde} \& {Viel}}{{Xia} et~al.}{2010}]{Xia2010sdssqsoctheta}
{Xia} J.-Q.,  {Bonaldi} A.,  {Baccigalupi} C.,  {De Zotti} G.,  {Matarrese} S.,
   {Verde} L.,    {Viel} M.,  2010, Journal of Cosmology and Astroparticle
  Physics, 8, 13, \eprint{1007.1969}

\bibitem[\protect\citeauthoryear{{Xia}, {Viel}, {Baccigalupi} \&
  {Matarrese}}{{Xia} et~al.}{2009}]{Xia2009highzisw}
{Xia} J.-Q.,  {Viel} M.,  {Baccigalupi} C.,    {Matarrese} S.,  2009, Journal
  of Cosmology and Astroparticle Physics, 9, 3, \eprint{0907.4753}

\end{thebibliography}
}
\normalsize

\appendix

\section{Pixelisation, band-limit and smoothing issues}\label{app:bandlimitandsmoothing}


The equations defining the PCL and QML estimators do not impose maximum multipoles for the reconstruction, the coupling or the covariance matrices. Such bounds are called band-limits, and are in practice imposed by the finite information content of $\vec{x}$, which depends on the resolution of the map and any additional operations such as smoothing. In particular, pixelising the signal $x$ into a (full or cut-sky) map $\vec{x}$ induces a distortion in the power spectrum estimates, which we parametrise as \equ{
	\mathcal{C}_\ell^{\rm pix} = b^2_\ell \mCl,
}
where $\mathcal{C}_\ell^{\rm pix}$ and $\mCl$ are the power spectra of $\vec{x}$ and $x$ respectively. The continuous map $x$ is usually not accessible, but can be approximated using a high-resolution pixelisation, which in practice will correspond to the highest resolution at which the data are available (and will only require a small correction of the pixelisation-induced bias as detailed below). $\vec{x}$ will then refer to a smoothed, lower-resolution map constructed from $x$, which will be used to estimate the power spectrum on a specific range of scales. This approach is motivated by the complexity of the PCL and QML pixel-space estimators which depends on the number of pixels in the mask. It is usually desirable to use the lowest resolution for which the power spectrum can be accurately estimated in the range of multipoles of interest.  

\begin{figure}\centering
\setlength{\unitlength}{.5in}
\begin{picture}(9,3)(0,0)
\put(0.3,0.1){\includegraphics[trim = 3.2cm 13.3cm 5.2cm 8.7cm, clip, width=8cm]{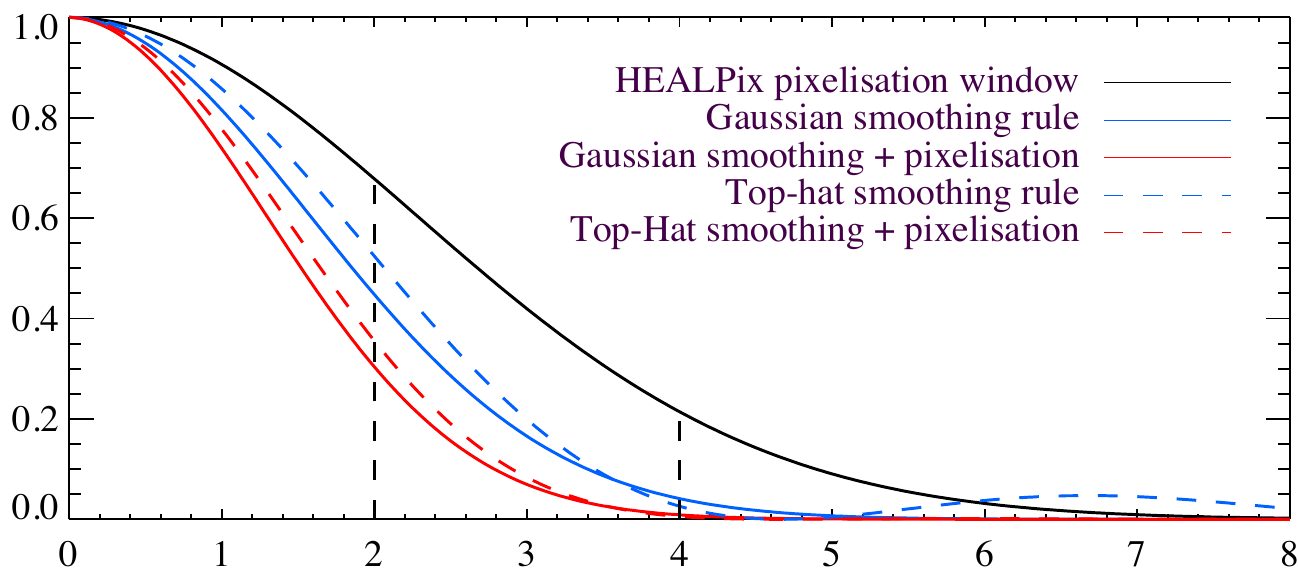}}
\put(3.1,-0.1){\footnotesize$\ell \ /\ \nside$}
\put(0.0,1.2){\rotatebox{90}{\footnotesize Beam $b_\ell^2$}}
\put(3.5,1.3){$\ellmax$ reconstruction $= 2\nside$}
\put(3.9,1.05){$\ellmax$ covariance $= 4\nside$}
\end{picture}
\caption{Beams that apply to the power spectrum estimates of pixelised, smoothed maps at \textsc{healpix} resolution $\nside$. The beam must be accounted for in the construction of the model covariance matrices, and inverted when comparing the estimates with theory.} 
\label{fig:beam}
\end{figure}

The beam $b_\ell$ is decomposed into a pixelisation-induced part $b_\ell^{\rm pix}$ and a smoothing part $b_\ell^{\rm pix}$. In the previous sections we have assumed that pixelisation distortions were negligible, i.e., $b_\ell^{\rm pix} \approx 1$ for all $\ell$, but this assertion is true for a mode $\ell$ only if the pixels of $\vec{x}$ are small compared with $180 / \ell$ degrees. We considered the \textsc{healpix} pixelisation of the sphere, where a map at resolution $\nside$ has $12{\rm N}^2_{\rm side}$ equal-area pixels. In terms of its effect on the power spectrum, pixelising a signal at resolution $\nside$ is well approximated by smoothing with a Gaussian kernel of full width at half maximum (FWHM) $41.7 / \nside$ degrees. This beam, shown in Fig.~\ref{fig:beam}, smoothly decays as $\ell$ increases, and imposes an effective band-limit of $\ellmax = 7\nside$ on the pixelised map.  However, it is well-known that the accessible multipoles for a map at \textsc{healpix} resolution $\nside$ lie within $\ell \in [0, 2\nside]$. Higher multipoles are not accessible because the integrals in the spherical harmonics and Legendre transforms are not accurately approximated by matrix multiplications at this resolution. It is essential to smooth the initial map before degrading it in order to avoid a mismatch between the band-limits, while insuring that the power spectrum can be reconstructed up to $\ellmax = 2\nside$. We investigated this issue and found that band-limiting the map at $\ellmax = 4\nside$ gives optimal performance. This can be realised by a Gaussian smoothing of FWHM $60 / \nside$ degrees, as illustrated in Fig.~\ref{fig:beam}. 

However, the smoothing procedure introduces information from outside the mask into the map, hence biasing the estimates \citep{Aurich2011smoothingbias, Copi2011smoothingbias, FPP11}. This effect depends on the resolution of the map and the shape of the mask, and can significantly affect the power spectrum on a wide range of scales. To avoid this smoothing-induced contamination, the mask must be extended. A systematic procedure is to smooth the complement of the mask, i.e., $1 - \vec{m}$, and keep the pixels below a certain threshold ($0.01$ in this work). But because the Gaussian kernel is band-limited in harmonic space but not compact in pixel space, the smoothing will always leak contamination signal into the data. Also, depending on the threshold, the mask extension might be large and thus significantly increase the variance of the estimates. This can be resolved by using a top-hat smoothing of diameter $90 / \nside$ degrees, which band-limits the data at $\ellmax = 4\nside$ similar to the Gaussian smoothing.  The top-hat kernel is compact and the mask extension is smaller and requires no threshold. However its extent in harmonic space is infinite (in other words it is not band-limited), as shown in Fig.~\ref{fig:beam}, which can introduce further contamination in the map due to approximated smoothing. This can be avoided by performing the top-hat smoothing at high-resolution with a large band-limit. Alternatively, the smoothing can easily be performed in pixel space through explicit convolution since the kernel has a simple shape and the procedure can be parallelised.  

In conclusion, the power spectrum of a high-resolution signal can be estimated from a map at lower resolution $\nside$ provided that the band-limits, the pixelisation and smoothing-induced biases are correctly handled. Before degradation, the initial map is smoothed with a Gaussian kernel of FWHM $60 / \nside$ degrees or a top-hat kernel of diameter $90 / \nside$ degrees, and the mask must be extended accordingly to minimise the contamination due to smoothing. The estimation can then be performed on the low resolution map in the range $\ell \in [0, 4\nside]$ and the signal covariance matrix created with a prior $b^2_\ell C_\ell$ up to $\ellmax = 4\nside$ in order to incorporate all the information in the data. Dividing by the beam $b^2_\ell$ leads to accurate unbiased band-power estimates of $C_\ell$ in the range $\ell \in [0, 2\nside]$.

\section{Karhunen-Lo\`eve compression demystified}\label{app:klcomp}

The Karhunen-Lo\`eve (KL) transform \citep{VS96klcomp, T97largeklcomp, THS1998future, T01df} refers to finding a basis in which the transformed data pixels are statistically orthogonal with respect to a prior on the power spectrum. The transformation matrix $\mat{B}$ satisfies
\equ{
	\vec{\tilde{x}} = \mat{B} \vec{\tilde{y}}.
}
such that 
\equ{
	\bra \tilde{y}_n \tilde{y}^*_{n'} \ket = \lambda_n \delta_{nn'},
}
and the columns $\vec{B}_n$ are the eigen-vectors of the pixel-pixel covariance matrix, i.e.,
\equ{
	\tilde{\mat{C}} \vec{B}_n = \lambda_n \vec{B}_n.
}
Hence $\vec{\tilde{x}}$ and $\vec{\tilde{y}}$ retain the same information and KL compression is equivalent to a principal component analysis performed through singular value decomposition (SVD) of the pixel-pixel covariance matrix. However, this approach is only optimal if the noise covariance is diagonal. Anisotropic noise requires a so-called prewhitening operation, changing the eigen-problem to be solved into
\equ{
	\tilde{\mat{S}} \vec{B}_n = \lambda_n \tilde{\mat{N}} \vec{B}_n,
}
where the eigen-vectors now diagonalise both the signal and the noise covariance matrices. If the transformation is renormalised such that $\vec{B}_n^\dagger \mat{N} \vec{B}_n = 1$ before transforming the map $\vec{x}$,  the  coefficients $\lambda_n$ can be interpreted as signal-to-noise ratios, i.e.,
\equ{
	\bra \tilde{y}_n \tilde{y}^*_{n'} \ket =  \delta_{nn'}(1+\lambda_n).
}
The power spectrum is invariant under rotation in the isotropic case, it can be estimated from $\vec{y}$ provided that the matrices $\tilde{\mat{C}}$ and $\Pl$ are in the same coordinate system. In the KL basis the most informative contributions to the spectrum explicitly come from the first KL modes, as they correspond to high SNR modes. Hence KL compression is often used to remove the noisiest modes and speed up power spectrum estimation, in particular for the matrix inversion in the QML estimator. However, this gain is usually offset by the cost of the preliminary SVD required to compute the KL transformation matrix. 

Moreover, although removing the noisiest modes leaves the PCL and QML estimates unbiased, it increases their variance and potentially impacts the estimates on a wide range of scales. In particular, assuming a constant diagonal noise,  the KL modes calculated from a full sky covariance matrix ${\mat{S}}$ directly relate to the theory spectrum. The largest modes correspond to the largest $C_\ell$'s with multiplicity $2\ell + 1$. On the cut sky, this degeneracy is broken and each mode uniquely relates to a linear combination of $C_\ell$'s with finite support peaking at the previous full sky value (thus conserving the order of the modes). Small masks increase the scope of this combination, and removing small KL modes can thus impact a wide range of multipoles. Logically, considering the CMB spectrum with a typical CMB mask shows that the lowest KL modes, the noisiest, relate to high $\ell$ multipoles, and removing them leaves the low $\ell$ power spectrum estimates unchanged. However for a galaxy survey the theory spectrum is flatter and the mask larger, and the noisiest modes then relate to a wide range of multipoles. Therefore, removing them also impacts the largest scales. Fortunately this effect is reduced when performing the estimation in bins, and the KL compression has a negligible impact on the quality and the variance estimates. 

We did not use KL compression in the context of our analyses because the gain in computer time is marginal. We were able to run the QML estimator at $\nside=64$ without any approximation.

\section{$\chi^2$ for band-power estimates}\label{app:chi2}

\begin{figure}\centering
\setlength{\unitlength}{.5in}
\begin{picture}(9,2.3)(0,0)
\put(0.2,0.0){{\includegraphics[trim = 2cm 1.6cm 9cm 14.5cm, clip, width=8.2cm]{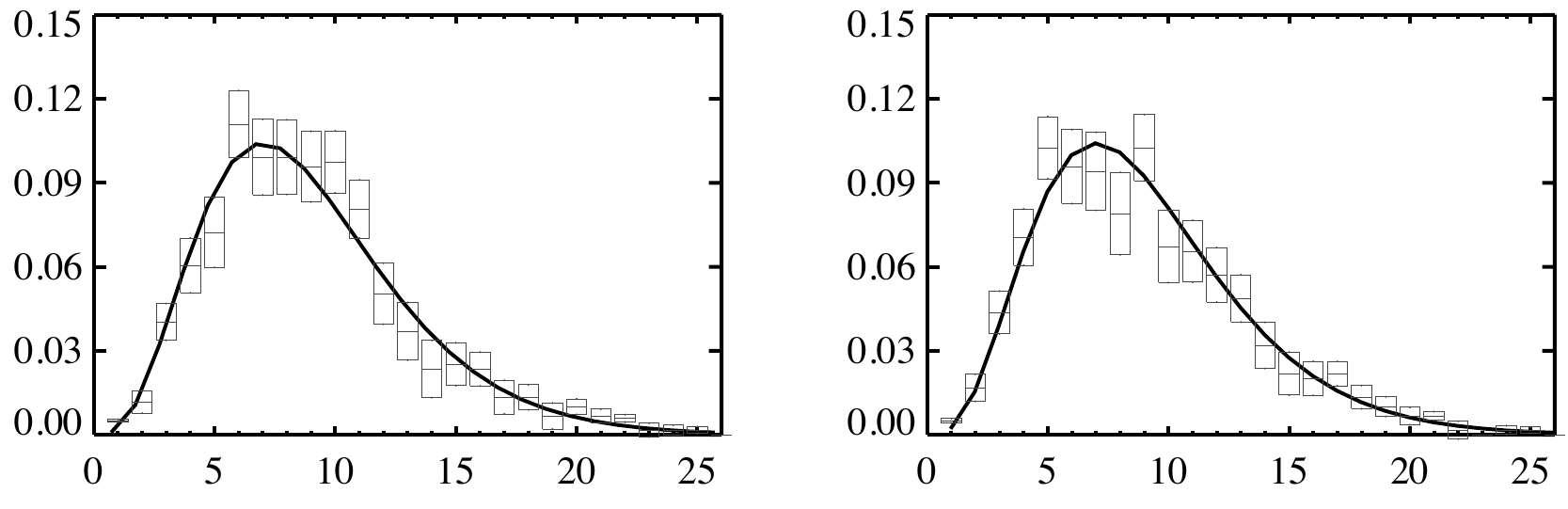}}}
\put(1.7,-0.1){\footnotesize{$\chi^2$}}
\put(-0.05,0.95){\rotatebox{90}{\footnotesize{$p(\chi^2)$}}}
\put(1.1,2.1){\footnotesize{PCL estimates}}
\put(5.05,-0.1){\footnotesize{$\chi^2$}}
\put(3.35,0.95){\rotatebox{90}{\footnotesize{$p(\chi^2)$}}}
\put(4.55,2.1){\footnotesize{QML estimates}}
\end{picture}
\caption{Histograms of the $\chi^2$ values of the PCL (left) and QML (right) estimates of the 600 CMASS mock catalogues. We used multipole bins of size $\Delta \ell =11$, and the theory prediction from \textsc{camb sources}. The boxes show the Poisson errors due to sample variance. The histograms were normalised and compared with the theoretical $\chi^2$ probability distributions (thick lines). Their good agreement demonstrates the validity of the $\chi^2$ given by Eq.~\ref{chi2meas}.} 
\label{fig:cmasschi2}
\end{figure}

To compare power spectrum measurements with theory predictions we used a $\chi^2$ defined as
\equ{
	\chi^2 = \sum_{bb'} (\hat{C}_b - C_b) \mat{V}^{-1}_{bb'} (\hat{C}_{b'} - C_{b'}) \label{chi2meas}
}
where $\hat{C}_b$ denotes the (PCL or QML) band-power estimates and $C_{b'}$ the theory band-powers, constructed from the theory power spectrum $C_\ell$ (calculated from \textsc{camb\_sources}) using the window functions $W_{b\ell}$ defined in Eqs.~(\ref{binnedexpectedvalue}) and (\ref{windowfcts}). The covariance matrix $\mat{V}$ is calculated using Eq.~(\ref{varianceestimates}) and is equal to the inverse of the Fisher matrix when the QML estimator is used. Note that the shot noise $\frac{1}{\bar{G}}$ must be subtracted from the auto-spectrum estimates $\hat{C}_b$.

It is well known that observed power spectra $\mCl$ as calculated by Eq.~(\ref{harmps}) are only described by Gaussian statistics at high-$\ell$ when the central limit theorem applies. For low-$\ell$ estimates, one must resort to alternative likelihood functions (see, e.g., \citealt{Jaffe1999cmblikelihoods, HamimecheAndLewis2009}). However, when estimating band-powers rather than individual multipoles, these effects can be neglected, and a Gaussian likelihood can be used to compare the estimates with theory band-powers. To illustrate this point, Fig.~\ref{fig:cmasschi2} shows the $\chi^2$ values for the PCL and QML estimates of the 600 CMASS mock catalogues with bin size $\Delta \ell = 11$ in the range [2,120], corresponding to $11$ degrees of freedom. The normalised histograms follow the theoretical $\chi^2$ distributions, demonstrating the validity of a Gaussian likelihood, as expected when using band-powers.

\end{document}